\documentclass[12pt]{article}
\usepackage{axodraw,cite,graphicx} % to put in axodraw
\usepackage{latexsym,amssymb,epsf} % don't remember if these are
\newcommand{\newc}{\newcommand}
\newc{\JC}{{\bf J}}
\newc{\gev}{\,GeV}
\newc{\ra}{\rightarrow}
\newc{\lra}{\longrightarrow}
\newc{\rpv}{$\mathrm{\not\!R_p}$}
\newc{\rp}{$\mathrm{R_p}$}
\newc{\real}{\mathcal{R}e}
\newc{\alsm}{{\displaystyle \sum_{\alpha=1,2}}}
\newc{\besm}{{\displaystyle \sum_{\beta=1,2}}}
\newc{\al}{\alpha}
\newc{\be}{\beta}
\newc{\ga}{\gamma}
\newc{\de}{\delta}
\newc{\cw}{\cos\theta_w}
\newc{\ssw}{\sin^2\theta_w}
\newc{\cbe}{\cos\beta}
\newc{\sbe}{\sin\beta}
\newc{\sh}{\hat{s}}
\newc{\uh}{\hat{u}}
\newc{\tha}{\hat{t}}
\newc{\sa}{\sin\al}
\newc{\ca}{\cos\al}
\newc{\eg}{{\it e.g.\ }}
\newc{\bv}{$\mathrm{\not\!B}$}
\newc{\lv}{$\mathrm{\not\!L}$}
\newc{\beq}{\begin{equation}}
\newc{\eeq}{\end{equation}}
\newc{\barr}{\begin{eqnarray}}
\newc{\earr}{\end{eqnarray}}
\newc{\ie}{{\it i.e.\/}\ }
\newc{\lam}{\lambda}
\newc{\cht}{\tilde{\chi}}
\newc{\upt}{\tilde{u}}
\newc{\elt}{\tilde{\ell}}
\newc{\nut}{\tilde{\nu}}
\newc{\dnt}{\tilde{d}}
\newc{\mr}{\mathrm}
\newc{\me}{\bf{M}}
\newc{\eq}[1]{Eqn.\,\ref{eqn:#1}}
\newc{\eqs}[2]{Eqns.\,\ref{eqn:#1},\ref{eqn:#2}}
\newc{\lab}[1]{\label{eqn:#1}}

\begin{document}
\begin{titlepage}
\vspace{-2cm}
\begin{flushright}
	OUTP-99-26P\\
	RAL-TR-1999-080\\
	hep-ph/9912407\\
\end{flushright}
\vspace{+2cm}
\begin{center}
 {\Large{\bf Parton-Shower Simulations of R-parity Violating
Supersymmetric  Models}}\\
        \vskip 0.6 cm {\large{\bf 
               H. Dreiner$^*$\footnote{E-mail address: dreiner@v2.rl.ac.uk},
               P. Richardson$^{\dagger}$\footnote{E-mail address: 
                                       P.Richardson1@physics.ox.ac.uk},
               and M.H. Seymour$^*$\footnote{E-mail address: 
                                M.Seymour@rl.ac.uk}}}
        \vskip 1cm
{{\it $^*$ Rutherford Appleton Laboratory, Chilton, Didcot OX11 0QX, U.K.}}\\
{{\it $^{\dagger}$ Department of Physics, Theoretical Physics,
                    University of Oxford}}\\
{\it 1 Keble Road, Oxford OX1 3NP, United Kingdom}\\
\vskip 0.2 cm
\end{center}
\begin{abstract}\noindent
  We study the colour connection structure of R-parity violating
  decays and production cross sections, and construct a Monte Carlo
  simulation of these processes including colour coherence effects. We
  then present some results from the implementation of
  these processes in the HERWIG Monte Carlo event generator. We
  include the matrix elements for the two-body sfermion and three-body 
  gaugino and gluino decays as well as the two-to-two resonant 
  hard production processes in hadron-hadron collisions.
\end{abstract}
\end{titlepage}
% reset the footnote numbers
\setcounter{footnote}{0}

\section{Introduction}
In the past few years there has been a large amount of interest in R-parity 
violating (\rpv) supersymmetric (SUSY) models, motivated by the possible 
explanations of various experimental discrepancies, \eg
\cite{Grant:1996bc,Dreiner:1996dd,Carena:1997xu,Chankowski:1996mx,Choudhury:1996gd,Dreiner:1997cd,Altarelli:1997ce,Choudhury:1997dt,Kalinowski:1997fk}.
It has become clear that if we are to explore all possible channels
for the discovery of supersymmetry then \rpv\    models must be
investigated. For a recent review on R-parity violation see \cite{Dreiner:1997uz}.

In the Minimal Supersymmetric Standard Model (MSSM) a discrete multiplicative symmetry, R-parity (\rp) is
imposed, 
\beq 
\mr{R_p} = (-1)^{3B+L+2S}, 
\eeq 
where $B$ is the baryon number, $L$ the lepton number and $S$ the spin of
the particle. All the Standard Model  particles have $\mr{R_p} =
+1$ and their super-partners have $\mr{R_p} = -1$. The conservation of R-parity
forbids the terms in the superpotential which violate baryon or
lepton number 
% Foot note on conventions
\footnote{It should be noted that some authors choose to
define this superpotential without the factors of one half in the LLE
and UDD terms. This will lead to differences in the Feynman rules, but
the results with this second convention can always be obtained by
taking $\lam$ or $\lam''$ to be twice their value in our convention.}
% End of footnote
\beq
{\bf W_{\not R_p}} =
\frac{1}{2}\lam_{ijk}\varepsilon^{ab}L_{a}^{i}L_{b}^{j}\overline{E}^{k}
+ \lam_{ijk}'\varepsilon^{ab}L_{a}^{i}Q_{b}^{j}\overline{D}^{k} +
\frac{1}{2}\lam_{ijk}''\varepsilon^{c_1c_2c_3}\overline{U}_{c_1}^{i}
\overline{D}_{c_2}^{j}\overline{D}_{c_3}^{k} + \kappa_iL_{i}H_2,
\label{eqn:super} 
\eeq 
where $i,j,k=1,2,3$ are the generation indices, $a,b=1,2$ are the
$SU(2)_L$ indices and $c_{1,2,3}=1,2,3$ are the $SU(3)_C$
indices. $L^i$ ($Q^i$) are the lepton (quark) $SU(2)$ doublet
superfields, $\overline{E}^{i}$ ($\overline{D}^{i}, \overline{U}^{i}$)
are the electron (down and up quark) $SU(2)$ singlet superfields, and
$H_n,$ $n=1,2$, are the Higgs superfields. We shall neglect the last
term in Eqn.\,\ref{eqn:super} which mixes the lepton and Higgs $SU(2)$
doublet superfields. For a recent summary of the bounds on the
couplings in Eqn.\,\ref{eqn:super} see \cite{Allanach:1999ic}.

This superpotential gives interactions which violate either lepton or
baryon number. For example the first term will give interactions of a
slepton and two leptons which violates lepton number. The third term
gives an interaction of two quarks and squark, which violates baryon
number. When combined with the MSSM superpotential there are also
terms involving the interactions of three sleptons/squarks and a Higgs
which violate either lepton or baryon number.

\rp\  is imposed in the MSSM to avoid the simultaneous presence of the
second two terms in Eqn.\,\ref{eqn:super}. These lead to fast proton
decay, in disagreement with the experimental lower bounds on the proton
lifetime. However in order to guarantee proton stability it is
sufficient to forbid only one set of these terms. This is achieved for
example by lepton parity 
\barr
 (L^i,\bar{E}^i) &\ra& - (L^i,\bar{E}^i), \\
 (Q^i,\bar{U}^i,\bar{D}^i,H_1,H_2) &\ra& (Q^i,\bar{U}^i,\bar{D}^i,H_1,H_2),
\earr
which allows the third term in Eqn.\,\ref{eqn:super} but forbids the
remaining terms. Thus baryon number is violated (\bv) but lepton
number is conserved and the proton is stable. Similarly there are
symmetries such that lepton number is violated and baryon number is
conserved. This also prevents proton decay. Both cases lead to very
different phenomenology from the MSSM.

In the MSSM the conservation of \rp\  implies that
\begin{enumerate}
 \item Sparticles are produced in pairs.
 \item The lightest supersymmetric particle (LSP) is stable.
 \item Cosmological bounds on electric- or colour-charged stable
 relics imply that a stable LSP must be a neutral colour singlet
 \cite{Ellis:1984ew}.
\end{enumerate}
  
However, in the case of \rpv\   models we can have
\begin{enumerate}
 \item Single sparticle production.
 \item The LSP can decay. As the LSP is unstable it does not have to
   be a neutral colour singlet. It can be any supersymmetric particle.
 \item Lepton or baryon number is violated.
\end{enumerate}

In the MSSM, as the LSP is stable, the experimental signatures of SUSY
processes typically involve missing transverse energy in collider
experiments. However, if R-parity is violated, and the LSP decays in
the detector, the missing energy signatures of the MSSM no
longer apply or are severely diluted.
 It therefore requires a different experimental search
strategy. In particular, in the \bv\  case, where the final state is
predominantly hadronic, it may be hard to extract a signal over the
QCD background in hadron colliders.

Despite the interest in \rpv\   and the potential experimental problems,
there have been few experimental studies at hadron colliders. The
first systematic study of \rpv\   signatures at hadron colliders was
presented in \cite{Dreiner:1991pe}. More recent overviews of the
search potential at the LHC and Run II of the Tevatron have been
presented in \cite{Barbier:1998fe,Allanach:1999bf}. These studies have
been limited by the fact that few simulations have been available. In
hadron-hadron collisions the only available Monte Carlo event
generator is ISAJET \cite{Paige:1998xm} where the \rpv\  decays can be
implemented using the \texttt{FORCE} command, \ie the decay mode of a given
particle, \eg the LSP, can be specified by hand.  However there has
been no simulation which includes all the decay modes and the single
sparticle production processes.

Here we present the calculations required to produce a Monte Carlo
event generator for the two-body sfermion and three-body gaugino and
gluino \rpv\  decay modes as well as all two-to-two \rpv\  resonant
production processes in hadron-hadron collisions. We have only
included those production processes where a resonance is possible, so
for example processes which can only occur via a $t$-channel diagram are
not included. However where a process can occur via a resonance all
the diagrams including non-resonant $s$-channel and $t$-channel diagrams
have been included. We also discuss colour coherence effects via the
angular ordering procedure (which we describe in detail below), and
some preliminary results from the implementation of these processes in
the HERWIG Monte Carlo event generator \cite{Marchesini:1991ch}.  
Details of the implementation of supersymmetric processes with and 
without \rpv\  can be found in \cite{SUSYimplement}.

After a general discussion of the angular ordering procedure in the
Standard Model 
in Section~\ref{sec-colour} we discuss the extension to the \rpv\   
decays and hard production processes in Section~\ref{sec-angular}.  In
Section~\ref{sec-hadron} we describe the hadronization procedure which we
adopt for the \rpv\   processes. We then present some preliminary results
 of the Monte Carlo
simulation in Section~\ref{sec-results}. We have made new calculations of all the
necessary matrix elements, and include them as an appendix.

%
%  Section on the monte-carlo event generator
%
\section{Monte Carlo Simulations}
\label{sec-colour}
  In general a Monte Carlo event generator, for a process involving at least
 one hadron, consists of three parts.
\begin{enumerate}
 \item A hard scattering process, either of the incoming fundamental
 particles in lepton collisions or of a parton extracted from a hadron
 in hadron initiated processes.
 \item A parton-shower phase where the partons coming into or leaving the
 hard process are evolved according to perturbative QCD.
 \item A hadronization model in which the partons left at the end of the
 parton-shower phase are formed into the hadrons which are observed. For
 processes with hadrons in the initial state after the removal of the
 partons in the hard process we are left with a hadron remnant. This remnant is
 also formed into hadrons by the hadronization model. 
\end{enumerate}
  We now discuss these three stages in turn.
%
% First subsection on the hard-process this should be longer??
%
\subsection{Hard Scattering}

The hard scattering process is described by a matrix element
calculated to a fixed order in perturbation theory, usually
only leading order. The momenta of
the particles involved in the collision can then be generated
according to the matrix element. The Monte Carlo event generator then
needs to take the results of this perturbative calculation, at a high
scale, and generate the hadrons which are observed.

%
% Second subsection on the parton-shower phase
% 
\subsection{Parton Shower}

In a scattering process the incoming or outgoing partons can emit QCD
radiation, \eg $q\ra gq$ and $g\ra gg$, or split into quark-antiquark
pairs, $g\ra q\bar{q}$. A full perturbative treatment of this part
of an event is not possible. (If it were possible it would be included
in the hard scattering matrix element.) We must therefore make an
approximation and focus on the dominant contributions in the showering
process. The emission of QCD radiation is enhanced for (a) Collinear
Emission and for (b) Soft Emission. We discuss these two below in more
detail. Our approximation will consist in focusing on these enhanced
regions of radiation. We then discuss this in an
explicit example.
% this is to get the labels a,b etc that Herbi appears to want
\renewcommand{\theenumi}{\alph{enumi}}
\renewcommand{\labelenumi}{(\theenumi)}
\begin{enumerate}
 \item Collinear Emission
\nopagebreak[3]

If we consider the emission of QCD radiation in the collinear limit
then, after azimuthal averaging, the {\it cross section\/} obeys a
factorization theorem \cite{Webber:1986mc}. This can be understood as
follows, the cross section for a process in which one parton pair is
much more collinear than any other pair can be written as the
convolution of a universal, \ie process independent, splitting
function and the cross section for the same process where the
collinear pair is replaced by a single parton of the corresponding
flavour. Due to this functional form we can then apply this procedure
to the next most collinear pair in the final state, and so on. We thus
have an iterative rule which leads to a description of multi-parton
final states as a Markov chain \cite{Webber:1986mc}.  This can be
viewed as an evolution in some energy-like scale, such as the
virtuality, where a parton at high scale is evolved by successive
branchings to a lower scale. However, the collinear factorization does
not specify what the evolution variable should be, \ie it has the same
form for any variable proportional to the virtuality, \eg the
transverse momentum. This iterative procedure then correctly
resums the leading collinear singularities to all orders in
perturbation theory \cite{Webber:1986mc}.

\item Soft Emission 
\nopagebreak[3]

For the emission of QCD radiation in the soft limit, a factorization
theorem exists for the {\it amplitude\/} of the process. The amplitude
for a process in which one gluon is much softer than the other energy
scales in the process can be written as a product of a universal
eikonal current and the amplitude for the same process without the
soft gluon. After we square the amplitude and sum over the spins of
the external partons, we obtain a result which depends on the momenta
of all the external partons. It therefore seems unlikely that a Markov
description based on sequential parton splittings can be recovered.
The surprising result \cite{Marchesini:1984bm,Marchesini:1988cf} is
that, after azimuthal averaging, these effects {\it can\/} be
incorporated into a collinear algorithm by using the correct choice
for the evolution scale, namely the opening angle.
\end{enumerate}
% this should put things back to normal after the a,b thing
\renewcommand{\theenumi}{\arabic{enumi}}
\renewcommand{\labelenumi}{\theenumi.}

%
% Feynman diagram and colour flow for e^+e^-\raq\bar{q}g
%
\begin{figure}[htp]
\begin{center} \begin{picture}(360,120)(50,0)
\SetScale{1.0}
% Feynman diagram
\SetOffset(50,-10)
\ArrowLine(5,30)(50,60)
\ArrowLine(50,60)(5,90)
\Photon(50,60)(100,60){5}{5}
\ArrowLine(100,60)(150,100)
\ArrowLine(150,20)(100,60)
\Gluon(125,80)(150,80){-3}{3}
% Colour Flow
\ArrowLine(250,65)(300,120)
\ArrowLine(300,65)(250,65)
\DashArrowLine(300,0)(250,55){5}
\DashArrowLine(250,55)(300,55){5}
% Labels
\Text(30,85)[]{$\mr{e}^+$}
\Text(30,35)[]{$\mr{e}^-$}
\Text(75,75)[]{$\mr{Z}^0/\gamma$}
\Text(130,95)[]{$\mr{q}$}
\Text(130,25)[]{$\mr{\bar{q}}$}
\Text(310,-5)[]{$\mr{\bar{q}}$}
\Text(310,60)[]{g}
\Text(310,125)[]{q}
\end{picture}
\end{center}\
\caption{Feynman diagram and colour flow for $\mr{e^+e^- \ra q \bar{q} g}$.}
\label{fig:qqg}
\end{figure}
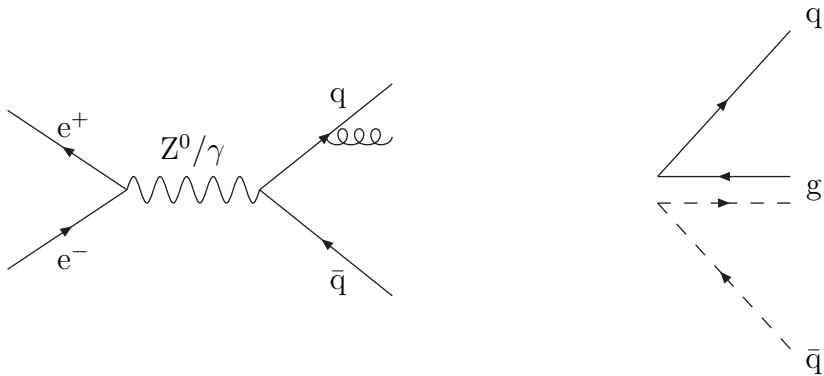
% End of the Figure

\subsubsection{Example: $\mathbf{e^+e^-\ra q{\bar q}gg}$}
\label{sec:example}
We can illustrate this with a simple example, \ie the process $\mr
{e^+e^- \ra q\, {\bar q}\, g_1}$, shown in Fig.\,\ref{fig:qqg}. The
semi-classical eikonal current can be used to study the emission of an
extra soft gluon in this process, \ie the process $\mr{e^+e^-\ra q \,
\bar{q} \, g_1 \, g_2}$ where the second gluon is much softer than the
other partons. The matrix element including the emission of the extra
soft gluon is given by
\beq
    {\bf {M}} (k_1,k_2,p_1,p_2,p_3;q) = g_s {\bf m}
   (k_1,k_2,p_1,p_2,p_3) \cdot {\bf J}(q)
\eeq
where
\begin{itemize}

  \item ${\bf m}(k_1,k_2,p_1,p_2,p_3)$ is the tree-level amplitude for the
        underlying process, \linebreak  
        $\mr{e^+}(k_1)\mr{e^-}(k_2) \ra  \mr{q}(p_1) 
        \mr{\bar{q}}(p_2) \mr{g_1}(p_3)$.

  \item ${\bf {M}} (k_1,k_2,p_1,p_2,p_3;q)$ is the matrix element for the
        process \linebreak $\mr{e^+}(k_1) \mr{e^-}(k_2) \ra 
                 \mr{q}(p_1) \mr{\bar{q}}(p_2) \mr{g_1}(p_3)  \mr{g_2}(q)$
        , \ie including the emission of an extra soft
        gluon, $\mr{g_2}$, with momentum $q$.

\item $\JC(q)$ is the non-Abelian semi-classical current for the
        emission of the soft gluon with momentum $q$, from the hard
        partons.

\item $g_s$ is the strong coupling constant.
\end{itemize}

Explicitly in our example, the current, ${\bf J}(q)$, is given by 
\cite{Marchesini:1988cf}
\beq 
    \JC(q) = \sum_{s=1,2}  \JC^{b,\mu}(q) \varepsilon_{\mu,s},
\eeq
where  $\varepsilon_{\mu,s}$ is the polarization vector of the
gluon and 
\beq  
   \JC^{b,\mu}(q) = 
                {\bf t}^{b,\mr{q}}_{c_1c_1'}{\bf t}^a_{c_1'c_2}
                                      \left(\frac{p^\mu_1}{p_1 \cdot q}\right)
                +{\bf t}^{a}_{c_1c_2'}{\bf t}^{b,\mr{\bar{q}}}_{c_2'c_2}
                                      \left(\frac{p^\mu_2}{p_2 \cdot q}\right)
                +i f^{aa'b} {\bf t}^{a'}_{c_1c_2} 
                                      \left(\frac{p^\mu_3}{p_3 \cdot q}\right)
\lab{current}
\eeq
where ${\bf t}^{b,q}$ and $f^{abc}$ are the $SU(3)$ colour generators
in the fundamental representation and adjoint representations
respectively.  

\subsubsection*{Radiation Functions}

We now define the radiation functions of \cite{Marchesini:1990yk}.
This will allow us to express the square of the current, \eq{current},
in a useful way. First we define the dipole radiation function
$W_{ij}(\Omega_q)$, which describes the radiation of a soft gluon,
with momentum $q$, from a pair of partons $i$ and $j$. Note that
$W_{ij}(\Omega_q)$ only depends on the direction of $q$, $\Omega_q$,
and not its energy.
\beq
\frac{2}{\omega^2}W_{ij}(\Omega_q) \equiv
- \left( \frac{p_{i}}{p_i
                   \cdot q}-\frac{p_{j}}{p_j\cdot q} \right)^2 =
                   \frac{2}{\omega^2} \left( \frac{\xi_{ij}}{\xi_{i}\xi_{j}}
                   -\frac{1}{2\gamma_{i}^2\xi_{i}^2}
   -\frac{1}{2\gamma_{j}^2\xi_{j}^2} \right)
\eeq
 where
\begin{itemize}
  \item $\omega$ is the energy of the soft gluon, 
  \item ${\xi_{ij}} = \frac{p_i \cdot p_j}{E_{i}E_{j}} = 1-v_{i}v_{j}
                \cos\theta_{ij}$,
  \item ${\xi_{i}} = 1-v_{i} \cos\theta_{i}$
  \item $\gamma_{i} = E_{i}/m_{i} = 1/\sqrt{1-v_{i}^2}$,
  \item $v_{i}$ is the velocity of parton $i$,  
  \item $\theta_{i}$ is the angle between the direction of motion of
         the soft gluon and the parton $i$,
  \item $\theta_{ij}$ is the angle between the partons $i$ and $j$.
\end{itemize}
We can now use the dipole radiation functions to express the current
squared for a given process in the following form,
\beq
   \JC^2(q) = \frac{ C_{\mathbf{m}}}{\omega^2}W(\Omega_q)
\label{eqn:sqcurrent}
\eeq
where $C_{\mathbf{m}}$ is the colour factor for the tree-level
process, and $W(\Omega_q)$ the radiation pattern, given below in terms
of the dipole radiation functions. For the process $\mr{e^+e^-\ra
q\bar{q}g}$ the colour factor $C_{\mathbf{m}} = C_FN_c$ and the
radiation pattern is given by
\beq
W_{q\bar{q}g}(\Omega_q) =   C_A\left[W_{qg}(\Omega_q)
                                +W_{\bar{q}g}(\Omega_q)\right]
                                -\frac1{N_c}W_{q\bar{q}}(\Omega_q)
\eeq 
where $C_F=\frac{N^2_c-1}{2N_c}$ and $C_A=N_c$ are the Casimirs of the
fundamental and adjoint representations respectively, with an
arbitrary number of colours $N_c$.  This corresponds to emission of
the soft gluon from colour dipoles, \ie $W_{qg}$ is emission from the
dipole formed by the quark and the anticolour line of the gluon,
$W_{\bar{q}g}$ emission from the colour line of the gluon and the
antiquark and $W_{q\bar{q}}$ emission from the quark and
antiquark. Note that the $q\bar{q}$ dipole is negative
which is a problem if we wish to use a probabilistic approach to treat
the soft gluon radiation.

The dipole radiation function can then be
split into two parts as was done in \cite{Marchesini:1990yk}, \ie
\beq
   W_{ij}(\Omega_q) = W_{ij}^{i}(\Omega_q) +W_{ij}^{j}(\Omega_q)
\eeq  
  where 
\beq
   W_{ij}^{i} = \frac{1}{2\xi_{i}} \left(
              1-\frac{1}{\gamma_{i}^2\xi_{i}}+\frac{\xi_{ij}-\xi_{i}}{\xi_{j}}
                 \right)
\eeq
  The function $W^i_{ij}$ has the following properties,
\begin{itemize} 
  \item After averaging over the azimuthal angle of the soft gluon about the
        parton $i$ it corresponds to emission in a  cone  about the direction
        of the parton $i$ up to the direction of $j$,
        \cite{Marchesini:1984bm,Marchesini:1988cf}.

  \item If the parton $i$ is massive then soft radiation in the 
        direction of the parton is reduced, 
        \ie emission within an angle of order $\theta \sim m_i/E_i$
        is suppressed, \cite{Marchesini:1990yk}.

  \item In the massless limit it  contains the collinear singularity as
        $\theta_{i} \rightarrow 0 $,\cite{Marchesini:1990yk}, but has no
        collinear singularity as $\theta_{j} \rightarrow 0 $.

  \item While $W_{ij}/\omega^2$ is Lorentz invariant the individual functions
        $W^i_{ij}/\omega^2$ and $W^j_{ij}/\omega^2$ are not.
\end{itemize}
This allows us to rewrite the square of the current, \eq{current}, 
in the following form, using these radiation functions, as
\barr
 W_{q\bar{q}g}(\Omega_q) = & 2 C_F \left( W^q_{qg} +  W^{\bar{q}}_{\bar{q}g} \right) 
                 +C_A \left( W^g_{g\bar{q}} +  W^{g}_{gq} \right)\nonumber\\
                &  +N_c^{-1} \left( W^{q}_{qg} -W^q_{q\bar{q}} 
               + W^{\bar{q}}_{\bar{q}g} -W^{\bar{q}}_{\bar{q}q}
                 \right),  
 \label{eqn:radqqg}
\earr
which should be inserted in Eqn.~\ref{eqn:sqcurrent}. This gives our
main result in this example. The last term in Eqn.\,\ref{eqn:radqqg},
and other terms of this type, can be neglected as it is suppressed by
$1/N^2_c$ with respect to the leading order term, as $C_F,\ C_A
\propto N_c$, and is also dynamically suppressed as it does not
contain a collinear singularity in the massless limit (\eg the
singularity in the quark direction cancels between the $W^q_{qg}$ and
$W^q_{q\bar{q}}$ terms.) Thus part of our approximation for the parton
shower will consist of dropping the $1/N_c^2$ terms.

\subsubsection*{Colour Connected Partons}

We can now define the concept of the colour connected parton. Two
partons are considered to be colour connected if they share the same
colour line. The colour flow, in the large $N_c$ limit, for the
process $\mr{e^+e^- \ra q {\bar q} g}$ is shown in Fig.\,\ref{fig:qqg}
with a dashed and a solid line.  Thus the $\mr{{\bar q}}$ and $\mr{g}$
are colour connected and the $\mr{q}$ and $\mr{g}$ are colour
connected, while the $\mr{{\bar q}}$ and $\mr{q}$ are not colour
connected. Each quark only has one colour connected partner in a given
Standard Model Feynman diagram and each gluon has two. Colour
connected partners are defined at each stage of the iterative parton
showering procedure. If the final state $\mr{q}$ were to emit another
gluon, $\mr{g_2}$, the new final state $\mr{q}$ would be colour
connected to $\mr{g_2}$ and no longer to $\mr{g}$. $\mr{g}$~and
$\mr{g_2}$ would then also be colour connected.

\subsubsection*{Angular Ordered Emission and Colour Coherence}

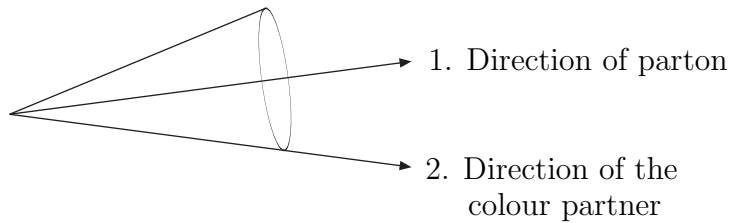
\begin{figure}[htp]
\begin{center} 
\begin{picture}(360,110)(0,0)
\SetOffset(100,-10)
\LongArrow(0,60)(150,79.75)
\LongArrow(0,60)(150,40.25)
\Line(0,60)(96.41,99.94)
%\Line(0,60)(96.41,20.06)
\Oval(100,73.17)(27.01,5)(7.5)
%\Oval(100,46.83)(27.01,5)(-7.5)
\Text(215,80)[]{1. Direction of parton}
\Text(205,40)[]{2. Direction of the}
\Text(210,25)[]{colour partner}
\end{picture}
\end{center}
\caption{Emission in angular ordered cones.}
\label{fig:cones}
\end{figure}

We see from Eqn.\,\ref{eqn:radqqg} that neglecting the final term,
using the properties of the function $W^i_{ij}$, and averaging over
the azimuthal angle of the gluon about a parton, the radiation can
only occur in a cone about the direction of the parton up to the
direction of its colour partner. This is shown in
Fig.\,\ref{fig:cones}. We can draw a cone around parton one with
half-angle given by the angle between the momenta of partons one and
two. The emission from parton one within the cone defined by its
colour connected partner, parton two, is called angular ordered
emission.

The angular ordering procedure is one way of implementing the
phenomenon of colour coherence.  The idea of colour coherence is that
if we consider a large angle gluon it can only resolve the total
colour charge of a pair of smaller angle partons, and not their
individual charges. It is therefore as if the larger-angle soft gluon
was emitted before the smaller angle branchings. There have been a
number of experimental studies of colour coherence effects. In
particular the string effect in $e^+e^-$ collisions 
\cite{Bartel:1984ij:Bartel:1985zx:Aihara:1985du:Akrawy:1991ag:Akers:1995xs},
where there is a suppression of soft QCD radiation between the two
quark jets in three jet events. There have also been studies of colour
coherence effects between the initial and final states in
hadron-hadron collisions, \cite{Abbott:1997bk,Abe:1994nj}.
It is now firmly established that event generators that do not
incorporate colour coherence cannot reliably predict the hadronic
final state.

Although we have averaged over the azimuthal angle of the emission of
the gluon in both the soft and collinear cases, azimuthal effects, \eg
due to spin correlations, can be included
\cite{Knowles:1988hu:Knowles:1988vs:Knowles:1988cu}, after the full
parton shower has been generated.

\subsubsection{Non-Planar Colour Flows}

We have explained in an example how the cross section for $n+1$
partons factorizes in both the collinear and soft limits into a
universal splitting term and the cross section for $n$ partons. Both
of these limits can be implemented by using angles as the evolution
variable in a Markov branching procedure.  We start at the hard
cross section, normally with a two-to-two process. The maximum angle
of emission from a parton is set by the direction of the colour
partner. We then generate some smaller angle parton, \eg a gluon from
a quark. Then we repeat the procedure, \ie the gluon's colour partner
is now the colour partner of the original quark, and its anticolour
partner the quark, and the colour partner of the quark is the
gluon. One of the partons will radiate with the maximum angle
given by the direction of the new colour partner and so on until the
cut-off is reached, of order 1\gev, below which emission does not
occur. This procedure then resums both the leading soft and collinear
singularities.

In processes where there is more than one Feynman diagram it is
possible for the colour flows in the diagrams to be different. This
leads to so called `non-planar' terms from the interference terms,
where the colour flows do not match. These are not positive definite
and hence cannot be interpreted in a probabilistic way for
implementation in the Monte Carlo procedure. They
are always suppressed by inverse powers of $N_c$. A procedure must be
adopted to split up the `non-planar' parts of the tree-level matrix
element to give redefined planar terms with positive-definite
coefficients that can be used in the Monte Carlo procedure. Such a
procedure was proposed in \cite{Marchesini:1988cf} and shown to work
correctly for all QCD processes. However, as shown in
\cite{Odagiri:1998ep}, this is inadequate for MSSM processes and hence
a new procedure was proposed, which we adopt here. In this procedure
the `non-planar' parts of the matrix element are split up according to
\beq |\overline{M}|^2_{full,i} =
  \frac{|\overline{M}|^2_i}{|\overline{M}|^2_{planar}} |
  \overline{M}|^2_{tot}
\label{eqn:kosuke}
\eeq
where $|\overline{M}|^2_i$ is the matrix element squared for the $i$th
colour flow, ${|\overline{M}|^2_{planar}}$ is the sum of the matrix
elements squared for the planar colour flows, and $|\overline{M}|^2_
{tot}$ is the total matrix element squared.  This ensures that the
terms are positive definite and the new full planar terms have the
correct pole structure and sum to the correct total cross section.
This can then be implemented numerically.

% Angular Ordering Figure

In this section we have explained how by using a Markov branching
procedure we can resum both the soft and collinear singularities in QCD.
%
% Sub section on hadronization
%
\subsection{Hadronization}
  
After the parton shower phase it is necessary to adopt some procedure
to combine the quarks and gluons into the observed hadrons. This is done
in the HERWIG event generator using the {\it cluster hadronization
model} \cite{Webber:1984if}. This model is based on the concept of
colour preconfinement. This implies that the
invariant mass of pairs of colour-connected partons has a spectrum
that is peaked at low values, a few times the cut-off used in the
parton-shower, and is universal, \ie independent of the hard scale and
type of the collision, as discussed below and shown in
Fig.\,\ref{fig:cluster}a.

In the cluster hadronization model \cite{Webber:1984if}, after the end
of the parton showering process we are left with gluons and quarks.
The gluons are non-perturbatively split into light quark-antiquark
pairs.
The final state then consists only of quarks and antiquarks which are,
in the planar approximation, uniquely paired in colour-anticolour pairs.
These pairs of colour-connected quarks do not necessarily have
the correct invariant mass to form a meson. Instead they are formed into
colour-singlet meson-like resonance called `clusters'.
 These clusters then decay
in their rest frame to a pair of hadrons (either two mesons or a baryon
and an antibaryon) with the type of hadron
determined by the available phase space. In the original model of
\cite{Webber:1984if} these decays were isotropic in the rest frame of
the cluster, however in the current implementation of the model
\cite{Marchesini:1991ch}, the hadrons containing the quarks from the
perturbative stage of the event continue in the same direction (in the
cluster rest frame) as the original quark.

It is reasonable to assume that the low mass clusters are
superpositions of hadron resonances and can be treated in this way
\cite{Webber:1984if}. However, a fraction of the clusters have higher
masses for which this assumption is not valid and these clusters are
first split using a string-like mechanism \cite{Webber:1984if} into
lighter clusters before they are decayed to hadrons.
   
A simple extension of this model is used for hadron remnants.  For
example in a collision in which a valence quark from a proton
participates in a hard process, the two remaining valence quarks are
left in the final state.  They are paired up as a `diquark' which, in
the planar approximation, carries an anticolour index and can be treated
like an antiquark.  The resulting cluster has baryonic quantum numbers
and decays into a baryon and a meson.
%
%
%
%  Section on angular ordering in RPV
%
\section{Angular Ordering in \boldmath{\rpv}}
\label{sec-angular}

  In Standard Model and MSSM processes apart from
complications involving processes where there are `non-planar' terms 
\cite{Odagiri:1998ep} the angular ordering procedure is relatively straightforward to
implement. However in \rpv\   SUSY there are additional
complications. 

  The lepton number violating processes, which come from
the first two terms in the superpotential, Eqn.\,\ref{eqn:super},
 have colour flows that are
the same as those which occur in the MSSM\@. 
On the other hand the baryon number violating interactions,
which come from the third  term in Eqn.\,\ref{eqn:super}, have a very
different colour structure involving the totally antisymmetric 
tensor, $\epsilon^{c_1c_2c_3}$. We look first at the colour structure of
the various baryon number violating decays which we include
in the Monte Carlo simulation and then at the structure of the various hard scattering
processes. 

\subsection{Decays}

 From the point of view of the colour structure there are three types of baryon
 number violating decays which we include in the Monte Carlo simulation.
\begin{enumerate} 
 \item Two-body \bv\  decay of an antisquark to two quarks or a
 squark to two antiquarks.
\item Three-body \bv\  decay of a colourless sparticle, \ie a
        neutralino or a chargino, to three quarks or antiquarks.
\item Three-body \bv\  decay of the gluino to three quarks or antiquarks.
\end{enumerate}

In general it is possible to consider for example the decay of a
neutralino to three quarks as either a three-body decay or two
sequential two-body decays, of the neutralino to an antisquark and a
quark and then of the antisquark to two quarks. If either of the two
sequential two-body decays are kinematically forbidden, \ie they can
only proceed if the internal particle in the three-body decay is
off-shell, then we consider the decay to be three-body, otherwise we
treat the decay as two sequential two-body decays.
 
The problem is then how to implement the angular ordering procedure
for these three processes. We shall consider them using the eikonal
current with an arbitrary number of colours as was done in
Section~\ref{sec:example} for the process $\mr{e^+e^-\ra q\bar{q}g}$.
So in these R-parity violating processes this means we need to
consider the decay of an antisquark to $(N_c-1)$ quarks and of the
neutralino, chargino and gluino to $ N_c$ quarks. We also have to use
the generalization to $N_c$ colours of the antisymmetric tensor, \ie
$\epsilon^{c_1 \ldots c_{N_c}}$.

\subsubsection{Squark Decays}

For the decay of an antisquark to $(N_c-1)$ quarks the leading
infrared contribution to the soft gluon distribution has the following
factorized form.
\beq
    {\bf M} (p_0,p_1,p_2, \ldots ,p_{N_c-1};q) = g_s {\bf m}
   (p_0,p_1,p_2,  \ldots ,p_{N_c-1}) \cdot \JC(q)
\eeq
where
\begin{itemize} 
  \item $ {\bf m} (p_0,p_1,p_2, \ldots ,p_{N_c-1})$ is the tree-level
         matrix element for an antisquark, with momentum $p_0$, to decay to
         $N_c-1$ quarks, with momentum $p_1,\ldots,p_{N_c-1}$. 
  \item ${\bf M} (p_0,p_1,p_2, \ldots ,p_{N_c-1};q)$ is the tree-level matrix
        element for the decay of an antisquark to $N_c-1$ quarks including
        the emission of an extra soft gluon, with momentum $q$.
  \item $ \JC(q)$ is the non-Abelian semi-classical current for
        the emission of the soft gluon with momentum $q$ from 
        the hard partons.
  \item $c_0$ is the colour of the decaying antisquark and $c_1,
        \ldots\, c_{N_c-1}$ are the colours of the quarks.
\end{itemize}
Again the current, $ \JC(q)$, is given by, $\JC(q) = \sum_{s=1,2}
\JC^{b,\mu}(q) \varepsilon_{\mu,s}$ where here
\beq  
   \JC^{b,\mu}(q) = -\left(\frac{p_{0}^{\mu}}{p_0 \cdot q}\right) {\bf
                             t}_{c_0,c_{0}'}^{b,\mr{\tilde{q}}^{*}}
                           \epsilon^{c_{0}',c_1, \ldots ,c_{N_c-1}}
                          +\sum_{i=1}^{N_c-1} \left(\frac{p_{i}^{\mu}}{p_i
                 \cdot q}\right) {\bf t}_{c_i,c_{i}'}^{b,\mr{q}_{i}}
                                 \epsilon^{c_{0}, \ldots
                                 ,c_{i}', \ldots ,c_{N_c-1}}.
\eeq
$b$ and $\mu$ are the colour and Lorentz indices of the emitted gluon;
${\bf t}^{b,\mr{\tilde{q}}^{*}}$, ${\bf t}^{b,\mr{q}_{i}}$ are the
colour matrices of the antisquark and quarks, respectively.

We obtain the soft gluon distribution simply by squaring the current
\beq
   \JC^{2}(q) =  -C_{F}N_{c}(N_{c}-2)!\left[
                      \displaystyle{\sum_{i=1}^{N_c-1}}
                      \left( \frac{p_{0}}{p_0
                   \cdot q}-\frac{p_{i}}{p_i\cdot q} \right)^2
                   + \displaystyle{\sum_{i=1}^{N_c-1}}
                     \displaystyle{\sum_{j>i}^{N_c-1}} \left(
                   \frac{p_{i}}{p_i
                   \cdot q}-\frac{p_{j}}{p_j\cdot q} \right)^2\right].
\eeq
This can be expressed in terms of the radiation functions as in
Eqn.\,\ref{eqn:sqcurrent}, where here the tree-level colour factor is
$C_{\mathbf{m}} = \epsilon^{c_{0}, \ldots,c_{N_c-1}} \epsilon^{c_{0},
\ldots,c_{N_c-1}} = N_{c}!$, where we have not averaged over the
initial colours, and the radiation pattern is given by
\beq  
   W(\Omega_q) = \frac{-\omega^2 C_{F}}{(N_c-1)} \left[ \sum_{i=1}^{N_c-1}
                   \left( \frac{p_{0}}{p_0
                   \cdot q}-\frac{p_{i}}{p_i\cdot q} \right)^2
                   + \sum_{i=1}^{N_c-2}\sum_{j>i}^{N_c-1} \left(
                   \frac{p_{i}}{p_i
                   \cdot q}-\frac{p_{j}}{p_j\cdot q} \right)^2\right].
\eeq
We can then re-express this result in terms of the functions given in 
\cite{Marchesini:1990yk}
\beq  
    W(\Omega_q) = \frac{2 C_{F}}{(N_c-1)}  \sum_{i=0}^{N_c-1}
                   \sum_{j \neq i}^{N_c-1} W_{ij}^{i}. 
\eeq
This is exactly the same result obtained in \cite{Gibbs:1995cw}, in
the context of baryon number violation in the Standard Model, except
that the massless radiation functions of their paper are now replaced
by the massive functions here.

This leads to the following approach for treating the soft gluon
radiation from this process. The quarks from the decay are randomly
colour connected to either the decaying antisquark or the other
quark. This then correctly treats the soft gluon radiation from the
decay products. 

In general, the QCD radiation from sparticles, which are here in the
{\it initial\/} state, is neglected in HERWIG\@. We would expect this
approximation to be valid for two reasons, firstly the sparticles will
usually have a short lifetime and secondly, due to their heavy masses
the QCD radiation will also be suppressed unless they have momenta
much greater than their masses. However for the decays we consider, we
can include the effects of radiation from the decaying sparticles.
This is done by treating the radiation in the rest frame of the
decaying squark where there is no radiation from the decaying
sparticle, which HERWIG would not generate anyway. However as stated
in Section~\ref{sec:example} while the radiation from individual
partons, \ie $W^i_{ij}$, is not Lorentz invariant the dipole radiation
functions are. Hence the total radiation pattern is Lorentz invariant
and therefore by treating the decay in the rest frame of the decaying
particle we correctly include the QCD radiation from the decaying
particle when we boost back to the lab frame.
 
\subsubsection{Chargino and Neutralino Decays}

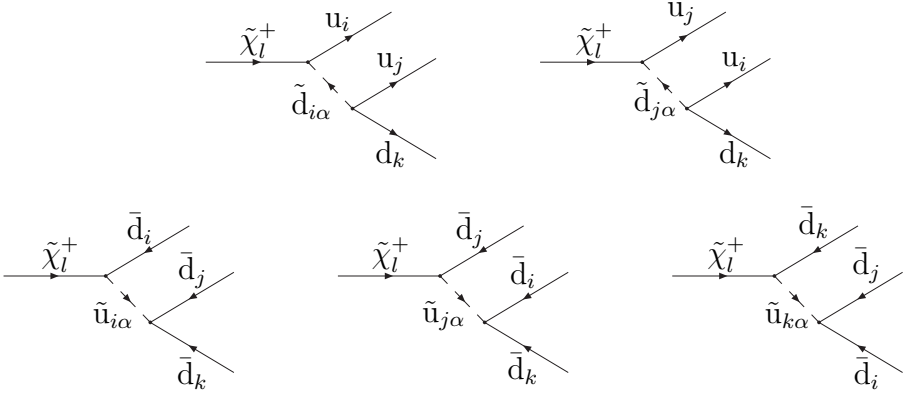
\begin{figure}[h!t]
\begin{center} 
\begin{picture}(360,80)(0,0)
\SetScale{0.7}
\SetOffset(-50,0)
\ArrowLine(185,78)(240,78)
\ArrowLine(240,78)(285,105)
\ArrowLine(264,53)(309,26)
\ArrowLine(264,53)(309,80)
\DashArrowLine(264,53)(240,78){5}
\Text(150,63)[]{$\cht^+_l$}
\Text(200,54)[]{$\mr{u}_j$}
\Text(200,20)[]{$\mr{d}_k$}
\Text(180,70)[]{$\mr{u}_i$}
\Text(170,40)[]{$\mr{\dnt}_{i\al}$}
\Vertex(240,78){1}
\Vertex(264,53){1}
\ArrowLine(365,78)(420,78)
\ArrowLine(420,78)(465,105)
\ArrowLine(444,53)(489,26)
\ArrowLine(444,53)(489,80)
\DashArrowLine(444,53)(420,78){5}
\Text(277,63)[]{$\cht^+_l$}
\Text(330,20)[]{$\mr{d}_k$}
\Text(310,73)[]{$\mr{u}_j$}
\Text(330,55)[]{$\mr{u}_i$}
\Text(300,40)[]{$\mr{\dnt}_{j\al}$}
\Vertex(420,78){1}
\Vertex(444,53){1}
\end{picture}
\begin{picture}(360,80)(0,0)
\SetScale{0.7}
\ArrowLine(5,78)(60,78)
\ArrowLine(105,105)(60,78)
\ArrowLine(129,26)(84,53)
\ArrowLine(129,80)(84,53)
\DashArrowLine(60,78)(84,53){5}
\Text(25,63)[]{$\cht^+_l$}
\Text(55,73)[]{$\mr{\bar{d}}_i$}
\Text(75,18)[]{$\mr{\bar{d}}_k$}
\Text(75,57)[]{$\mr{\bar{d}}_j$}
\Text(45,40)[]{$\mr{\upt}_{i\al}$}
\Vertex(60,78){1}
\Vertex(84,53){1}
\ArrowLine(185,78)(240,78)
\ArrowLine(285,105)(240,78)
\ArrowLine(309,26)(264,53)
\ArrowLine(309,80)(264,53)
\DashArrowLine(240,78)(264,53){5}
\Text(150,63)[]{$\cht^+_l$}
\Text(200,56)[]{$\mr{\bar{d}}_i$}
\Text(200,20)[]{$\mr{\bar{d}}_k$}
\Text(180,72)[]{$\mr{\bar{d}}_j$}
\Text(170,40)[]{$\mr{\upt}_{j\al}$}
\Vertex(240,78){1}
\Vertex(264,53){1}
\ArrowLine(365,78)(420,78)
\ArrowLine(465,105)(420,78)
\ArrowLine(489,26)(444,53)
\ArrowLine(489,80)(444,53)
\DashArrowLine(420,78)(444,53){5}
\Text(277,63)[]{$\cht^+_l$}
\Text(330,18)[]{$\mr{\bar{d}}_i$}
\Text(310,75)[]{$\mr{\bar{d}}_k$}
\Text(330,59)[]{$\mr{\bar{d}}_j$}
\Text(300,40)[]{$\mr{\upt}_{k\al}$}
\Vertex(420,78){1}
\Vertex(444,53){1}
\end{picture}
\end{center}
\caption{UDD decays of the ${\tilde\chi}^+$.}
\label{fig:UDDchar}
\end{figure}

\begin{figure}[h!b]
\begin{center} 
\begin{picture}(360,80)(0,0)
\SetScale{0.7}
\ArrowLine(5,78)(60,78)
\ArrowLine(60,78)(105,105)
\ArrowLine(84,53)(129,26)
\ArrowLine(84,53)(129,80)
\DashArrowLine(84,53)(60,78){5}
\Text(25,63)[]{$\cht^{0}_l$}
\Text(55,70)[]{$\mr{u}_i$}
\Text(75,20)[]{$\mr{d}_k$}
\Text(75,56)[]{$\mr{d}_j$}
\Text(45,40)[]{$\mr{\upt}_{i\al}$}
\Vertex(60,78){1}
\Vertex(84,53){1}
\ArrowLine(185,78)(240,78)
\ArrowLine(240,78)(285,105)
\ArrowLine(264,53)(309,26)
\ArrowLine(264,53)(309,80)
\DashArrowLine(264,53)(240,78){5}
\Text(150,63)[]{$\cht^{0}_l$}
\Text(200,54)[]{$\mr{u}_i$}
\Text(200,20)[]{$\mr{d}_k$}
\Text(180,72)[]{$\mr{d}_j$}
\Text(170,40)[]{$\mr{\dnt}_{j\al}$}
\Vertex(240,78){1}
\Vertex(264,53){1}
\ArrowLine(365,78)(420,78)
\ArrowLine(420,78)(465,105)
\ArrowLine(444,53)(489,26)
\ArrowLine(444,53)(489,80)
\DashArrowLine(444,53)(420,78){5}
\Text(277,63)[]{$\cht^{0}_l$}
\Text(330,20)[]{$\mr{u}_i$}
\Text(310,74)[]{$\mr{d}_k$}
\Text(330,57)[]{$\mr{d}_j$}
\Text(300,40)[]{$\mr{\dnt}_{k\al}$}
\Vertex(420,78){1}
\Vertex(444,53){1}
\end{picture}
\end{center}
\caption{UDD decays of the ${\tilde\chi}^0$.}
\label{fig:UDDneut}
\end{figure}
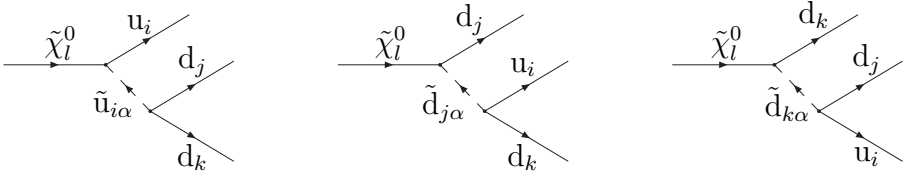

  The charginos decay via the process shown in
Fig.\,\ref{fig:UDDchar}, and the neutralinos via the process in
Fig.\,\ref{fig:UDDneut}. If we consider the QCD radiation from the
decay of a colour neutral
object which decays, for an arbitrary number of colours $N_c$, to $N_c$
quarks, then we see that there is only one possible colour flow for
this process.  The squarks appearing in these processes,
$\mr{\tilde{q}}_{i\al}$, can be either of the states $\al=1,2$ resulting
from the mixing of $\mr{\tilde{q}}_{iL}$ and $\mr{\tilde{q}}_{iR}$, as
discussed in more detail in the appendix.

In fact, the colour structure of this process is very similar to that
of the squark decay and the matrix element in the soft limit can be
written in the same factorized form as before. Again, we can express
the current as in Eqn.\,\ref{eqn:sqcurrent} where the tree-level
colour factor $C_{\mathbf{m}}=\epsilon^{c_{0}, \ldots,c_{N_c-1}}
\epsilon^{c_{0}, \ldots,c_{N_c-1}}=N_c!$ and the radiation function is
given by
\beq  
    W(\Omega_q) = \frac{2 C_{F}}{(N_c-1)}  \sum_{i=1}^{N_c}
                   \sum_{j \neq i}^{N_c} W_{ij}^{i} 
\eeq
This result can be interpreted as saying that a quark in the final
state should be randomly connected to any of the other quarks from the
neutralino or chargino decays.

\subsubsection{Gluino Decays}

\begin{figure}[htp]
\begin{center} 
\begin{picture}(360,50)(0,0)
\SetScale{0.7}
\SetOffset(0,-30)
\ArrowLine(5,78)(60,78)
\ArrowLine(105,105)(60,78)
\ArrowLine(84,53)(129,26)
\ArrowLine(129,80)(84,53)
\DashArrowLine(84,53)(60,78){5}
\Text(25,63)[]{$\mr{\tilde{g}}$}
\Text(55,70)[]{$\mr{u}_i$}
\Text(75,20)[]{$\mr{d}_k$}
\Text(75,56)[]{$\mr{d}_j$}
\Text(45,40)[]{$\mr{\upt}_{i\al}$}
\Vertex(60,78){1}
\Vertex(84,53){1}
\ArrowLine(185,78)(240,78)
\ArrowLine(285,105)(240,78)
\ArrowLine(264,53)(309,26)
\ArrowLine(309,80)(264,53)
\DashArrowLine(264,53)(240,78){5}
\Text(150,63)[]{$\mr{\tilde{g}}$}
\Text(200,54)[]{$\mr{u}_i$}
\Text(200,20)[]{$\mr{d}_k$}
\Text(180,72)[]{$\mr{d}_j$}
\Text(170,40)[]{$\mr{\dnt}_{j\al}$}
\Vertex(240,78){1}
\Vertex(264,53){1}
\ArrowLine(365,78)(420,78)
\ArrowLine(420,78)(465,105)
\ArrowLine(489,26)(444,53)
\ArrowLine(489,80)(444,53)
\DashArrowLine(444,53)(420,78){5}
\Text(277,63)[]{$\mr{\tilde{g}}$}
\Text(330,18)[]{$\mr{u}_i$}
\Text(310,75)[]{$\mr{d}_k$}
\Text(330,57)[]{$\mr{d}_j$}
\Text(300,40)[]{$\mr{\dnt}_{k\al}$}
\Vertex(420,78){1}
\Vertex(444,53){1}
\end{picture}
\end{center}
\caption{UDD decays of the $\mr{{\tilde{g}}}$.}
\label{fig:UDDgluino}
\end{figure}
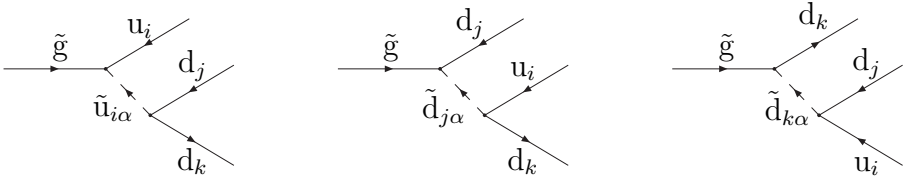

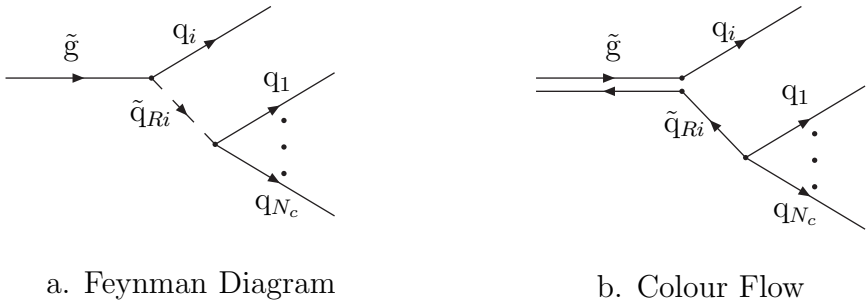
\begin{figure}[htp]
\begin{center} \begin{picture}(360,130)(0,40)
\SetScale{1.0}
\ArrowLine(5,128)(60,128)
\ArrowLine(60,128)(105,155)
\ArrowLine(84,103)(129,76)
\ArrowLine(84,103)(129,130)
\DashArrowLine(60,128)(84,103){5}
\Text(30,140)[]{$\mr{\tilde{g}}$}
\Text(73,145)[]{$\mr{q}_i$}
\Text(108,80)[]{$\mr{q}_{N_c}$}
\Text(108,127)[]{$\mr{q}_1$}
\Text(60,115)[]{$\mr{\tilde{q}}_{Ri}$}
\Vertex(60,128){1}
\Vertex(84,103){1}
\SetOffset(20,0)
\ArrowLine(185,128)(240,128)
\ArrowLine(240,123)(185,123)
\ArrowLine(240,128)(285,155)
\ArrowLine(264,98)(309,71)
\ArrowLine(264,98)(309,125)
\ArrowLine(264,98)(240,123)
\Text(215,140)[]{$\mr{\tilde{g}}$}
\Text(284,122)[]{$\mr{q}_1$}
\Text(284,77)[]{$\mr{q}_{N_c}$}
\Text(257,146)[]{$\mr{q}_i$}
\Text(242,111)[]{$\mr{\tilde{q}}_{Ri}$}
\Vertex(240,128){1}
\Vertex(240,123){1}
\Vertex(264,98){1}
% Dots for missing particles
\Vertex(90,112){1}
\Vertex(90,102){1}
\Vertex(90,92){1}
\Vertex(290,107){1}
\Vertex(290,97){1}
\Vertex(290,87){1}
% Diagram labels
\Text(55,50)[]{a. Feynman Diagram}
\Text(248,50)[]{b. Colour Flow}
\end{picture}
\caption{Baryon number violating decay of $\tilde{g}$.}
\label{fig:gluino}
\end{center}
\end{figure}

  The colour structure of the gluino decay is very different from that
of the colourless objects or the squarks which we have already
considered, the diagrams for this process are shown in 
Fig.\,\ref{fig:UDDgluino}. Again if we consider an arbitrary number of
colours, $N_c$, the gluino will decay to $N_c$ quarks. In this case
there will be
$N_c$ possible colour flows, corresponding to the Feynman diagrams and
colour flows shown in Fig.\,\ref{fig:gluino}. These different colour
flows will lead to `non-planar' terms which must be dealt with.

The leading infrared contribution to the soft gluon distribution can
be written in the following factorized form.
\beq
    {\bf M} (p_0,p_1,p_2, \ldots ,p_{N_c};q) = g_s \sum_{i=1}^{N_c}{\bf m}_i
   (p_0,p_1,p_2,  \ldots ,p_{N_c}) \cdot \JC_i(q)
\eeq
   where 
\begin{itemize}
 \item $ {\bf m}_i (p_0,p_1,p_2, \ldots ,p_{N_c})$ is the tree-level
   matrix element of the three-body gluino decay for the $i$th possible 
   colour flow.
  \item$ {\bf M} (p_0,p_1,p_2, \ldots ,p_{N_c};q)$ is the tree-level
   matrix element for the three-body gluino decay including the extra 
  emission of a soft gluon of momentum  $q$.
 \item $ \JC_i(q)$ is the non-Abelian semi-classical
  current for the emission of the soft gluon, momentum $q$, from the
  hard partons for the $i$th possible colour flow.
\end{itemize}

Again the current has the form $\JC_i(q) = \displaystyle{\sum_{s=1,2}}
  \JC_i^{b,\mu}(q) \varepsilon_{\mu,s}$, where in this case 
\barr  
   \JC_i^{b,\mu}(q) = & i\left(\frac{p_0^\mu}{p_0 \cdot q}\right) 
                        {\bf f}^{ba'a}{\bf t}_{c_ic'_i}^{a'}
                        \epsilon^{c_1 \ldots c'_i \ldots c_{N_c}}
                         + \left(\frac{p_i^\mu}{p_i \cdot q}\right)
                        {{\bf t}_{c_ic'_i}^b\bf t}_{c'_ic''_i}^a
                        \epsilon^{c_1 \ldots c''_i \ldots c_{N_c}} \nonumber \\
&    +{\displaystyle \sum_{j=1,j \neq i}^{N_c}} \left(\frac{p_j^\mu}{p_j 
    \cdot q}\right)
                         {\bf t}_{c_jc'_j}^b{\bf t}_{c_ic'_i}^a
                        \epsilon^{c_1 \ldots c'_i  \ldots c'_j
                         \ldots c_{N_c}} \\ \nonumber
\earr

  We can write the matrix element squared for this process as
\barr
   |{\bf M} (p_0,p_1,p_2, \ldots ,p_{N_c};q)|^2 = & g^2_s \displaystyle 
    \sum_{i=1}^{N_c} |{\bf m}_i
 (p_0,p_1,p_2,  \ldots ,p_{N_c})|^2 \cdot |\JC_i(q)|^2 \nonumber \\
     & + g^2_s  
   \displaystyle \sum_{i=1}^{N_c} \displaystyle \sum_{j=1,j \neq i}^{N_c}
 {\bf m}_i{\bf m}_j^* \cdot \JC_i(q)\cdot \JC_j^*(q) \\ \nonumber
\earr

% Gluino Decay

The procedure of \cite{Odagiri:1998ep}, which was described in
Section~\ref{sec-colour} can be used with the matrix elements for this
process given in the appendix to deal with the `non-planar' terms.  So
we now consider the radiation pattern of the planar terms.  The
current can be written as in Eqn.\,\ref{eqn:sqcurrent} where here the
tree-level colour factor
\beq
C_{\mathbf{m}} = {\bf
 t}_{c_ic'_i}^b\epsilon^{c_1 \ldots c'_i \ldots c_{N_c}}{\bf
 t}_{c_i''c_i}^b \epsilon^{c_{0} \ldots c''_i \ldots c_{N_c}} = C_F
 N_{c}!.
\eeq
We have not averaged over the initial colours,
and the radiation function is given by
\barr  
   W(\Omega_q) = & C_A W_{0i}^{0} +2 C_F W_{0i}^{i} 
                   +\frac{2 C_{F}}{(N_c-1)}{\displaystyle
                         \sum_{j \neq i, k \neq
                        i,j}^{N_c}} W^{j} _{jk}  +\frac1{(N_c-1)}{\displaystyle \sum_{j 
                        \neq i}^{N_c}} 
                        \left( C_{A}W_{0j}^{0} + 2C_{F}W_{0j}^{j}
                        \right) \nonumber \\
                    & + \frac1{N_c} W_{0i}^{i} 
                        +\frac1{N_c(N_c-1)}{\displaystyle 
                        \sum_{j \neq i}^{N_c}}
                         \left(W_{0j}^{j}-W_{ij}^{i}-W_{ij}^{j}
                     \right). \label{eqn:planar} \\ \nonumber
\earr
This planar piece of the soft radiation pattern gives us the result
we would na\"{\i}vely expect. This pattern can be thought of as saying
the $i$th quark should be connected to the colour line of the gluino as
in an MSSM process and the anti-colour line of the gluino and the
remaining quarks should be randomly connected in the same way as for
the baryon number violating squark decay. This radiation pattern,
Eqn.\,\ref{eqn:planar}, also contains pieces which are of order
$1/N_c^2$ with respect to the leading order pieces that we will
neglect as in Section~\ref{sec-colour}.  We therefore connect the
$i$th quark to the gluino in the standard MSSM way with probability
given by $\frac{|M|^2_{full,i}}{|M|^2_{tot}}$, where $|M|^2_{full,i}$
is given by Eqn.\,\ref{eqn:kosuke}. We can then treat the anti-colour
line and the remaining quarks as a decaying antisquark.

\subsection{Hard Processes}

  In addition to the decays which we have already discussed, there are a
number of baryon number violating hard subprocesses we 
include in the simulation. All of the colour structures of the
hard processes that actually violate baryon number have already been
discussed as these processes are merely crossed versions of the various
decays discussed above. However in addition to these processes there are
some hard processes that occur via the third term in the
superpotential but involve no net baryon number violation, \eg
Fig.\,\ref{fig:bump}.

  We will therefore only discuss this type of process which cannot be
  obtained by crossing the previous results.

\subsubsection{Resonant Squark production followed by \boldmath{\bv}\  Decay}

\begin{figure}[htp]
\begin{center} \begin{picture}(180,70)(0,40)
\SetScale{1.3}
% Feynman Diagram
\ArrowLine(5,26)(60,52)
\ArrowLine(5,78)(60,52)
\DashArrowLine(90,52)(60,52){5}
\ArrowLine(90,52)(145,26)
\ArrowLine(90,52)(145,78)
% Labels
\Text(195,35)[l]{$\mr{q}_{N_c-1}(k_{N_c-1})$}
\Text(195,105)[l]{$\mr{q}_1(k_1)$}
\Text(2,105)[r]{$\mr{q}_1(p_1)$}
\Text(2,35)[r]{$\mr{q}_{N_c-1}(p_{N_c-1})$}
\Text(100,80)[]{$\mr{\tilde{q}}_{R{N_c}}$}
% Vertex and Dots
\Vertex(60,52){1}
\Vertex(90,52){1}
\Vertex(25,43){1}
\Vertex(25,53){1}
\Vertex(25,63){1}
\Vertex(125,43){1}
\Vertex(125,53){1}
\Vertex(125,63){1}
\end{picture}
\end{center}
\caption{Resonant squark production followed by \bv\  decay for an
arbitrary number of colours $N_c$.}
\label{fig:bump}
\end{figure}
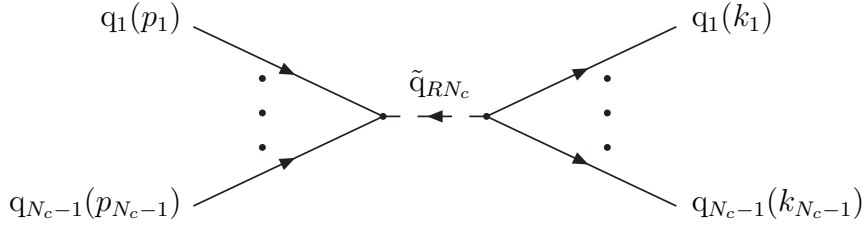

  As before, we will consider the process in Fig.\,\ref{fig:bump}
for an arbitrary number of colours,~$N_c$. 

  We can write the matrix element for the emission of an
extra soft gluon in the following form.
\beq
    {\bf M} (p_1,\ldots ,p_{N_c-1}: k_1,\ldots ,k_{N_c-1} ;q) = 
    g_s {\bf m}(p_1,\ldots ,p_{N_c-1}:k_1,\ldots ,k_{N_c-1})
   \cdot \JC(q)
\eeq
where
\begin{itemize}
   \item ${\bf m}(p_1,\ldots ,p_{N_c-1}:k_1,\ldots ,k_{N_c-1}) $
         is the tree-level matrix element for the $(N_c-1)$ quarks to
          $(N_c-1)$ quarks scattering.
   \item $ {\bf M} (p_1,\ldots ,p_{N_c-1}:k_1,\ldots ,k_{N_c-1} ;q)$
         is the tree-level matrix element for the $(N_c-1)$ quarks to
          $(N_c-1)$ quarks scattering with the additional emission of a
         soft gluon with momentum $q$.
   \item $\JC(q)$ is the non-Abelian semi-classical current for 
         the emission of the soft gluon, momentum $q$, from the hard partons.
   \item $p_1,\ldots ,p_{N_c-1}$ are the momenta of the partons in the
         initial state.
   \item $k_1,\ldots ,k_{N_c-1}$ are the momenta of the partons in the
         final state.
\end{itemize}

  Again the current has the form $\JC(q) = \displaystyle{\sum_{s=1,2}}
  \JC^{b,\mu}(q) \varepsilon_{\mu,s}$, where in this case 
\barr  
   \JC^{b,\mu}(q) =& -\displaystyle \sum_{i=1}^{N_c-1} 
\left(\frac{p_{i}^{\mu}}{p_i
    \cdot q}\right) {\bf t}_{c'_ic_i}^b \epsilon^{c_1 \ldots c'_i 
    \ldots c_{N_c}}  \epsilon^{d_1 \ldots d_{N_c-1} c_{N_c}}\nonumber \\ 
   &    +\displaystyle\sum_{i=1}^{N_c-1} \left(\frac{k_{i}^{\mu}}{k_i
    \cdot q}\right) {\bf t}_{d_id_{i}'}^b \epsilon^{c_1 \ldots c_{N_c-1} 
    d_{N_c}}  \epsilon^{d_1 \ldots d'_i \ldots d_{N_c}}, \\ \nonumber
\earr
where $b$ and $\mu$ are the colour and Lorentz indices of the emitted
gluon, respectively.

We can now obtain the soft gluon distribution by squaring the
current. This can be rewritten using Eqn.\,\ref{eqn:sqcurrent} where
here the tree-level colour factor is given by $C_{\mathbf{m}}=N_c!
(N_c-1)!$, again we have not averaged over the initial state colours,
and the radiation function is given by
\barr  
   W(\Omega_q)& = & \frac{2 C_{F}}{(N_c-1)}
                    \sum_{i=1}^{N_c-1}
                   \sum_{j \neq i} W_{ij}^{i}
                   +\frac{2 C_{F}}{(N_c-1)}
                     \sum_{l=1}^{N_c-1}
                   \sum_{m \neq l} W_{lm}^{l} \nonumber \\
 &&                + \frac{2 C_{F}}{(N_c-1)^2}
                     \sum_{i=1}^{N_c-1}
       \sum_{l=1}^{N_c-1}\left( W_{il}^{i}+ W_{il}^{l} \right), \\ \nonumber
\earr
  where the partons $i$ and $j$ are in the initial state and the partons $l$
and $m$ are in the final state.

  This radiation pattern gives quite an unusual angular ordering
procedure. If we consider one of the quarks in the initial state, this
quark should be randomly connected to any of the other quarks in the
initial state or to the final state, \ie the probability of connecting to a given
quark in the initial state and the final state as a whole is equal. If the quark
is connected to the final state it must then be connected at random to one of the
final state quarks. Similarly the final state quarks are connected at random to any
of the other final state quarks or the initial state, and again quarks connected to
the initial state are then randomly connected to any of the initial state quarks.

\section{Hadronization in \boldmath{\rpv}}
\label{sec-hadron}

  As  we saw in   Section~\ref{sec-angular}, it is possible to  angular
  order  the baryon  number violating decays and hard  processes.
  It is then necessary to decide how to hadronize these events using
  the cluster hadronization model \cite{Webber:1984if} for a full simulation
  of these processes. The procedure described in Section~\ref{sec-colour}
  also works in the MSSM provided that the lifetime of
  the coloured sparticles does not exceed the hadronization time-scale. 
  However some modifications  to this model are required for \rpv\  processes.
   
In the Standard Model and MSSM cases the colour partner for the colour
coherence effects and for the hadronization phase are always the
same. In the \bv\  decays and hard processes we see for the first time
cases where the colour connection for the angular ordering and for the
hadronization can be different.  This is because while the colour
connection for the angular ordering procedure is determined by the
eikonal current, the colour connection for the hadronization phase is
defined by the colour flow in the leading order diagram. When baryon
number is conserved these are identical, however when baryon number is
violated, there are cases where the two are different.

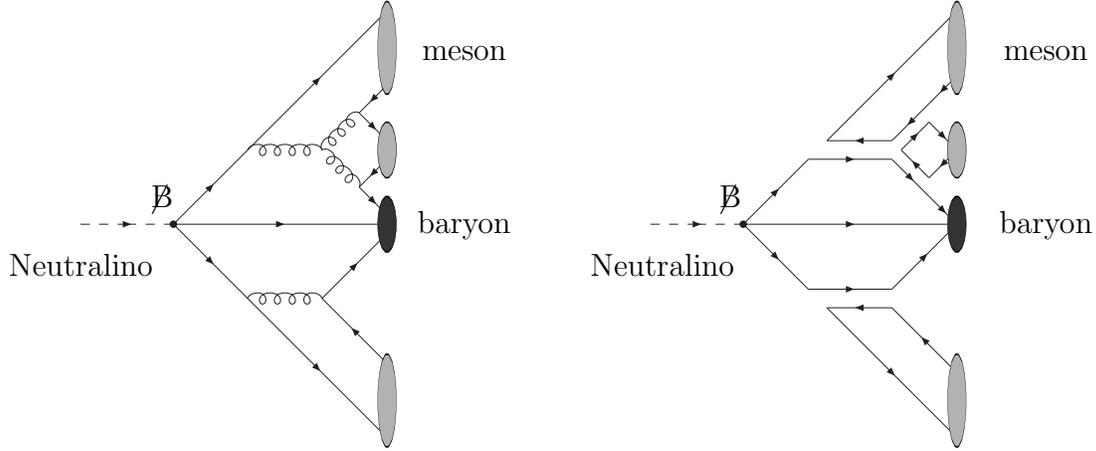
\begin{figure}[htp]
\begin{center} \begin{picture}(360,200)(0,0)
\SetScale{0.7}
\SetOffset(50,0)
% Feynman Diagram
% Incoming Line
\DashArrowLine(-50,150)(0,150){5}
% Vertex and the label
\Vertex(0,150){2}
\Text(-5,115)[]{\bv}
% Decay
\ArrowLine(0,150)(40,190)
\ArrowLine(0,150)(115,150)
\ArrowLine(0,150)(40,110)
% Gluon Radiation
\Gluon(40,190)(80,190){3}{4}
\Gluon(40,110)(80,110){3}{4}
\Gluon(80,190)(100,210){3}{3}
\Gluon(80,190)(100,170){3}{3}
% Gluon Splitting
\ArrowLine(40,190)(115,265)
\ArrowLine(40,110)(115,35)
\ArrowLine(80,110)(115,145)
\ArrowLine(115,75)(80,110)
\ArrowLine(115,225)(100,210)
\ArrowLine(100,210)(115,195)
\ArrowLine(115,185)(100,170)
\ArrowLine(100,170)(115,155)
% Ovals for clusters
\GOval(115,245)(25,5)(0){0.7}
\GOval(115,150)(15,5)(0){0.2}
\GOval(115,190)(15,5)(0){0.7}
\GOval(115,55)(25,5)(0){0.7}
% Labels
\Text(-35,90)[]{Neutralino}
\Text(110,170)[]{meson}
\Text(110,105)[]{baryon}
%%%%%%%%%%%%% Colour Flows
\SetOffset(125,0)
% Incoming Line
\DashArrowLine(150,150)(200,150){5}
% Vertex and the label
\Vertex(200,150){2}
\Text(135,115)[]{\bv}
% Decay
\ArrowLine(200,150)(235,185)
\ArrowLine(200,150)(315,150)
\ArrowLine(200,150)(235,115)
% Gluon Radiation
\ArrowLine(235,185)(280,185)
\ArrowLine(280,195)(245,195)
\ArrowLine(235,115)(280,115)
\ArrowLine(280,105)(245,105)
\ArrowLine(280,185)(300,165)
\ArrowLine(300,175)(285,190)
\ArrowLine(285,190)(300,205)
\ArrowLine(300,215)(280,195)
% Gluon Splitting
\ArrowLine(245,195)(315,265)
\ArrowLine(245,105)(315,35)
\ArrowLine(315,190)(300,175)
\ArrowLine(300,165)(315,150)
\ArrowLine(300,205)(315,190)
\ArrowLine(315,230)(300,215)
\ArrowLine(315,70)(280,105)
\ArrowLine(280,115)(315,150)
% Ovals for clusters
\GOval(315,245)(25,5)(0){0.7}
\GOval(315,150)(15,5)(0){0.2}
\GOval(315,190)(15,5)(0){0.7}
\GOval(315,55)(25,5)(0){0.7}
% Labels
\Text(110,90)[]{Neutralino}
\Text(255,170)[]{meson}
\Text(255,105)[]{baryon}
\end{picture}\end{center}
\caption{The Feynman diagrams and colour flows for the 
          hadronization of a \bv\  neutralino decay.}
\label{fig:neuthad}
\end{figure}

First we consider the simplest type of decay, \ie a neutralino or
chargino decaying to three quarks. The method described in
Section~\ref{sec-angular} correctly implements the angular ordering
procedure. After the parton-showering phase (and the splitting of
remaining gluons to quark anti-quark pairs) we will be left with pairs
of colour connected partons forming colour singlets as well as {\it
three\/} further quarks. An example of this is shown in
Fig.\,\ref{fig:neuthad}. These three remaining quarks form a colour
singlet with baryonic quantum numbers, a baryonic cluster. To handle
baryonic clusters HERWIG needs the constituents to be labelled as one
quark and one diquark rather than three quarks, so we randomly pair
two of them up as a diquark. In our example in Fig.\,\ref{fig:neuthad}
the three quarks in the middle together form a colour singlet which
is combined into a baryon.

% Showering and hadronization of a neutralino decay
% End of the Figure

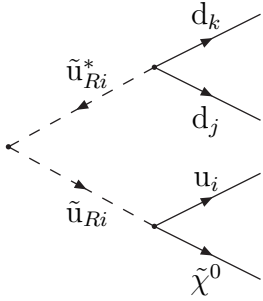
\begin{figure}[htp]
\begin{center} \begin{picture}(180,100)(0,20)
\SetScale{1.0}
\SetOffset(40,0)
\DashArrowLine(60,100)(5,70){5}
\DashArrowLine(5,70)(60,40){5}
\ArrowLine(60,40)(100,60)
\ArrowLine(60,40)(100,20)
\ArrowLine(60,100)(100,120)
\ArrowLine(60,100)(100,80)
\Text(80,80)[]{$\mr{d}_j$}
\Text(80,120)[]{$\mr{d}_k$}
\Text(80,57)[]{$\mr{u}_i$}
\Text(80,20)[]{$\cht^{0}$}
\Text(35,98)[]{$\mr{\upt}^*_{Ri}$}
\Text(35,45)[]{$\mr{\upt}_{Ri}$}
\Vertex(5,70){1}
\Vertex(60,100){1}
\Vertex(60,40){1}
\end{picture}\end{center}
\caption{Hadronization with one \bv\  decay.}
\label{fig:had1}
\end{figure}

This procedure is relatively easy to implement in the case of
electroweak gaugino decays. However it becomes more difficult in the
case of the \bv\  decay of a squark to two quarks. If the colour
partner of the decaying squark is a particle which decays via a baryon
number conserving process then the two quarks and the particle which
gets the colour of the second decaying particle can be clustered as in
the neutralino case, \eg in Fig.\,\ref{fig:had1} the $\mr{u}_i$,
$\mr{d}_j$, $\mr{d}_k$ should be formed into a baryonic cluster.

\begin{figure}[htp]
\begin{center} \begin{picture}(180,100)(0,30)
\SetScale{1.0}
\SetOffset(40,0)
\DashArrowLine(60,100)(5,70){5}
\DashArrowLine(5,70)(60,40){5}
\ArrowLine(100,60)(60,40)
\ArrowLine(100,20)(60,40)
\ArrowLine(60,100)(100,120)
\ArrowLine(60,100)(100,80)
\Text(80,80)[]{$\mr{d}_j$}
\Text(80,120)[]{$\mr{d}_k$}
\Text(80,57)[]{$\mr{\bar{d}}_l$}
\Text(80,20)[]{$\mr{\bar{d}}_m$}
\Text(35,98)[]{$\mr{\upt}^{*}_{Ri}$}
\Text(35,45)[]{$\mr{\upt}_{Ri}$}
\Vertex(5,70){1}
\Vertex(60,100){1}
\Vertex(60,40){1}
\end{picture}\end{center}
\caption{Hadronization with two \bv\  decays.}
\label{fig:had2}
\end{figure}
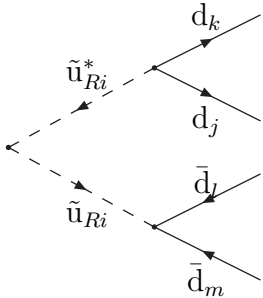

However, if this second particle decays via \bv\  as well, then the
procedure must be different, as shown in Fig.\,\ref{fig:had2}. Here,
instead of forming one baryonic cluster, we form two mesonic
clusters. This is done by pairing the $\mr{d}_k$ randomly with either
the $\mr{\bar{d}}_l$ or $\mr{\bar{d}}_m$ into a standard colour
singlet, the remaining quark and antiquark are also paired into a
colour singlet. This is not the colour connection for the angular
ordering but the colour connection for the hadronization phase, which
is different in this case and determined by the colour flow in the
tree-level diagram.

This leaves the case of the gluino decay for which it is easiest to
consider the two colour lines separately. The colour line should be
treated as normal and the anticolour line like a decaying
antisquark. So if the anticolour partner of the gluino is a Standard
Model particle or decays via a baryon number conserving MSSM decay
mode we form the three quarks into a baryonic cluster. However if the
anticolour partner decays via a \bv\  mode we then form two mesonic
clusters.

% Hadronization if one BV decay of a squark
% End of the figure

There is one further type of colour flow to be considered which is the
production of a resonant squark via \bv\  which then also decays via
\bv. The correct hadronization procedure in this case is similar to
that adopted for the case of two colour connected \bv\  decays. We
randomly connect the final state quarks to the colour partners of
either of the initial state quarks to form a colour singlet. The
remaining final state quark can then be paired with the colour partner
of the other initial state quark. This gives two colour-singlet
clusters. Again the colour partner for hadronization is determined by
the colour flow in the tree-level diagram.

% Hadronization with two BV squarks decays
% End of the figure

% Cluster Mass Spectrum
\begin{figure}[htp]
\centering
\vskip 5mm
\includegraphics[angle=90,width=0.4\textwidth]{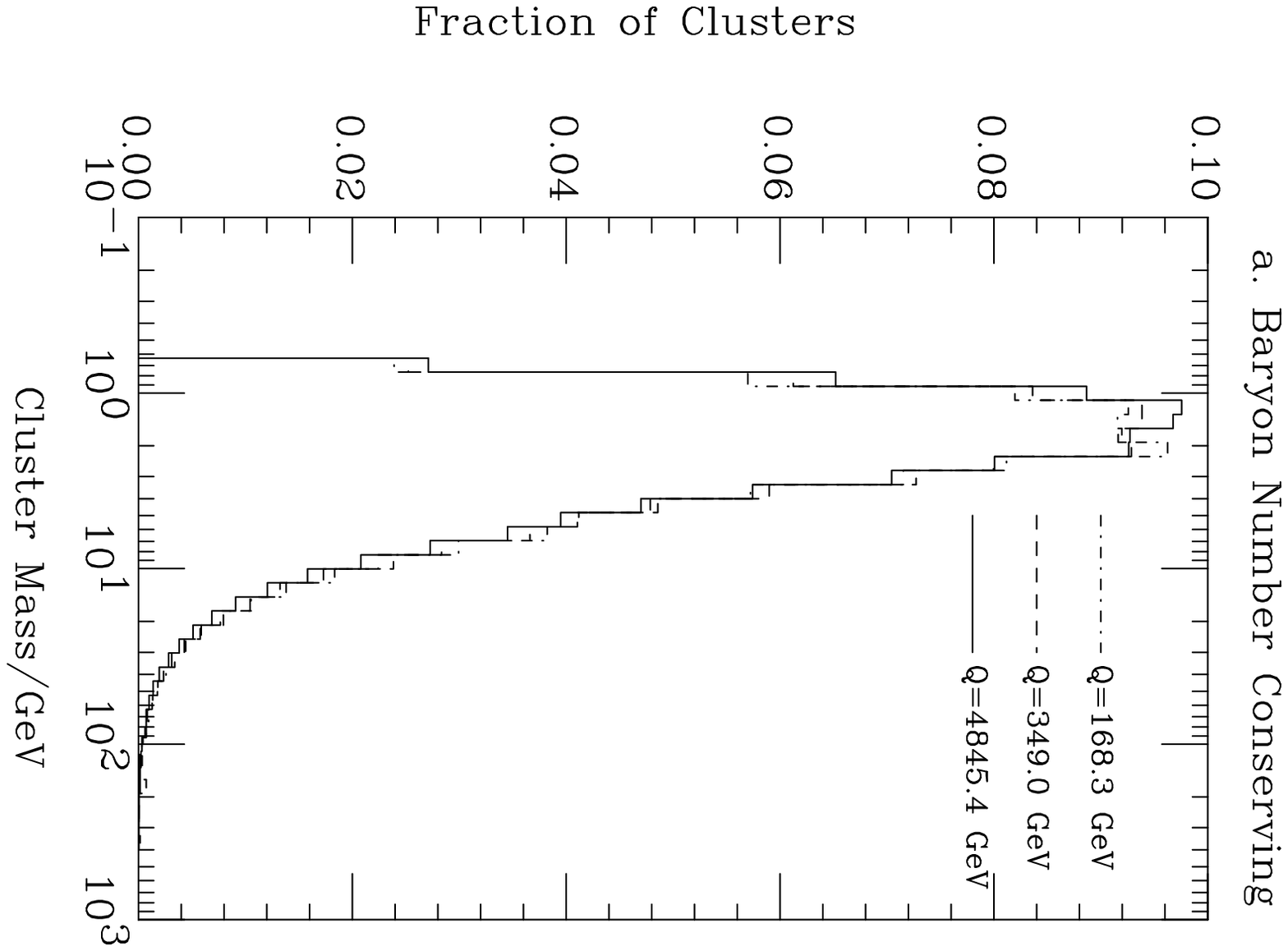}
\hfill
\includegraphics[angle=90,width=0.4\textwidth]{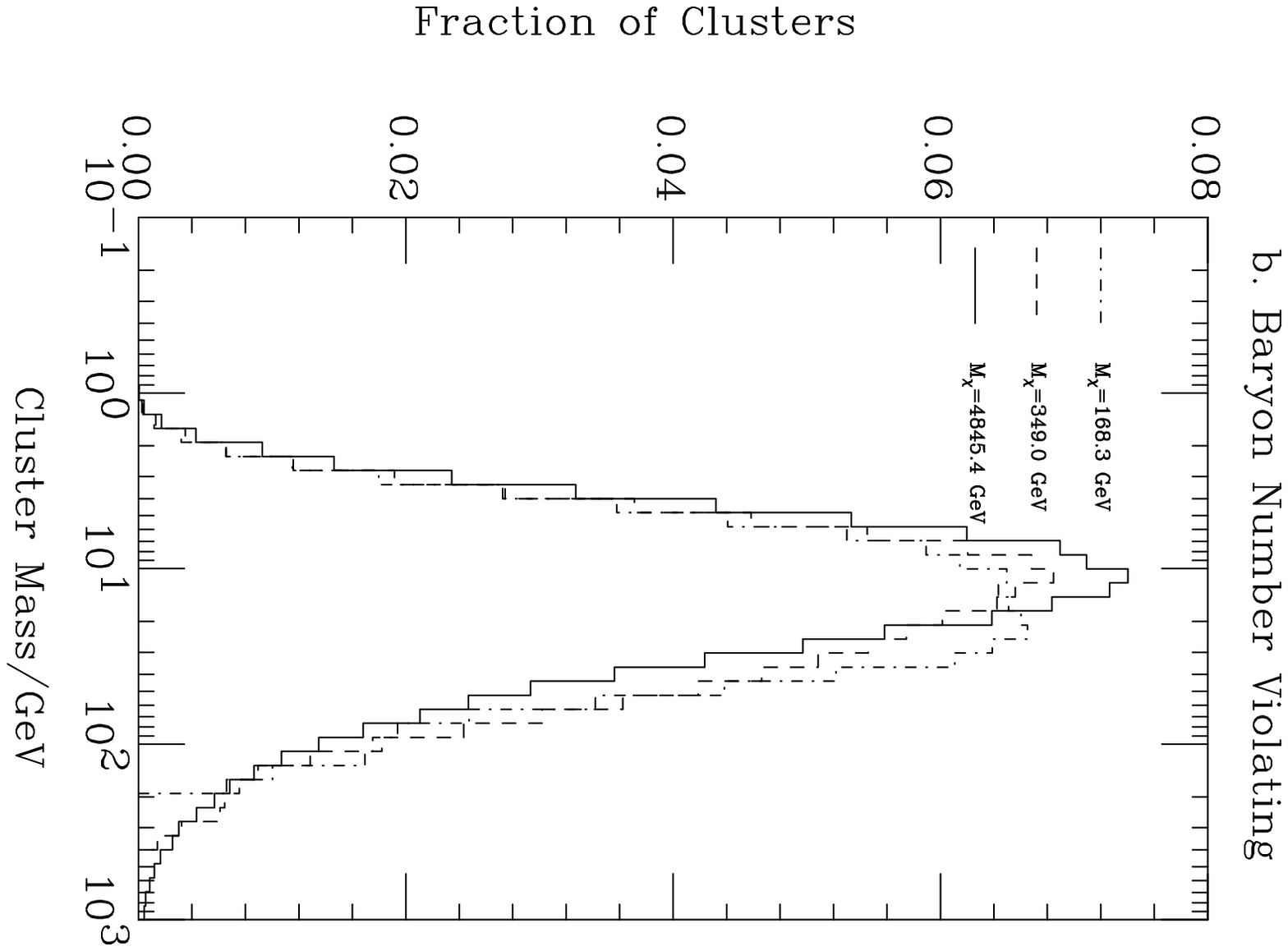}
\\
\caption{Distributions of the colour singlet cluster masses.
  The  baryon number conserving clusters come from $\mr{e^+e^-}$ events
  at the given scale whereas the baryon number violating clusters come
  from decays of neutralinos at the given masses.}
\label{fig:cluster}
\end{figure}
% End of the Figure

Using the procedures we have outlined above it is possible to
hadronize any of the \bv\  decays or hard processes. There is however
one potential problem.  The cluster model is based on the idea of colour
preconfinement.  In baryon number violating processes we see a very
different spectrum for the baryonic clusters formed from the baryon
number violation to that seen for clusters in the hitherto studied
Standard Model events.  Fig.\,\ref{fig:cluster} shows the spectra for
both Standard Model and baryon number violating clusters. The spectrum
for the baryon number violating clusters peaks at a much higher mass
than the baryon number conserving clusters and has a large tail at
high masses. This therefore means that before these clusters are
decayed to hadrons most of them must be split into lighter
clusters. The spectrum of the baryon number conserving clusters in
these events has the same spectrum as in Standard Model events.
  
  Fig.\,\ref{fig:cluster} contains the mass spectrum of pairs of colour connected
  partons after the parton-shower phase and the non-perturbative splitting of the
  gluons into quark-antiquark pairs. The baryon number conserving clusters, 
  Fig.\,\ref{fig:cluster}a, contains all the clusters in $\mr{e^+e^-}$ events at the 
  given centre-of-mass energies, whereas the baryon number violating clusters,
  Fig.\,\ref{fig:cluster}b only contains those clusters which contain the three quarks
  left after all the other quarks are paired into colour-singlets from neutralino decays
  at the given mass.

\begin{figure}[htp]
\centering
\vskip 5mm
\includegraphics[angle=90,width=0.4\textwidth]{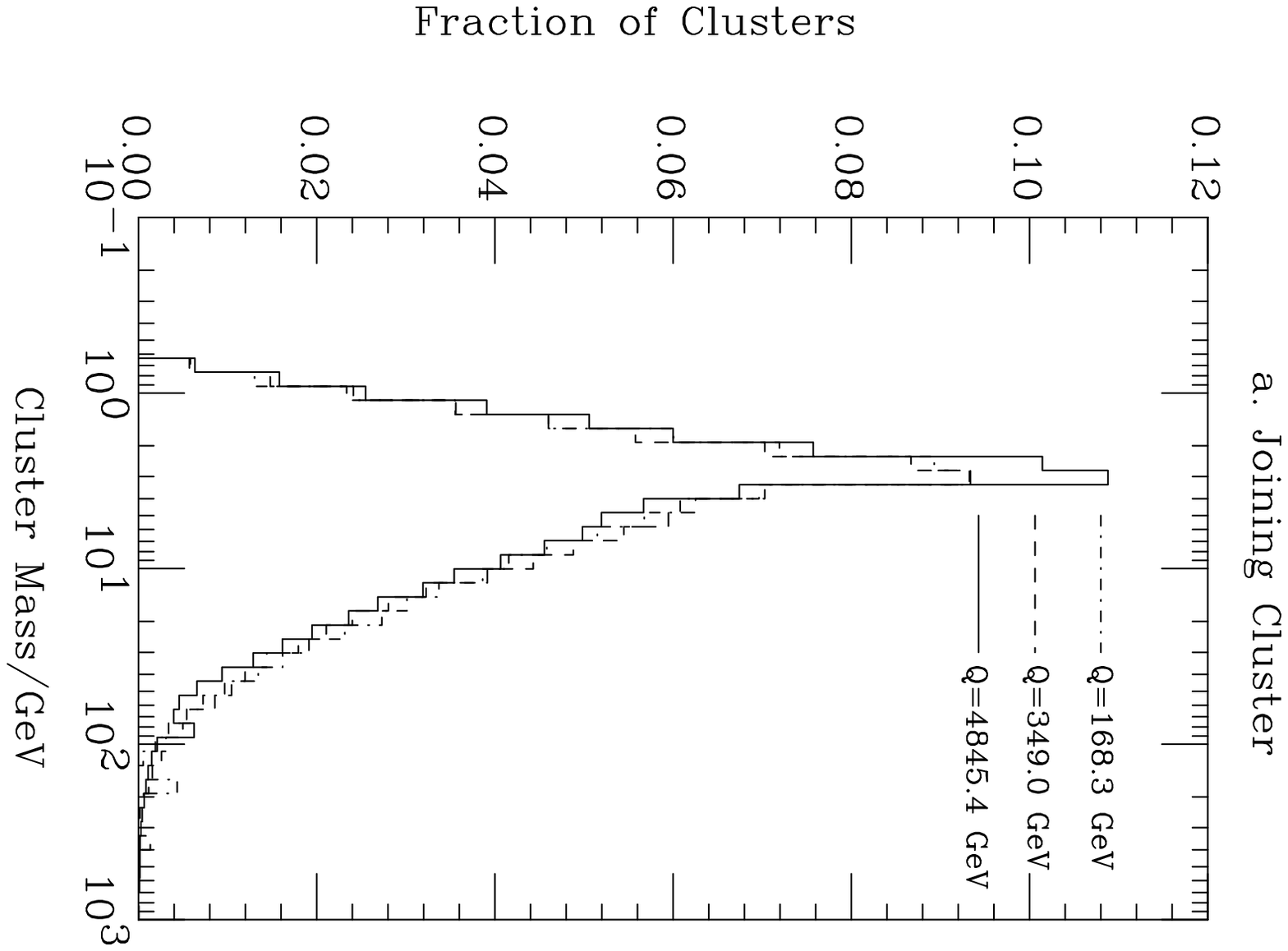}
\hfill
\includegraphics[angle=90,width=0.4\textwidth]{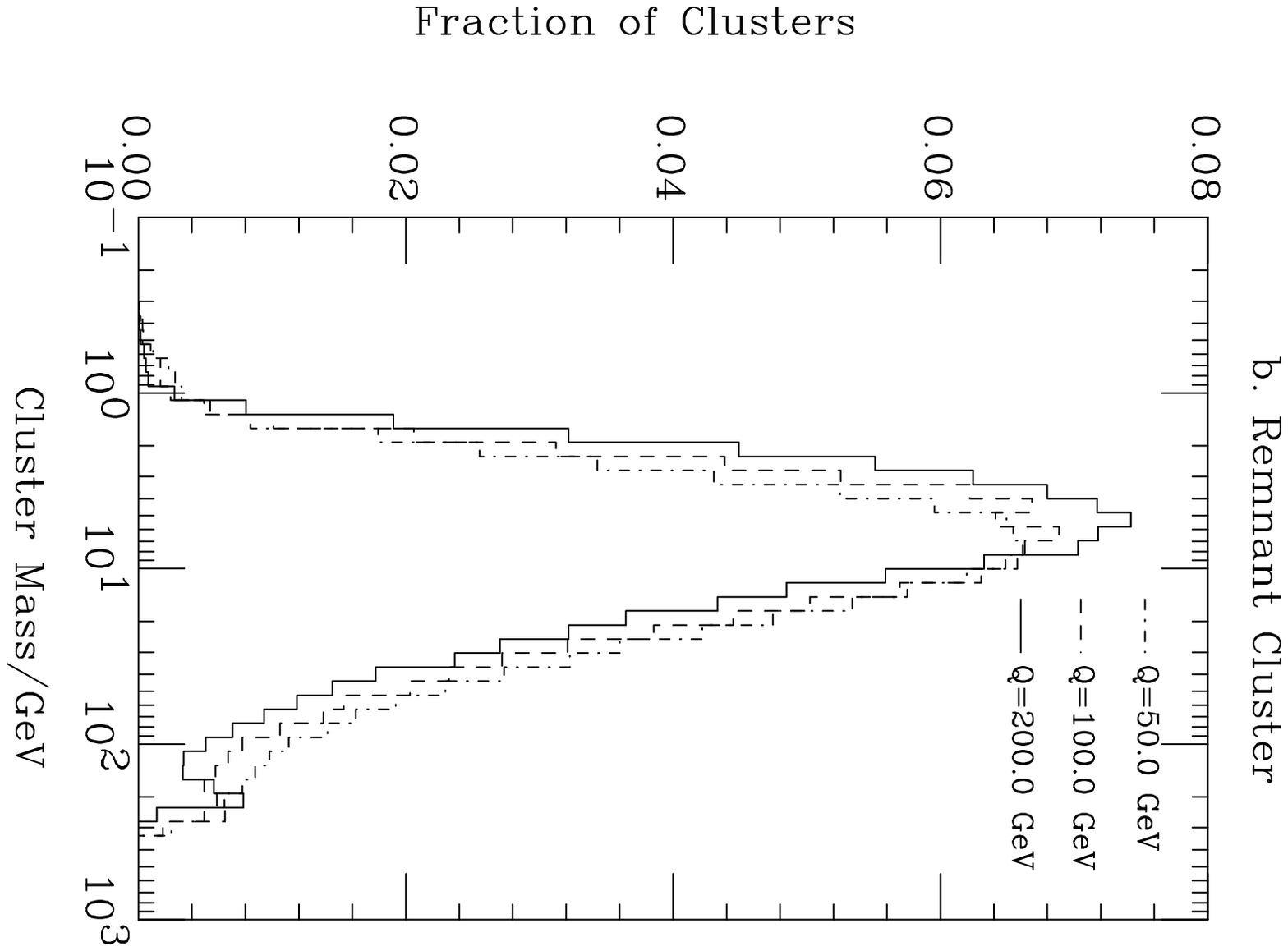}
\\
\caption{Masses of the joining and remnant clusters. The joining clusters 
are from $\mr{e^+e^-}$ events at the given scale. The remnant  clusters 
were generated in $\mr{e^-p}$ events with 30 GeV electrons and 820 GeV 
protons with the given scale as the minimum value of Q.}
\label{fig:cluster2}
\end{figure}

  The joining clusters in Fig\,\ref{fig:cluster2}a shows the clusters from 
  $\mr{e^+e^-}$ events with one quark from the parton-shower of the
  quark, and an antiquark from the parton-shower of the antiquark. The remnant clusters,
  Fig.\,\ref{fig:cluster2}b come from the cluster in DIS events which contains the 
  diquark, formed from two of the valence quarks.

% Cluster Mass Spectrum for joining and DIS clusters
% End of the Figure

We would expect the baryon-number violating clusters to be heavier
than the standard baryon number conserving clusters because
\begin{enumerate}

\item The baryonic cluster is formed from three quarks originating from three
{\it different\/} jets, as also shown in Fig.\,\ref{fig:cluster} of the
neutralino decay. In normal $\mr{e^+e^-\ra hadrons}$ events the
clusters joining partons from different jets are heavier than the
clusters which come entirely from partons from one jet,
Fig.\,\ref{fig:cluster2}a.

\item The new cluster contains a diquark and in general the clusters 
containing diquarks in for example the hadron remnant in deep
inelastic scattering are heavier than the normal quark-antiquark
clusters, Fig.\,\ref{fig:cluster2}b.
\end{enumerate}

As these clusters are heavier they will be more sensitive to the fine
details of the hadronization model. In particular these clusters are
sensitive to the maximum cluster mass before the clusters are split
and the details of this splitting mechanism. It is worth noting that
the same is true of the joining clusters in $\mathrm{e^+e^-\ra 3 \  jet}$
events, and it is precisely these clusters that contribute to the
`string effect', which is well described by HERWIG.

\section{Preliminary Results}
\label{sec-results}

We have implemented the R-parity violating decays and hard processes
into the HERWIG Monte Carlo event generator according to the
algorithms given in Sections~\ref{sec-angular} and~\ref{sec-hadron}.
They are available in the latest version
6.1 \cite{Marchesini:1991ch}. Having taken care to implement colour coherence
effects, it is of immediate interest to see whether they have a
significant influence on observable final-state distributions. To this
end we have studied some jet production processes and compared the
final-state distributions with those from standard QCD two-jet events.
It was observed in Ref.~\cite{Abe:1994nj} that certain variables can be
constructed that are particularly sensitive to colour coherence
effects. In particular these variables are sensitive to the presence
of colour connections that link the initial and final states.  To
investigate these effects for the different colour connection
structures of the \rpv\  models, we will study these variables for
two-jet production via resonant sparticle production in hadron-hadron
collisions. We essentially follow the details of the analysis of
Ref.~\cite{Abe:1994nj}.

\begin{figure}[htp]
\begin{center} \begin{picture}(360,80)(20,0)
\SetScale{0.7}
\SetOffset(30,0)
\ArrowLine(5,26)(60,52)
\ArrowLine(60,52)(5,78)
\DashArrowLine(90,52)(60,52){5}
\ArrowLine(90,52)(145,26)
\ArrowLine(145,78)(90,52)
\Text(85,20)[]{$\mr{u}$}
\Text(85,55)[]{$\mr{\bar{d}}$}
\Text(23,55)[]{$\mr{\bar{d}}$}
\Text(23,20)[]{$\mr{u}$}
\Text(55,46)[]{$\tilde{\tau}_{\alpha}$}
\Vertex(60,52){1}
\Vertex(90,52){1}
\SetOffset(155,0)
\ArrowLine(5,26)(60,52)
\ArrowLine(60,52)(5,78)
\DashArrowLine(90,52)(60,52){5}
\ArrowLine(90,52)(145,26)
\ArrowLine(145,78)(90,52)
\Text(85,18)[]{$\mr{d}$}
\Text(85,55)[]{$\mr{\bar{d}}$}
\Text(23,55)[]{$\mr{\bar{d}}$}
\Text(23,20)[]{$\mr{d}$}
\Text(55,46)[]{$\nut_\tau$}
\Vertex(60,52){1}
\Vertex(90,52){1}
\SetOffset(280,-35)
\ArrowLine(60,128)(5,128)
\ArrowLine(115,128)(60,128)
\ArrowLine(5,76)(60,76)
\ArrowLine(60,76)(115,76)
\DashArrowLine(60,76)(60,128){5}
\Text(65,45)[]{$\mr{d}$}
\Text(65,98)[]{$\mr{\bar{d}}$}
\Text(23,98)[]{$\mr{\bar{d}}$}
\Text(23,45)[]{$\mr{d}$}
\Text(50,71)[]{$\nut_\tau$}
\Vertex(60,128){1}
\Vertex(60,76){1}
\end{picture}
\end{center}
\caption{Feynman diagrams for $\mr{\bar{u} d \longrightarrow \bar{u} d}$ and
 $\mr{\bar{d} d \longrightarrow \bar{d} d}$.}
\label{fig:LQDbump}
\end{figure}
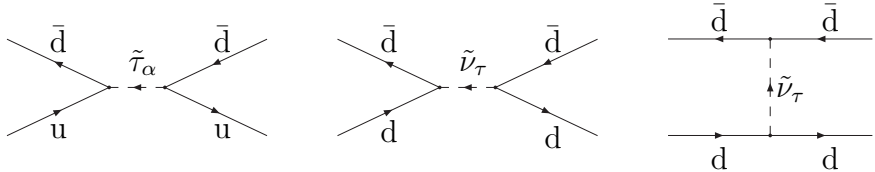

As  examples we study the processes 
\begin{itemize}
\item $\mr{\bar{u} d \longrightarrow \bar{u} d}$ via a resonant
stau, and $\mr{\bar{d}d \longrightarrow \bar{d} d}$ via a resonant tau
sneutrino, this occurs via the coupling ${\lam'}_{311}$.
The Feynman diagrams for this process are shown in
Fig.\,\ref{fig:LQDbump}.  This process involves lepton-number
violating couplings but no baryon number violating vertices.

\item Resonant squark production via the coupling ${\lam''}_{212}$. This leads to
 resonant down, strange and charm squark production.  
 The resonant diagram for this process is shown in Fig.\,\ref{fig:bump}.
\end{itemize}

%
% Feynman diagrams of the processes 
%
% end of the figure
These couplings were chosen to try and maximize the cross section given the  
experimental constraints on the couplings. The coupling ${\lam'}_{311}$
has an upper
bound, at the $2\sigma$ level, given by \cite{Allanach:1999ic}
\beq
   {\lam'}_{311} < 0.11\left(\frac{M_{\mr{\dnt}_R}}{100\mathrm{\gev}}\right).
\eeq
While the bounds on other LQD couplings are weaker they involve higher generation
quarks and hence the cross sections will be suppressed by the parton luminosities.

Similarly for the resonant squark production the couplings that couple two
first-generation quarks would in principle give the highest cross
sections. However, the limits on these couplings are so strict that we
used ${\lam''}_{212}$, which is only limited by perturbativity
\cite{Allanach:1999ic}.

We would expect very different results for the variables that are
sensitive to the initial-final state colour connections for these two
processes. The first process only has such colour connection in the
relatively suppressed $t$-channel sneutrino diagram, whereas the second
process has initial-final state colour connections in the resonant
diagram due to the random colour structure at the \bv\  vertex. This is
the effect we wish to demonstrate explicitly below. 

\subsection{Resonant Slepton Production}

For resonant slepton production we consider both the signal and the
background which were generated using the program described in
Ref.~\cite{Marchesini:1991ch}. The only cut made was to require the presence of
at least one jet with $E_T$ of greater than 200\gev\  in the event. A
parton level cut requiring the $p_T$ of the two final-state partons to
be greater than 150\gev\  each, was used to reduce the number of events
we needed to simulate, however this should not affect the results. The
signal points were generated using the following SUGRA parameters,
$M_0=600$\gev, $M_{\frac{1}{2}}=200$\gev, $A_0=0$\gev, $\tan\beta=10$,
and $\mathrm{sgn\mu=+}$. At this SUGRA point the right down squark mass
is $728$\gev\  which corresponds to a limit of ${\lam'}_{311}<0.80$.
The third generation lepton masses are all close to 600~GeV,
$M_{\tilde{\tau}_1}=599$~GeV, $M_{\tilde{\tau}_2}=617$~GeV,
$M_{\tilde{\nu}_\tau}=610$~GeV. These processes have previously been considered in
\cite{Hewett:1998fu,Allanach:1999bf}.

The results in all the graphs correspond to the
number of events at Run II of the Tevatron, with centre of mass energy
of 2 TeV, and integrated luminosity of 2$\mathrm{fb}^{-1}$.

\begin{figure}[htp]
\includegraphics[angle=90,width=0.4\textwidth]{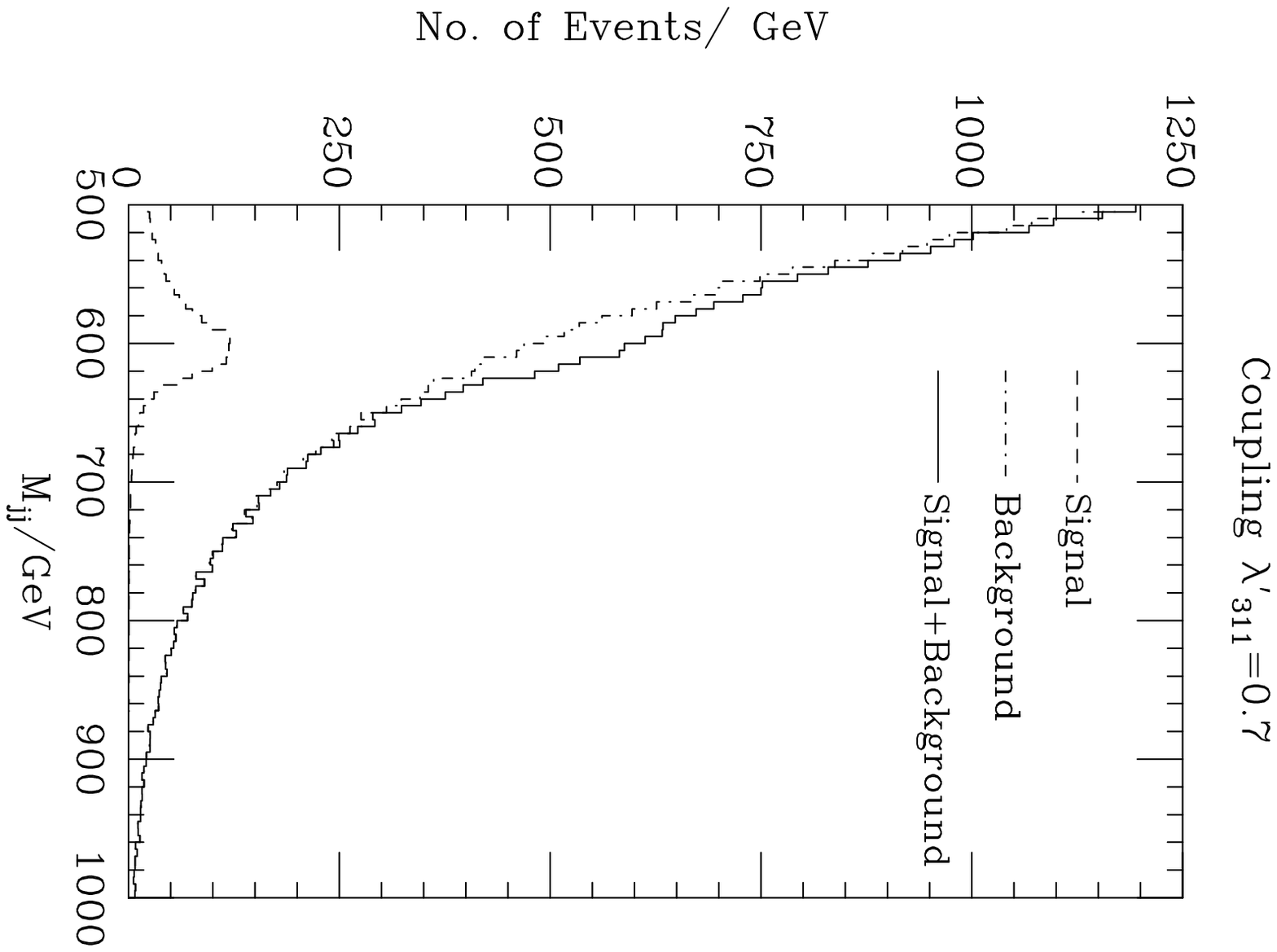}
\hfill
\includegraphics[angle=90,width=0.4\textwidth]{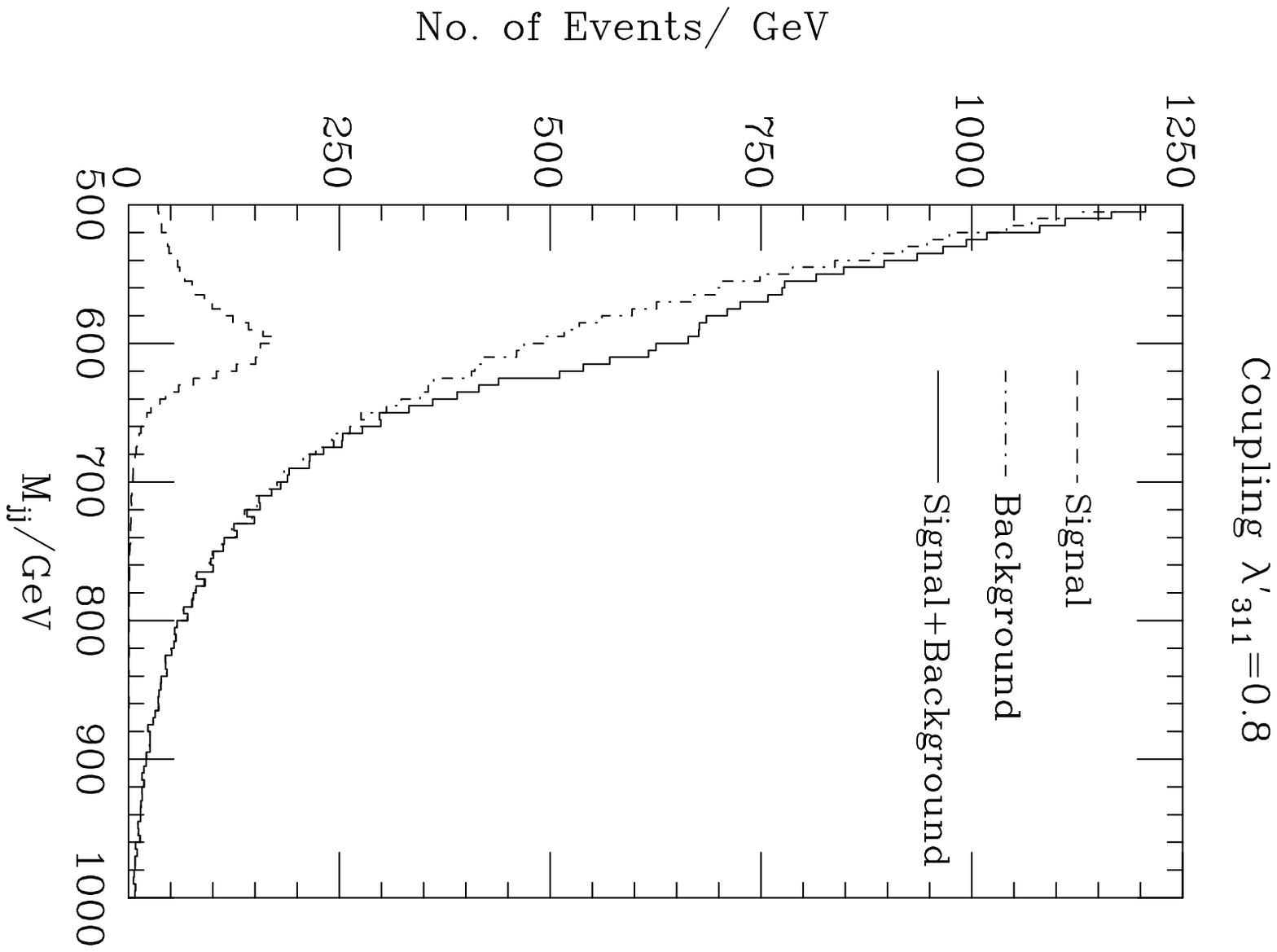}
\caption{Di-jet invariant mass distribution for ${\lam'}_{311}=
0.7$ and ${\lam'}_{311}=0.8$.}
\label{fig:invm}
\end{figure}

As can been seen in Fig.\,\ref{fig:invm}, the results for two
different values of the coupling show that there is a bump in the
di-jet invariant mass distribution, $M_{jj}$, from the resonant
particle production for large values of the coupling.
%
% Invariant mass distribution of slepton
%
% End of the Figure

We can now study the events around the bump, $580 \mathrm{\gev}\leq M_{jj}
\leq 640 \mathrm{\gev}$, 
in the distribution and plot the variables that
are sensitive to angular ordering for these events \cite{Abe:1994nj}.
These variables depend on the distribution of a third jet in the
events which is generated in the simulation by the parton-shower
algorithm. The three relevant variables are: $\eta_3$, $R$ and $\al$.
They are defined in the following way: (1) If we define the jets in
the order of their $E_T$, with jet 1 being the hardest jet in the
event, then $\eta_3$ is the pseudo-rapidity of the third jet. (2)
Defining $\Delta\eta \equiv\eta_3-\eta_2$ and the difference in polar
angles $\Delta \phi \equiv \phi_3-\phi_2$, then the variable
$R\equiv\sqrt {\Delta\eta^ 2+\Delta \phi^2}$. This is the distance
between the second and third jets in the $(\eta,\phi)$ space.
(3) If we first define $\Delta H \equiv \mr{sgn}(\eta_2) \Delta\eta$, 
we can then consider the polar angle in the $(|\Delta\phi|,\Delta H)
$ space. This is $\al\equiv\tan^{-1}(\Delta H/|\Delta\phi|)$.

In the analysis in Ref.~\cite{Abe:1994nj} additional cuts were imposed
in terms of these new variables, which we also implement in our
analysis
\begin{enumerate}
  \item A pseudo-rapidity cut on the two highest $p_T$ jets in the
                event, $|\eta_1|,|\eta_2|<0.7$.

  \item Requiring the two leading jets in $E_T$ to be back to back 
        $||\phi_1-\phi_2|-\pi|<20^0$.

  \item We require the transverse energy of the third jet, $E_{T3}
        >10 \mathrm{\gev}$ to avoid background from the underlying event.

  \item For the study of $\al$ only, we make the additional cut $1.1<R
        <\pi$ to avoid problems with overlapping cones in the jet
        clustering algorithm.
\end{enumerate}

\begin{figure}[htp]
\includegraphics[angle=90,width=0.3\textwidth]{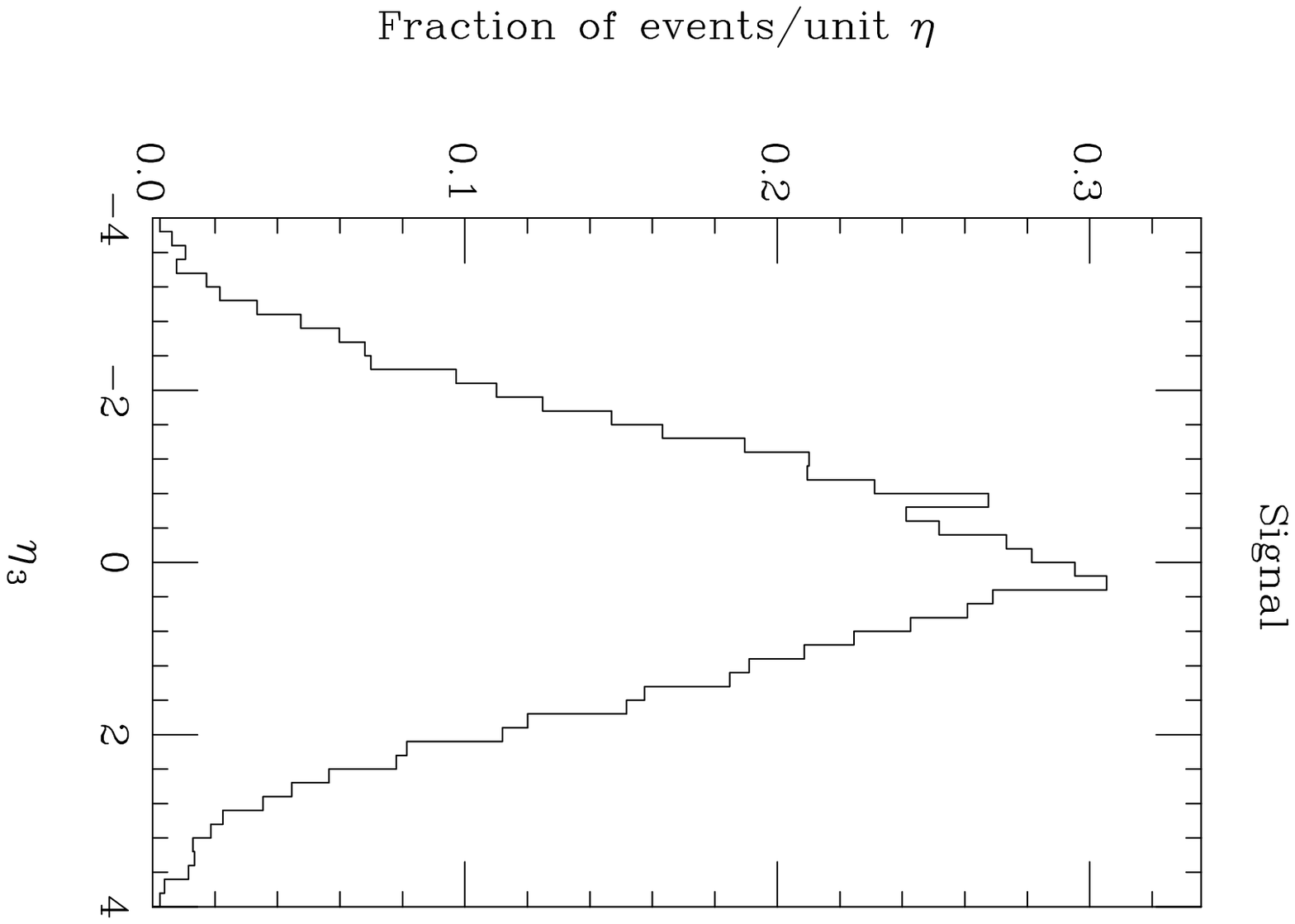}
\hfill
\includegraphics[angle=90,width=0.3\textwidth]{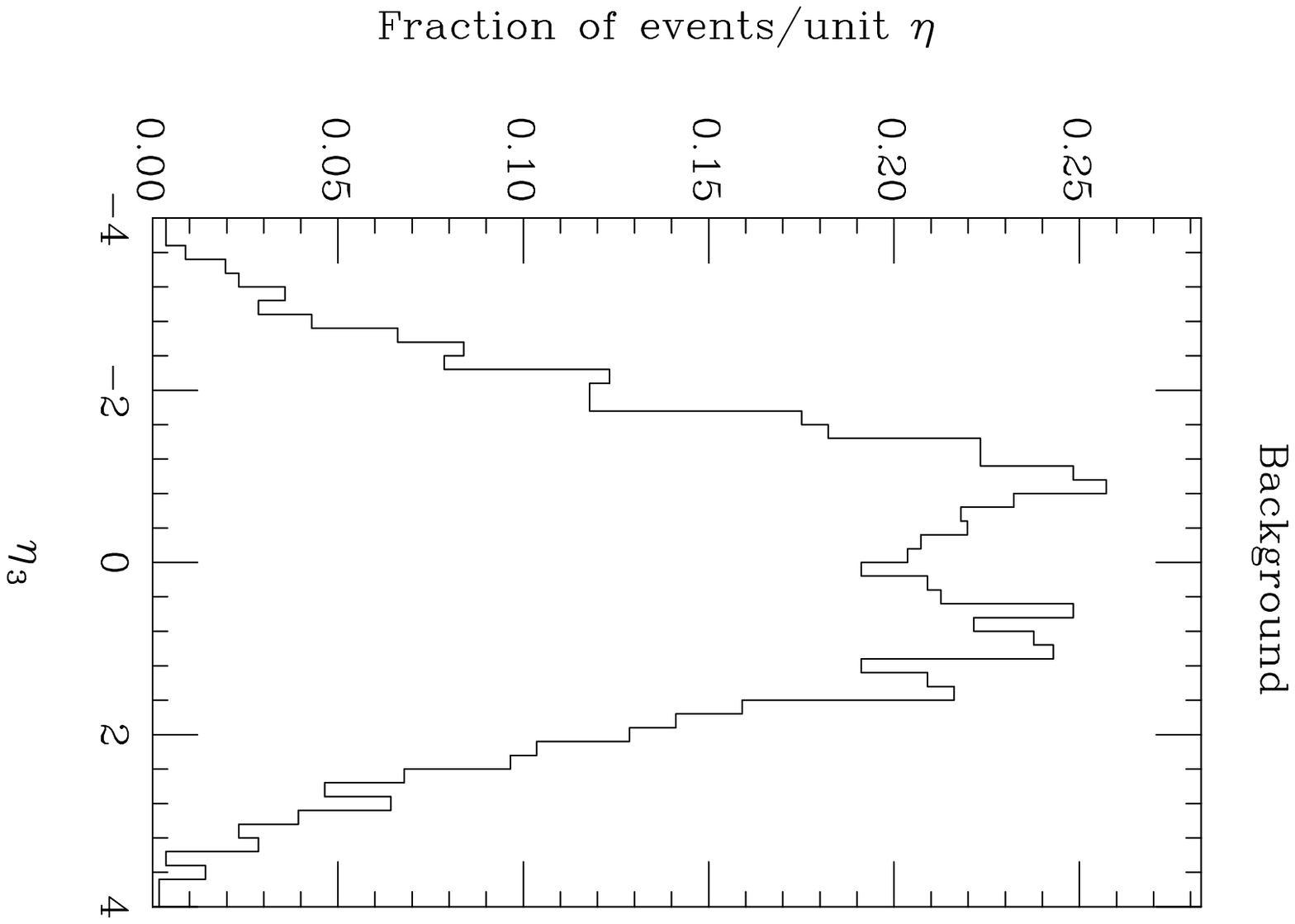}
\hfill
\includegraphics[angle=90,width=0.3\textwidth]{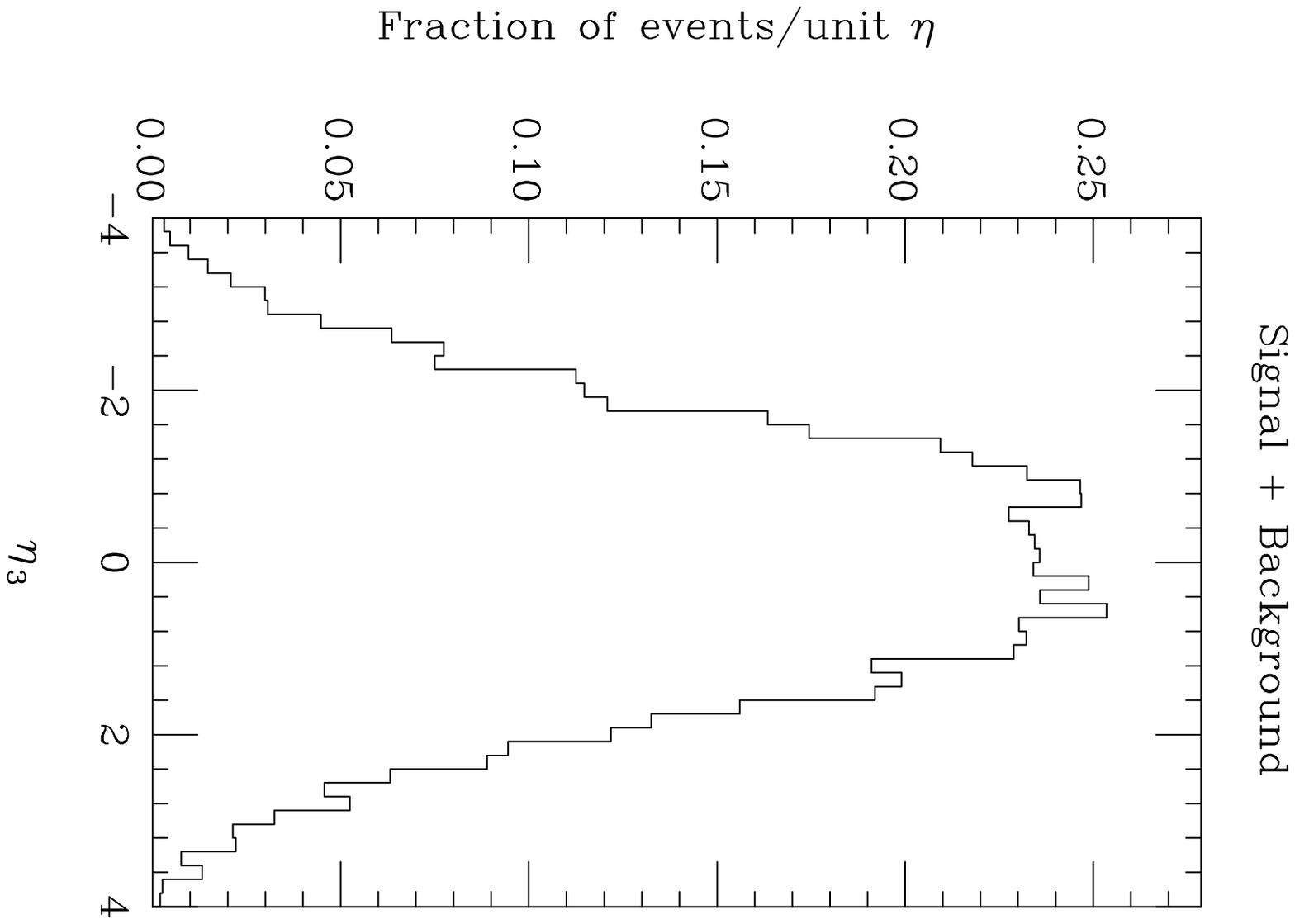} 
\\
\caption{The distribution of events in $\eta_3$ for (a) the resonant slepton 
production, (b) the QCD background, and (c) the combination of the two.}
\label{fig:LQDeta}
\end{figure}

We can now study the distributions for the signal, the background and
the signal plus background for the resonant slepton production with
coupling ${\lam'}_{311}=0.8$.  There are significant differences
between the signal and the background for this process. In the $\eta_3$
distributions, Fig.\,\ref{fig:LQDeta}, instead of a dip in the background at $
\eta_3=0$ there is a bump in the signal. This dip in the QCD
background was observed in \cite{Abe:1994nj}, and is a feature of the
initial-final state colour connection. In our study it is present in
the background, but not the signal.

%
%  Distribution of eta_3 for the LQD processes
%
% end of the figure

\begin{figure}[htp]
\includegraphics[angle=90,width=0.3\textwidth]{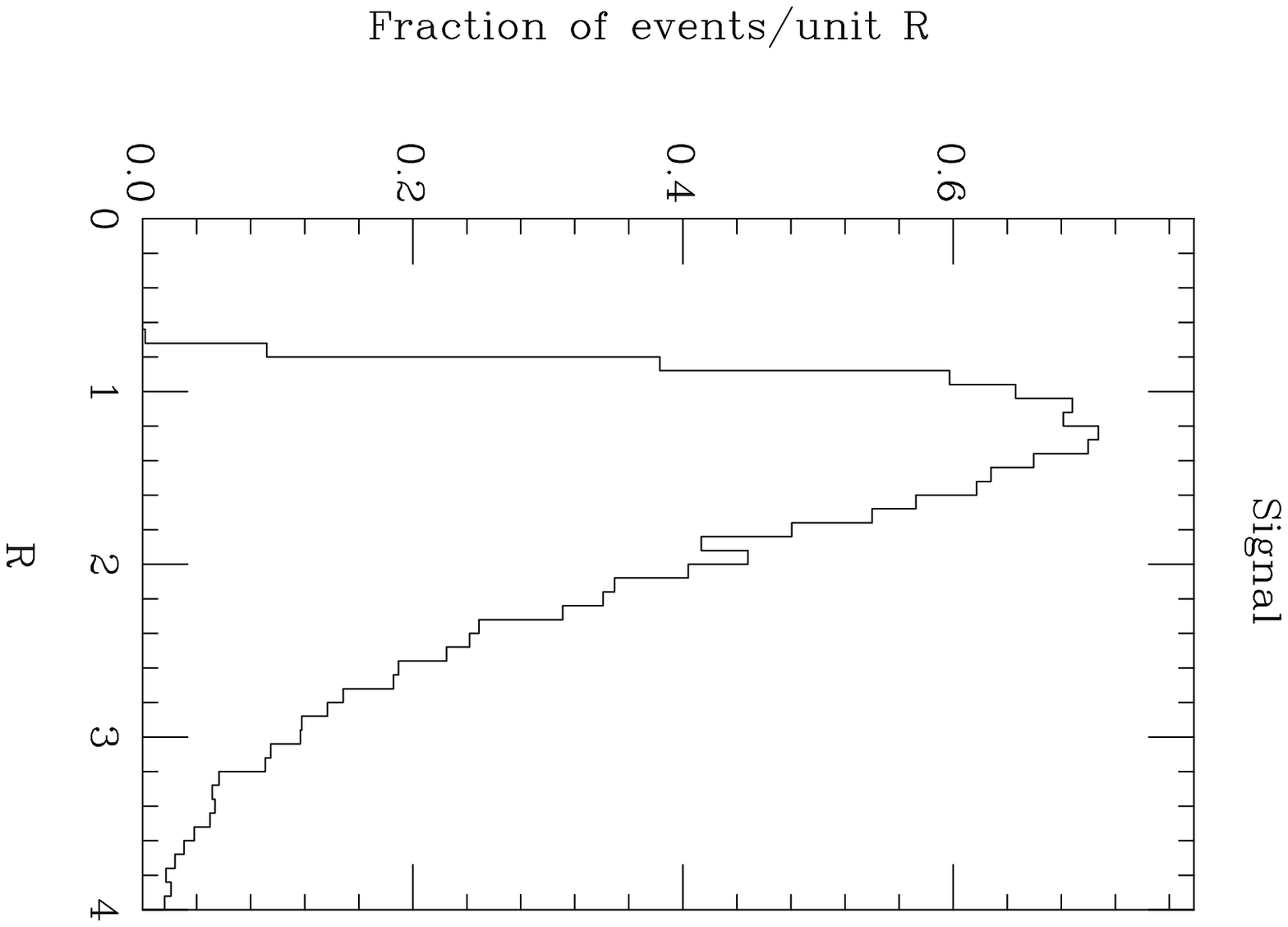}
\hfill
\includegraphics[angle=90,width=0.3\textwidth]{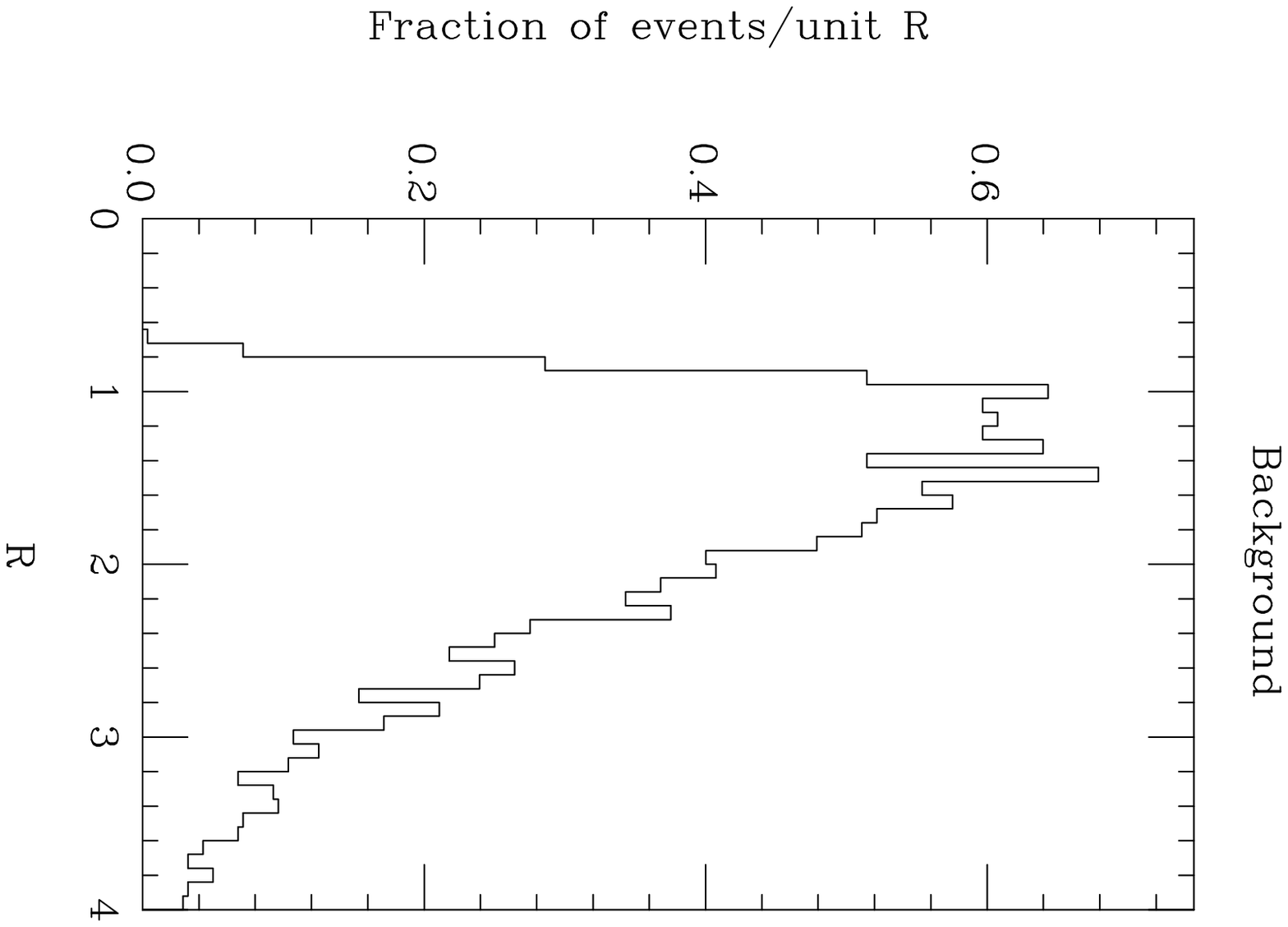}
\hfill
\includegraphics[angle=90,width=0.3\textwidth]{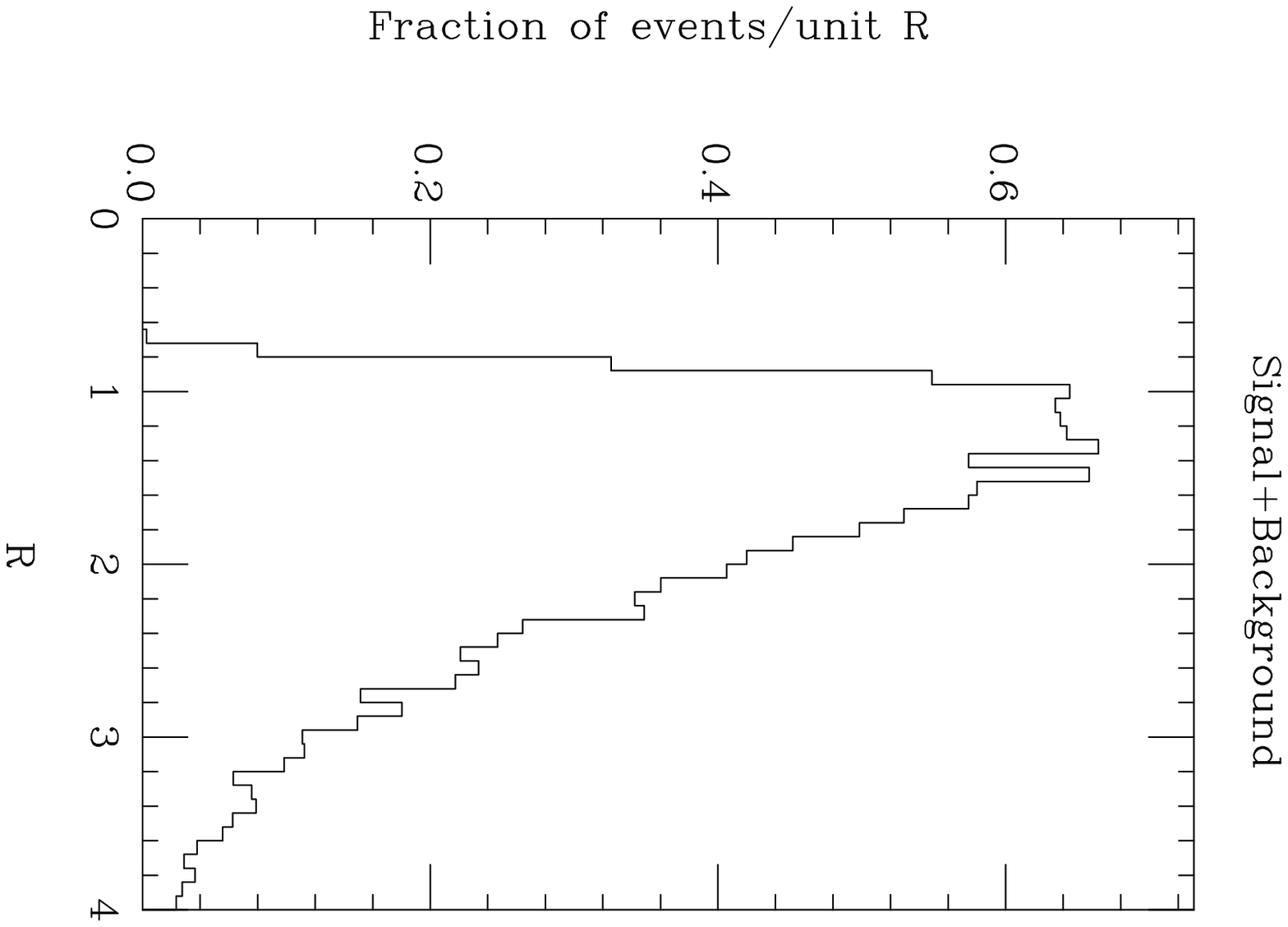} 
\\
\caption{The distribution of events in $R$ for (a) the resonant slepton 
production, (b) the QCD background, and (c) the combination of the two.}
\label{fig:LQDR}
\end{figure}

\begin{figure}[htp]
\includegraphics[angle=90,width=0.3\textwidth]{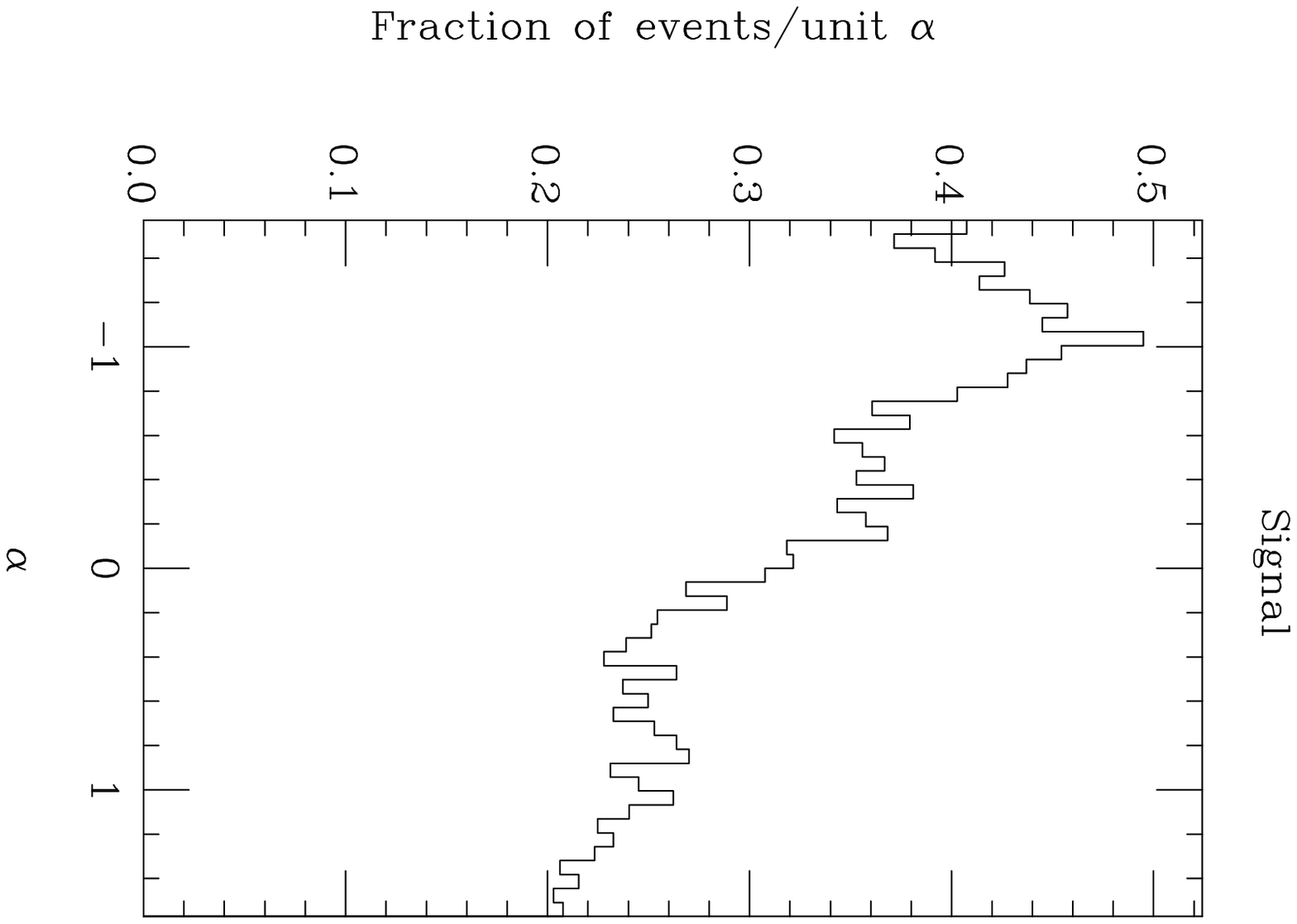}
\hfill
\includegraphics[angle=90,width=0.3\textwidth]{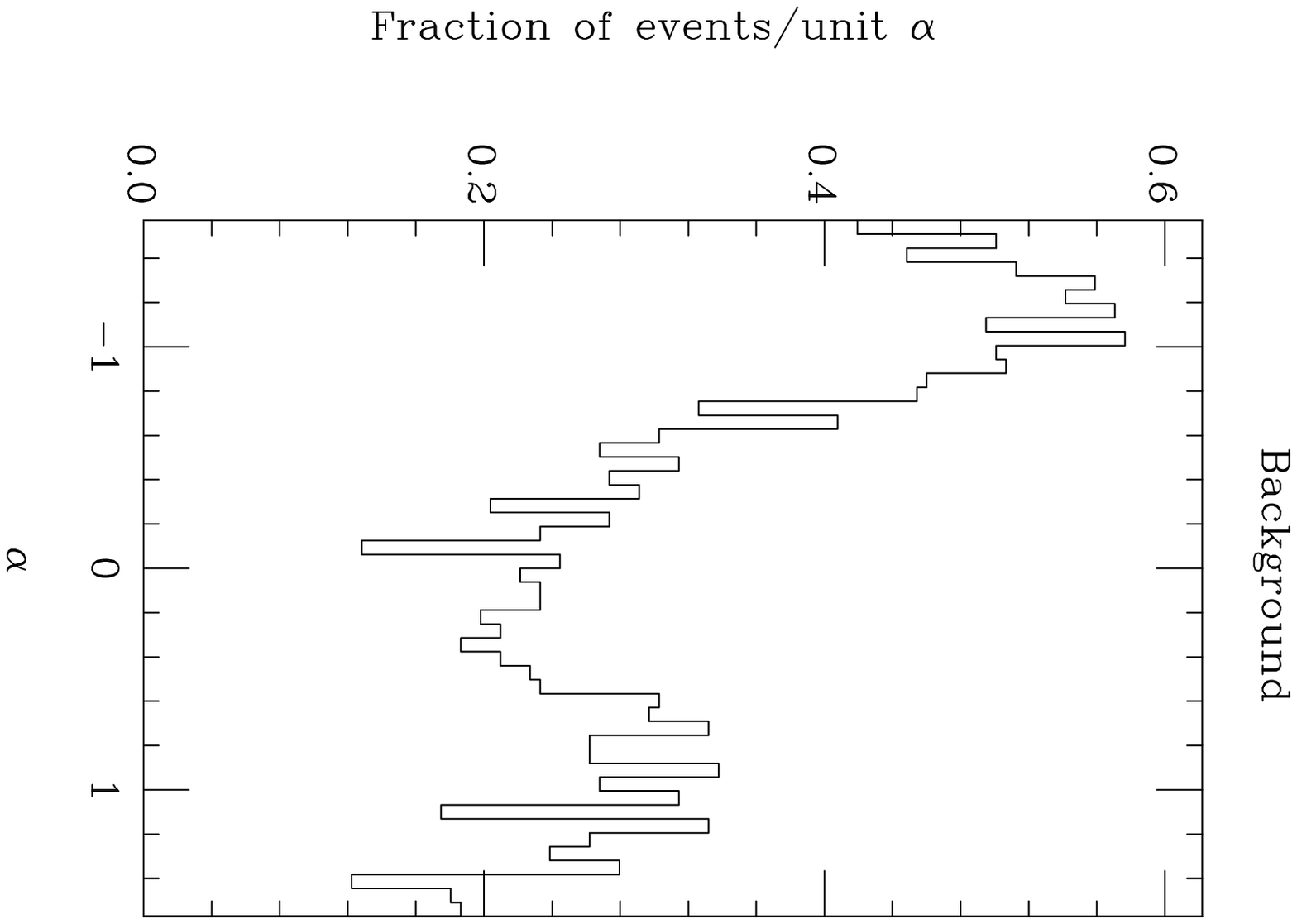}
\hfill
\includegraphics[angle=90,width=0.3\textwidth]{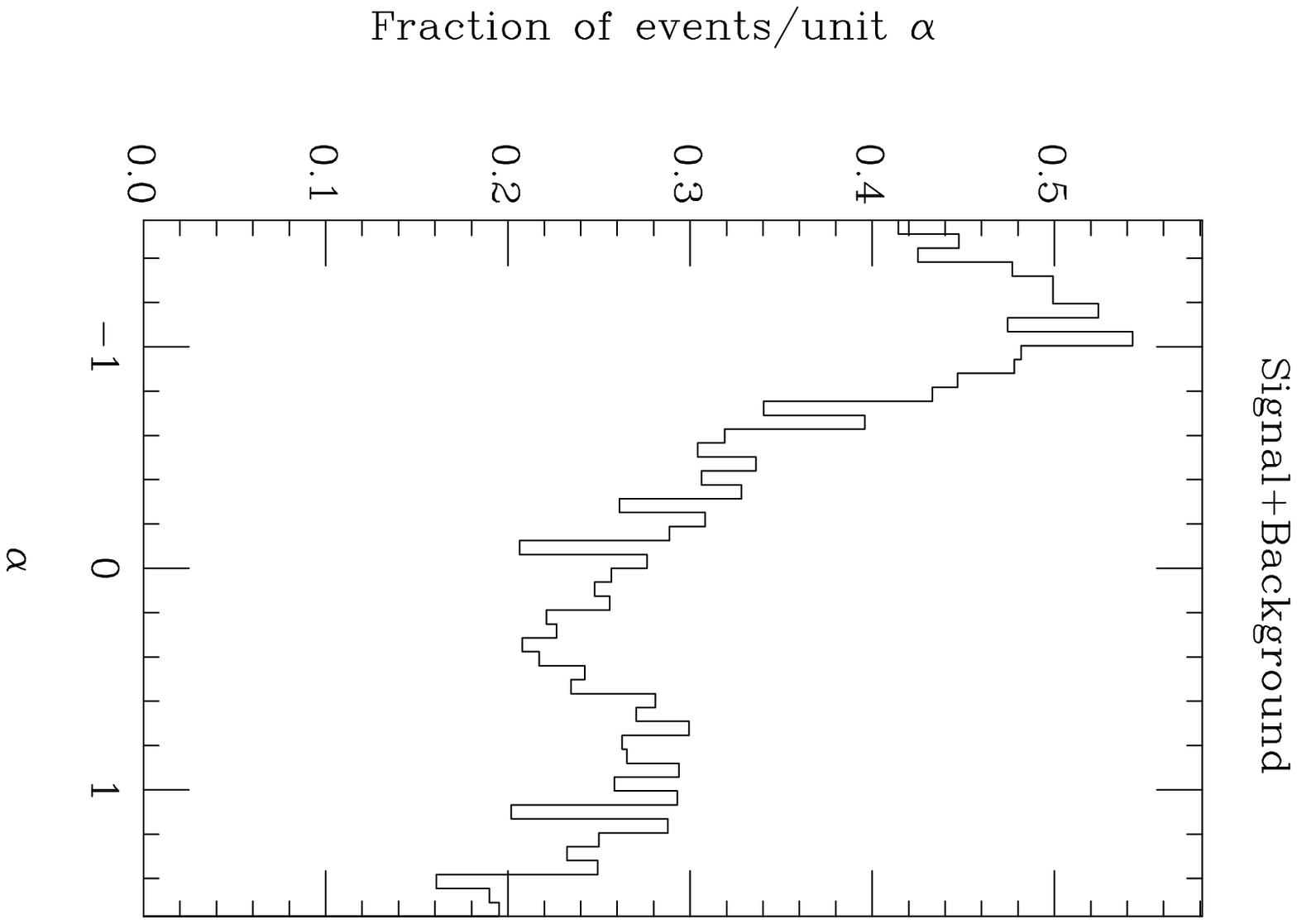} 
\\
\caption{The distribution of events in $\al$ for (a) the resonant slepton
production, (b) the QCD background and (c) the combined events.}
\label{fig:LQDalpha}
\end{figure}

The distribution of events in $R$, shown in Fig.\,\ref{fig:LQDR}, is very similar for
both the signal and background. In fact in the study of \cite{Abe:1994nj} all
the event generators, even those which do not include angular ordering, 
gave good agreement with the data for this observable. The distribution
of events in $\al$, Fig.\,\ref{fig:LQDalpha}, also shows a difference
between the signal and the background, with the signal not showing the
dip in the middle.  This is again an effect of the initial-final state
colour connection which is present in the background but not in the
signal.

%
%  Distribution of R for the LQD processes
%
% end of the figure

%
%  Distribution of alpha for the LQD processes
%

\begin{figure}[htp]
\includegraphics[angle=90,width=0.4\textwidth]{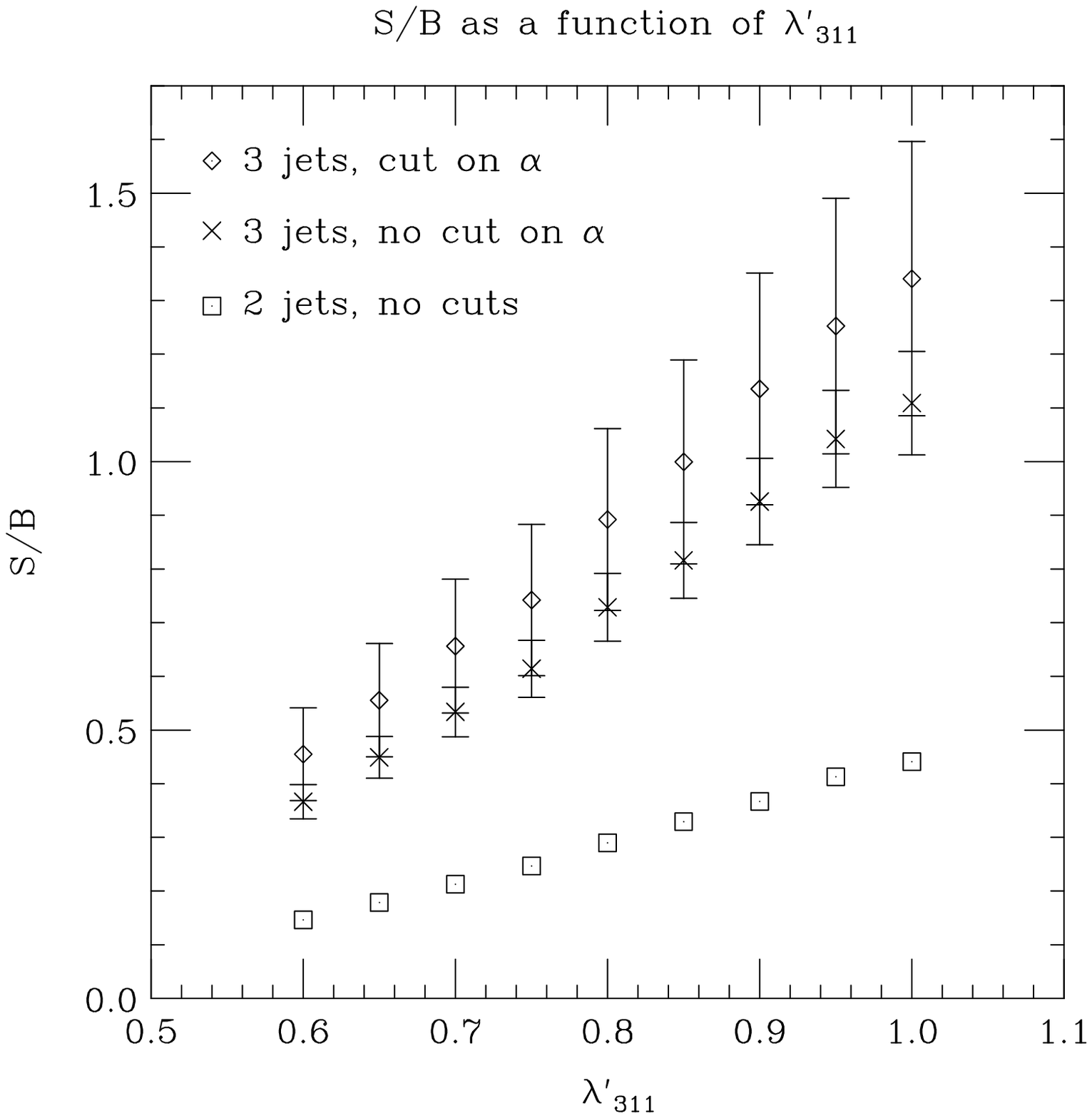}
\hfill
\includegraphics[angle=90,width=0.4\textwidth]{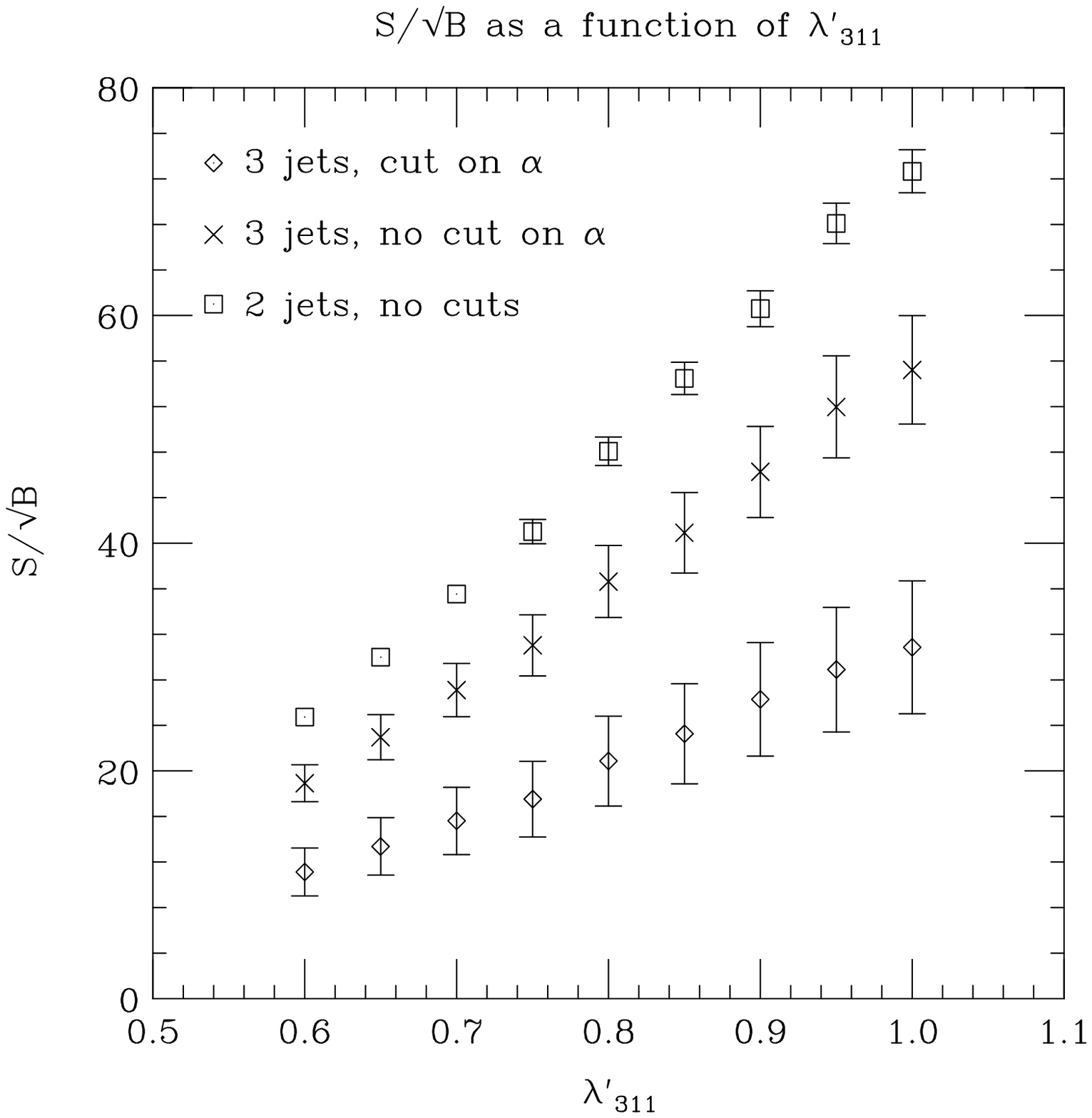}
\caption{Effect of the cuts on the angular ordering variables as a
function of ${\lam'}_{311}$.}
\label{fig:cuts}
\end{figure}

\begin{figure}[htp]
\includegraphics[angle=90,width=0.4\textwidth]{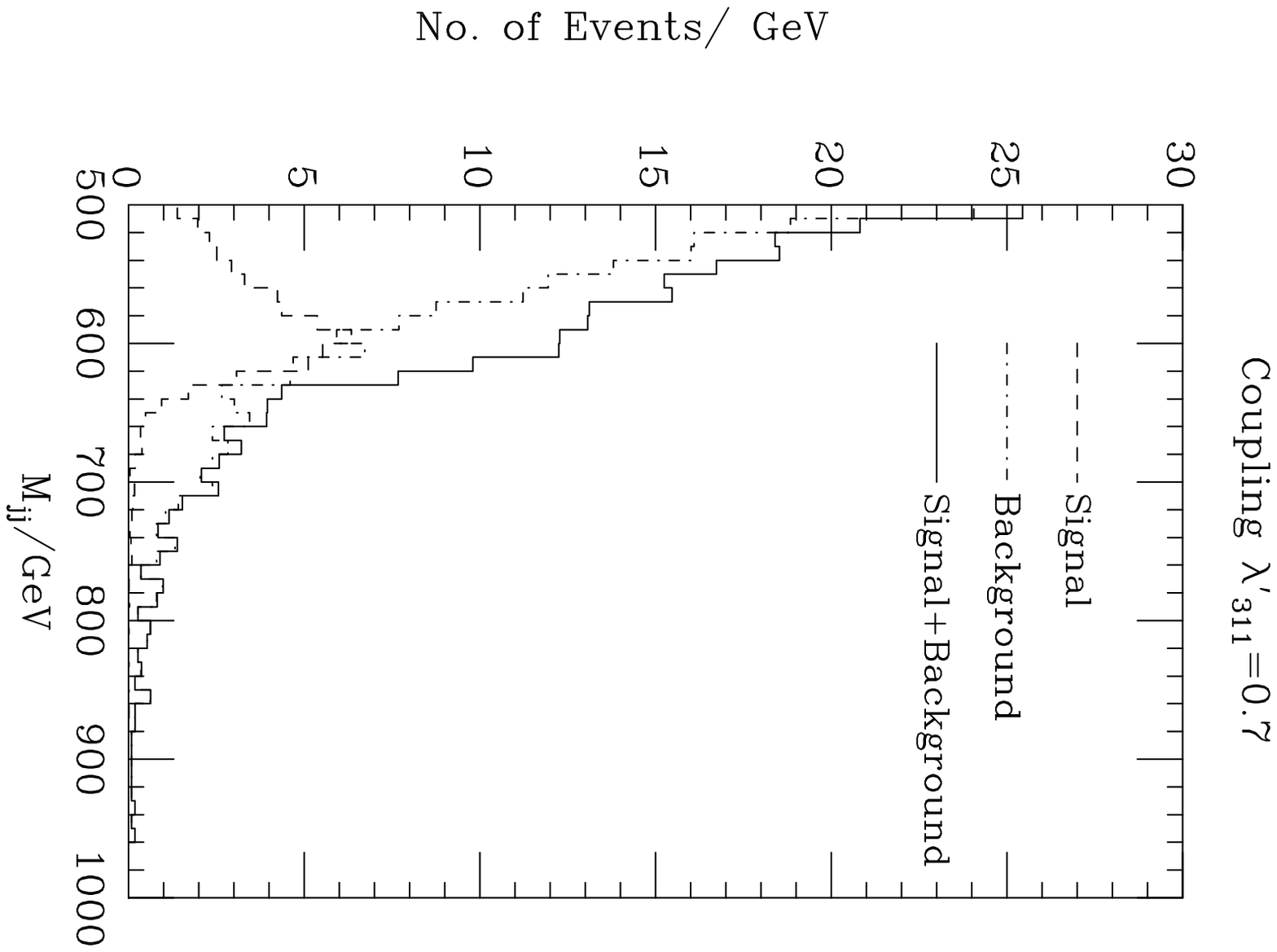}
\hfill
\includegraphics[angle=90,width=0.4\textwidth]{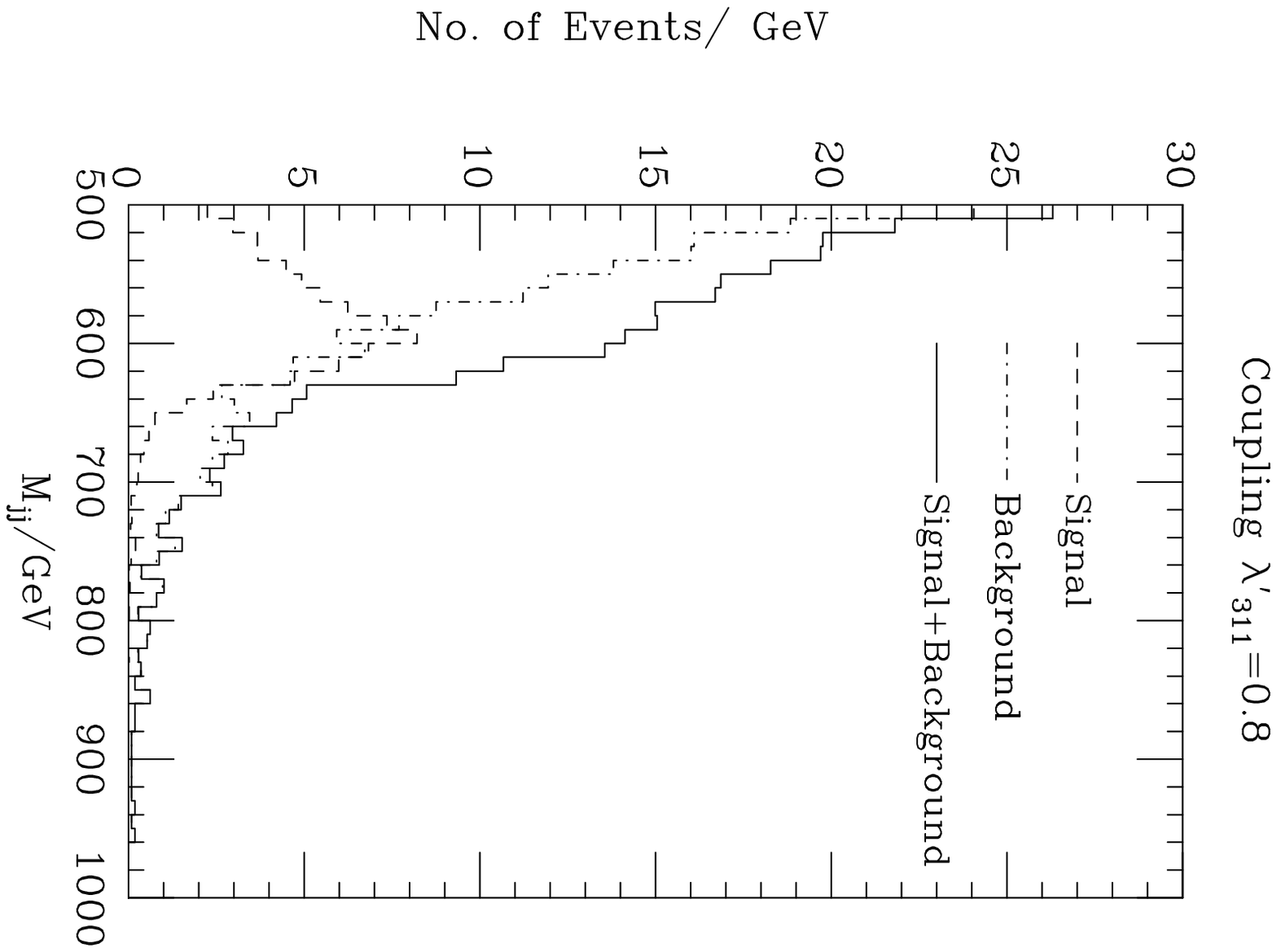}
\caption{Invariant mass distribution for ${\lam'}_{311}=0.7$ and
${\lam'}_{311}=0.8$ after cuts on the angular ordering variables.}
\label{fig:invm2}
\end{figure}

As can be seen in all the distributions apart from the disappearance of
the dip at $\eta_3=0$, once the signal and
background are added the effect of the signal is minimal.
While there are differences between the signal and background 
it is hard to see how cuts can be applied on these variables to improve the extraction
of a signal over the QCD background. The only major difference which can be cut on
is the difference in the distribution of $\al$.
  We consider two approaches to increase the 
ratio of signal to background $S/B$.
\begin{enumerate}
 \item Accept all the events with at least three jets, provided they
       pass the cuts described above from the analysis of \cite{Abe:1994nj}.

 \item Reject all the two jet events and only accept the events with
 more than two jets provided that $|\al|\le\al_{cut}$. We apply a cut
 of $\al_{cut}=0.4$ for the these jet events.
\end{enumerate}
These cuts were chosen to maximize $S/B$ while not reducing $S/\sqrt
{B}$ below five. As can been seen in Fig.\,\ref{fig:cuts} both of these cuts 
significantly increase the $S/B$. This can be seen in the effect on the
invariant mass distribution with the second cut, Fig.\,\ref{fig:invm2}.
In the invariant mass distribution 
the signal is now more visible over the background.

This shows that by using the colour coherence effects we can improve
the extraction of a signal. We would expect obtaining a large $S/B$ to
be important for this process because we do not have an accurate
prediction for the QCD background. However given the limits on this coupling
this signal will only be visible above the background at the highest couplings 
currently allowed by low energy experiments. In \cite{Hewett:1998fu,Allanach:1999bf}
it was suggested that by using the sidebands to normalize the background
that resonant slepton production could be probed to much smaller values of the couplings.
Indeed the $S/\sqrt{B}$ numbers in Fig.\,\ref{fig:cuts} suggest that without 
any of our additional cuts the signal is visible at much lower couplings. However their
results were obtained using the narrow-width limit for the production cross section and
did not included the effects of QCD radiation. Our results suggest that after including
these effects the signal will only be visible for large values of the coupling.
It may be possible  to use the
sidebands which we have removed with our cuts to normalize this
background, as in \cite{Hewett:1998fu,Allanach:1999bf}, to improve the extraction of the
signal. However this may not be possible due to the increased width of the resonance, 
Fig.\,\ref{fig:invm}, due to QCD radiation.
The situation will hopefully improve with the availability of
a next-to-leading order calculation for the QCD background.

%
%  Effect of the cuts for varying couplings
%
% end of figure
%
%  Effect of the cuts on the LQD mass distribution
%
%  end of figure

\subsection{Resonant Squark Production}  

The cross section at the Tevatron for resonant squark
 production\footnote{Resonant squark production via $U_iD_jD_k$
 has previously been considered in \cite{Dimopoulos:1990fr}.} is
much lower than the resonant slepton production due to the reduced
parton luminosity.

\begin{figure}[htp]
\includegraphics[angle=90,width=0.3\textwidth]{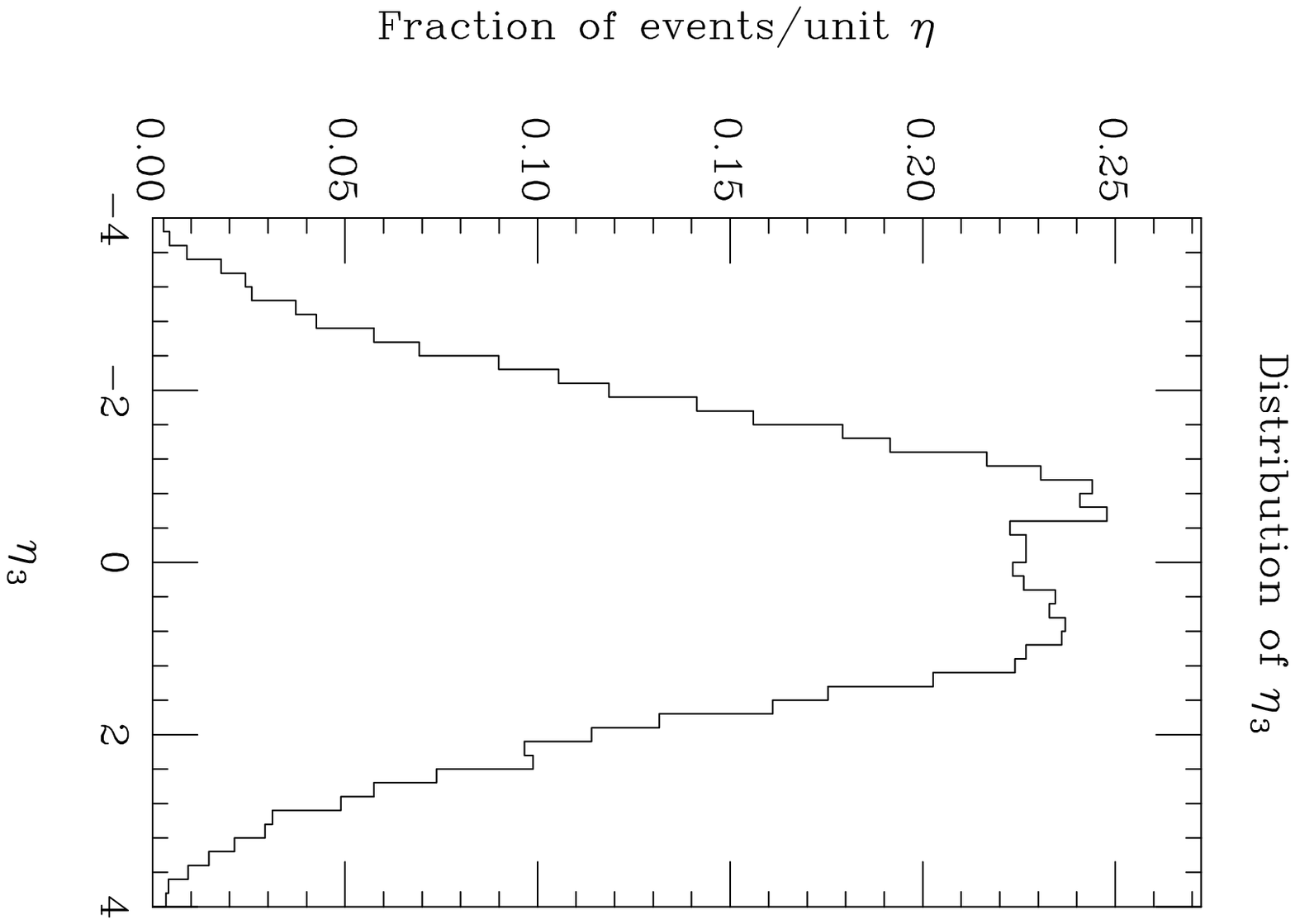}
\hfill
\includegraphics[angle=90,width=0.3\textwidth]{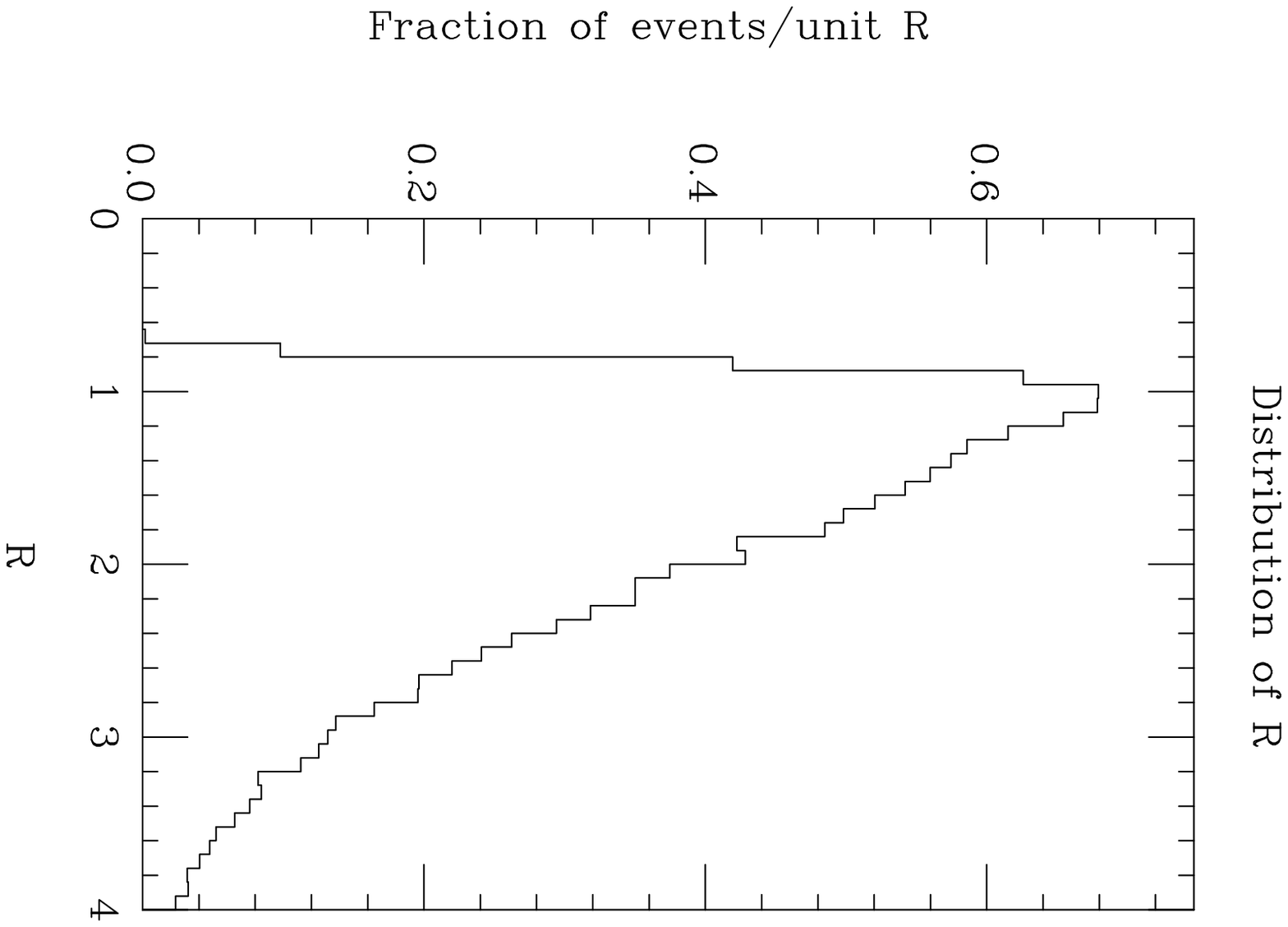}
\hfill
\includegraphics[angle=90,width=0.3\textwidth]{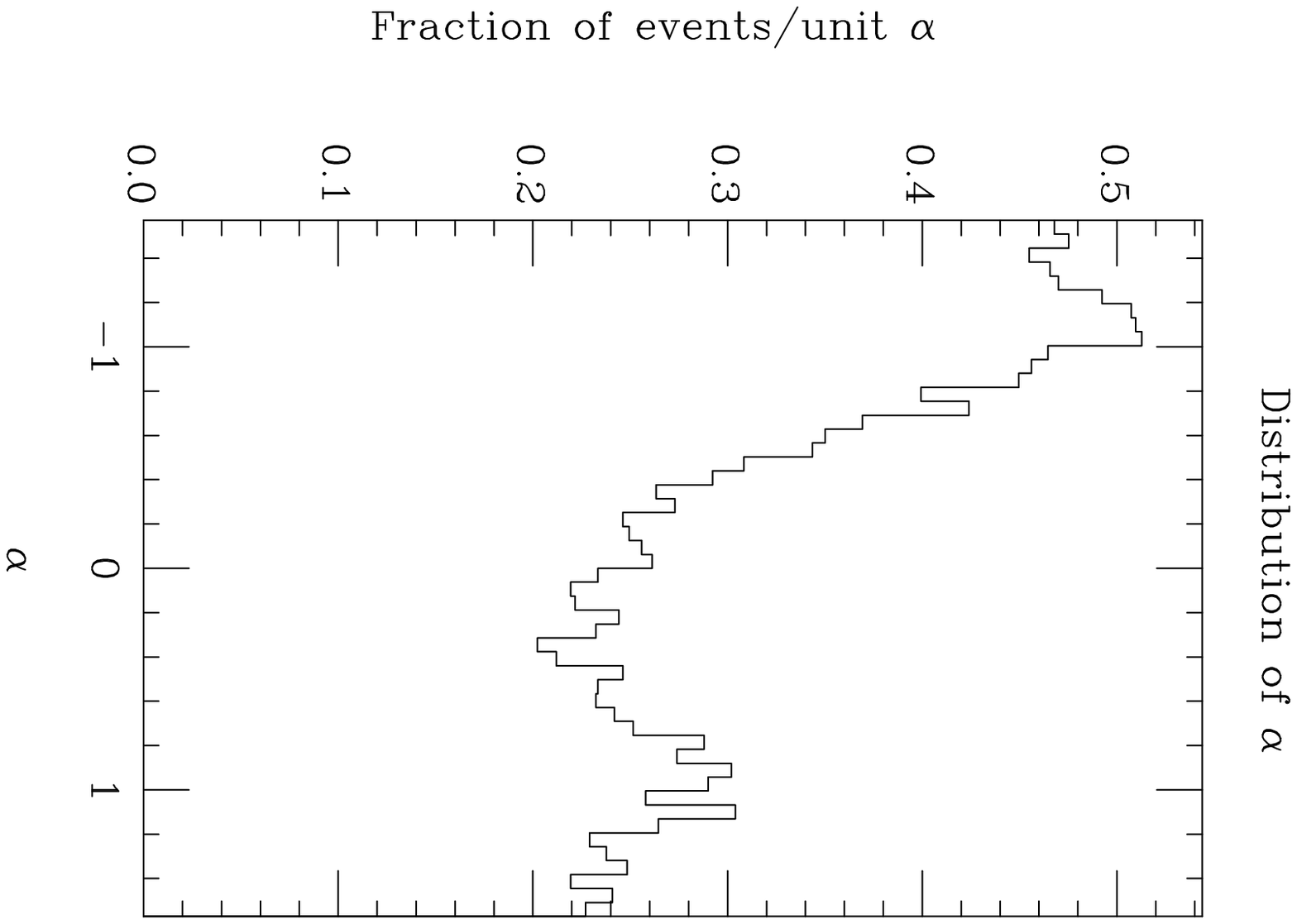} 
\\
\caption{Distributions of the signal for resonant squark production.}
\label{fig:baryon}
\end{figure}

 (This will be reversed at the LHC.) It is therefore
unlikely that an excess of events can be seen over the QCD background.
It is however still interesting to look at the distributions of the
angular ordering variables.  The
signal point was generated using the following SUGRA parameters,
$M_0=500$\gev, $M_{\frac{1}{2}}=200$\gev, $A_0=0$\gev, $\tan\beta=10$,
and $\mathrm{sgn\mu=+}$ and ${\lam''}_{212}=1$. At this point the
squark mass $M_{\mr{\dnt}_R,\mr{\tilde{s}}_R,\mr{\tilde{c}}_R}=601$~GeV.
As can be seen in Fig.\,\ref{fig:baryon},
there is now less difference in the shape of the distribution between
the signal and the background ({\it c.f.\/}  Figs.\,\ref{fig:LQDeta},
\ref{fig:LQDR},~\ref{fig:LQDalpha}). The resonant squark production
shows a dip at $\eta_3=0$ and a rise as $\alpha\ra\frac{\pi}{2}$, which
is due to the colour connection between the initial and final
states. The effect is slightly less than for the QCD background as
there are combinatorially fewer such connections.

  The fact that the final state distributions of the resonant slepton
  and resonant squark production processes are so different, despite
  the identities and kinematics of the jets themselves being so
  similar, clearly shows that colour coherence plays an important role
  in determining the properties of R-parity violating processes. Even
  if this is not used as a tool to enhance the signal, it is likely
  that it will affect the efficiency of any cuts that are applied, so
  it is essential that any experiments looking for R-parity violating
  processes take into account colour coherence in their simulations of
  the signal.

  Even if R-parity violating hard processes were added to ISAJET
  \cite{Paige:1998xm}, it would not be expected to describe the final
  state well, as it is based on the incoherent parton shower and
  independent fragmentation models. Thus, in our case for example, the
  resonant slepton and resonant squark processes would have very
  similar properties. It is worth noting that ISAJET gives a 
  poor description of the CDF data  \cite{Abe:1994nj} on $\eta_3$, $R$
  and $\al$ in standard QCD two-jet events.
%
%  Angular Ordering Variables for the squark production
%
% end of this figure
%
%  Conclusions
%
\section{Conclusion}

  We have presented a procedure for implementing colour coherence
effects via the angular ordering procedure in R parity violating SUSY
models. We find that the baryon number violating processes have a random
colour connection structure
for angular ordering. In these processes we see for the first time
differences in the colour partners for the colour coherence effects
and those used with the idea of colour preconfinement for
hadronization in the cluster model. 

  A full set of decays and hadron-hadron cross sections have now been 
implemented in the HERWIG Monte Carlo event generator 
\cite{Marchesini:1991ch,SUSYimplement}. The first preliminary
results for these processes show that the inclusion of colour
coherence is important and that for some processes we can can
use the colour coherence properties of the processes to help extract an
R parity violating SUSY signature. 

  The availability of a full simulation should allow a more
  detailed experimental study of these processes for the first time.
%
%   Acknowledgment
%

\section*{Acknowledgments}

  P. Richardson would like to thank PPARC for a research studentship,
  number \linebreak PPA/S/S/1997/02517. We thank P. Gondolo for
  helpful discussions.
%
%   Appendix
%
\appendix
\section{Conventions}

Here we present all the matrix elements for R-parity violating two-
and three-body decays of sparticles as well as the the matrix elements
for single sparticle production via $2\ra2$ scattering processes. We
disregard those possibilities where the sfermion resonance
kinematically is not probed, \eg
\beq
d_j+{\bar d}_k\ra {\tilde\nu}_i\ra{\tilde\nu}_i+Z^0.
\eeq
We also do not consider the processes generated from quark-photon
scattering as discussed in \cite{Allanach:1997sa}.  First we present
the decay matrix elements of the sfermions, neutralinos, charginos and
the gluino. We will then give the matrix elements for the cross
sections most of which can simply be obtained by crossing the various
decay matrix elements. Throughout we allow for more than one \rpv\ 
coupling to be non-zero.
 
We follow the conventions of \cite{Haber:1985rc,Gunion:1986yn} for the
neutralino and chargino mixing matrices and the convention of
\cite{Baer:1994xr} for the mixing of the sfermions. For the current
eigenstates $\tilde{q}^i_{L,R}$ and the mass eigenstates
$\tilde{q}^i_{1,2}$ the mixing is 
\beq
\left( \begin{array}{c} \tilde{q}^i_L \\ \tilde{q}^i_R \end{array}\right) 
= \left( \begin{array}{cc} \cos\theta^i_q & \sin\theta^i_q \\
-\sin\theta^i_q & \sin\theta^i_q\end{array}\right)
\left( \begin{array}{c} \tilde{q}^i_1 \\ \tilde{q}^i_2 \end{array}\right).
\eeq
We denote the mixing matrix above as $Q^i_{jk}$ where $i=u,\,d,\,s,\,
c,\,b,\,t$ is the quark flavour index. The analogous slepton mixing
matrix is denoted $L^i_{jk}$, where $i=\mr{e^-},\,\mr{\nu_e},\,\mr{\mu^-},
\,\mr{\nu_\mu},\,\mr{\tau^-},\,\mr{\nu_\tau}$ is the lepton flavour index. 
We neglect inter-generational sfermion mixing. As we do not consider
the right-handed neutrino we also neglect the lepton mixing. We give
the formulae below for general generation indices. However, in HERWIG,
the mixing for the first two generations of sleptons and squarks is
not included as it is expected to be small.

In order to simplify the notation for the matrix elements we introduce
the following functions
\barr
R(\tilde{a},m_{bc}^2) &\equiv&
\frac{1}{(m_{bc}^2-M_{\tilde{a}}^2)^2 +\Gamma_{\tilde{a}}^2 
M_{\tilde{a}}^2},\\
S(\tilde{a},\tilde{b},m_{cd}^2,m_{ef}^2) &\equiv& R(\tilde{a},m_{cd}^2)
   R(\tilde{b},m_{ef}^2) 
 \left[(m_{cd}^2-M_{\tilde{a}}^2)(m_{ef}^2-M_{\tilde{b}}^2) + 
   \Gamma_{\tilde{a}} \Gamma_{\tilde{b}} M_{\tilde{a}}  M_{\tilde{b}}\right].
\earr
Here $m_{bc}^2 = (p_b+p_c)^2$, and $M_{\tilde a}$, $\Gamma_{\tilde a}$
are the mass and the decay width of the sfermion ${\tilde a}$,
respectively. The various terms in the matrix elements can be more easily
expressed in terms of 
\barr 
\Psi(\tilde{a},1,2,3) &\equiv &
R(\tilde{a},m^2_{12}) \left(m^2_{12}-m^2_1-m^2_2\right) \nonumber \\
&& \left[\left(a^2(\tilde{a})+b^2(\tilde{a})\right)
\left(M^2_0+m^2_3-m^2_{12}\right)+4a(\tilde{a})b(\tilde{a})m_3 M_0
\right],\\ 
\Upsilon(\tilde{a},1,2,3) &\equiv&
S(\tilde{a}_1,\tilde{a}_2,m_{12}^2,m_{12}^2)\left(m^2_{12}-m^2_1-m^2_2\right)
\nonumber \\ &&
\left[\left(a(\tilde{a}_1)a(\tilde{a}_2)+b(\tilde{a}_1)b(\tilde{a}_2)\right)
\left(M^2_0+m^2_3-m^2_{12}\right)\right.\nonumber\\
&&\left. +2\left(a(\tilde{a}_1)b(\tilde{a}_2)
+a(\tilde{a}_2)b(\tilde{a}_1)\right) m_3 M_0 \right],\\
\Phi(\tilde{a},\tilde{b},1,2,3)& \equiv &
S(\tilde{a},\tilde{b},m_{12}^2,m_{23}^2) \left[
m_1m_3a(\tilde{a})a(\tilde{b})
\left(m^2_{12}+m^2_{23}-m^2_1-m^2_3\right) \right.\nonumber \\ &&
+m_1M_0b(\tilde{a})a(\tilde{b})\left(m^2_{23}-m^2_2-m^2_3\right)
\nonumber \\ &&
+m_3M_0a(\tilde{a})b(\tilde{b})\left(m^2_{12}-m^2_1-m^2_2\right)
\nonumber \\ && \left.  +b(\tilde{a})b(\tilde{b})
\left(m^2_{12}m^2_{23}-m^2_1m^2_3-M^2_0m^2_2\right) \right].
\earr 
Here $\tilde{a}_1$ and $\tilde{a}_2$ are the mass eigenstates of the
relevant SUSY particle. The functions $a$ and $b$ are
gaugino-sfermion-fermion coupling constants and are given in the
tables below: Table\,\ref{tab:neutcp} for the neutralino,
Table\,\ref{tab:chargecp} for the chargino and
Table\,\ref{tab:gluinocp} for the gluino.  The couplings are defined
such that $a(\tilde{c}^*)=b(\tilde{c})$, and $b(\tilde{c}^*)=a(\tilde
{c})$.  In all the above expressions $M_0$ is the mass of the decaying
sparticle and 1, 2, 3, are the decay products.

\section{Decays}
\label{sec-DecayME}
\subsection{Sfermions}

Here we present the matrix elements for the two-body sfermion decays
including left/right mixing. In general the spin and colour averaged
matrix elements have the form
\beq
  |\overline{\me}(a \rightarrow b,c)|^2 =  C^a_{bc} (M_a^2-m_b^2-m_c^2), 
\eeq
  where $C^a_{bc}$ is the colour factor and the coupling for the
process. These factors are tabulated for the various sfermion decays
in Table\,\ref{tab:scalarcp}. $N_c$ denotes the number of colours.

\renewcommand{\arraystretch}{1.5}
\begin{table}[htp]
\begin{center}
\begin{tabular}{|l|l|l|}
\hline
Operator &  Process & Colour Factor and Coupling $C^a_{bc}$ \\
\hline
 LLE     & $\tilde{e}_{j\al}^- \longrightarrow \bar{\nu}_i \ell^-_k$ &
            $|\lam_{ijk}|^2 |L^{2j-1}_{1\al}|^2$ \\
\hline
 LLE     & $\tilde{e}_{k\al}^- \longrightarrow \nu_i \ell^-_j$ &
            $|\lam_{ijk}|^2 |L^{2k-1}_{2\al}|^2$ \\
\hline
 LLE     & $\nut_j \longrightarrow \ell^+_i \ell^-_k$ &
             $|\lam_{ijk}|^2$ \\
\hline
 LQD     & $\tilde{e}^-_{i\al} \longrightarrow \bar{u}_j d_k$ &
             $N_c|\lam_{ijk}'|^2 |L^{2i-1}_{1\al}|^2$ \\
\hline
 LQD     & $\nut_i \longrightarrow \bar{d}_j d_k$ & 
            $N_c|\lam_{ijk}'|^2$  \\
\hline
 LQD     & $\dnt_{j\al} \longrightarrow \bar{\nu}_i d_k$ &
            $|\lam_{ijk}'|^2 |Q^{2j-1}_{1\al}|^2$ \\
\hline
 LQD     &  $\upt_{j\al} \longrightarrow e_i^+ d_k$ &
            $|\lam_{ijk}'|^2 |Q^{2j}_{1\al}|^2$ \\
\hline
 LQD     &  $\dnt_{k\al} \longrightarrow \nu_i d_j$ &
            $|\lam_{ijk}'|^2 |Q^{2k-1}_{2\al}|^2$ \\
\hline
 LQD     &  $\dnt_{k\al} \longrightarrow e_i^- u_j$ &
            $|\lam_{ijk}'|^2 |Q^{2k-1}_{2\al}|^2$ \\
\hline
 UDD     &  $\upt_{i\al} \longrightarrow \bar{d}_j \bar{d}_k$ &
            $(N_c-1)! |\lam_{ijk}''|^2 |Q^{2i}_{2\al}|^2$ \\
\hline
 UDD     &  $\dnt_{k\al} \longrightarrow \bar{u}_i \bar{d}_j$ &
            $(N_c-1)! |\lam_{ijk}''|^2 |Q^{2k-1}_{2\al}|^2$ \\
\hline
\end{tabular}
\caption{Coefficients for the Scalar Decays.}
\label{tab:scalarcp}
\end{center}
\end{table}

  In all these terms the Roman indices represent the generation of the
particle and the Greek indices the mass eigenstate of the sfermions
when there is mixing. The decay rate can be obtained by
integrating over the two body phase space. This gives
\beq
  \Gamma(a \rightarrow b,c) = \frac{|\overline{\me}(a 
\rightarrow b,c)|^2 p_{cm}}  {8 \pi M_a^2},
\eeq
where $p_{cm}$ is the final-state momentum in the rest frame of the
decaying particle
\beq
   p_{cm}^2 = \frac{1}{4M_a^2}
           \left[M_a^2-(m_b+m_c)^2\right]
           \left[M_a^2-(m_b-m_c)^2\right] \nonumber.
 \eeq

\subsection{Neutralinos}

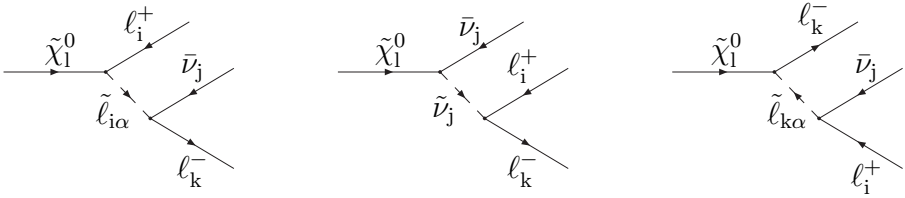
\begin{figure}[htp]
\begin{center} 
\begin{picture}(360,80)(0,0)
\SetScale{0.7}
\ArrowLine(5,78)(60,78)
\ArrowLine(105,105)(60,78)
\ArrowLine(84,53)(129,26)
\ArrowLine(129,80)(84,53)
\DashArrowLine(60,78)(84,53){5}
\Text(25,63)[]{$\mr{\cht^{0}_l}$}
\Text(55,73)[]{$\mr{\ell^+_i}$}
\Text(75,18)[]{$\mr{\ell^-_k}$}
\Text(75,56)[]{$\mr{\bar{\nu}_j}$}
\Text(45,40)[]{$\mr{\elt_{i\al}}$}
\Vertex(60,78){1}
\Vertex(84,53){1}
\ArrowLine(185,78)(240,78)
\ArrowLine(285,105)(240,78)
\ArrowLine(264,53)(309,26)
\ArrowLine(309,80)(264,53)
\DashArrowLine(240,78)(264,53){5}
\Text(150,63)[]{$\mr{\cht^{0}_l}$}
\Text(200,57)[]{$\mr{\ell^+_i}$}
\Text(200,18)[]{$\mr{\ell^-_k}$}
\Text(180,70)[]{$\mr{\bar{\nu}_j}$}
\Text(170,40)[]{$\mr{\nut_j}$}
\Vertex(240,78){1}
\Vertex(264,53){1}
\ArrowLine(365,78)(420,78)
\ArrowLine(420,78)(465,105)
\ArrowLine(489,26)(444,53)
\ArrowLine(489,80)(444,53)
\DashArrowLine(444,53)(420,78){5}
\Text(277,63)[]{$\mr{\cht^{0}_l}$}
\Text(330,15)[]{$\mr{\ell^+_i}$}
\Text(310,76)[]{$\mr{\ell^-_k}$}
\Text(330,56)[]{$\mr{\bar{\nu}_j}$}
\Text(300,40)[]{$\mr{\elt_{k\al}}$}
\Vertex(420,78){1}
\Vertex(444,53){1}
\end{picture}
\end{center}
\caption{LLE decays of the $\mr{{\tilde\chi}}$.}
\label{fig:LLEneut}
\end{figure}

\begin{figure}[htp]
\begin{center} 
\begin{picture}(360,80)(0,0)
\SetScale{0.7}
\ArrowLine(5,78)(60,78)
\ArrowLine(105,105)(60,78)
\ArrowLine(84,53)(129,26)
\ArrowLine(129,80)(84,53)
\DashArrowLine(60,78)(84,53){5}
\Text(25,63)[]{$\mr{\cht^{0}_l}$}
\Text(55,72)[]{$\mr{\bar{\nu}_i}$}
\Text(75,20)[]{$\mr{d_k}$}
\Text(75,56)[]{$\mr{\bar{d}_j}$}
\Text(45,40)[]{$\mr{\nut_{i\al}}$}
\Vertex(60,78){1}
\Vertex(84,53){1}
\ArrowLine(185,78)(240,78)
\ArrowLine(285,105)(240,78)
\ArrowLine(264,53)(309,26)
\ArrowLine(309,80)(264,53)
\DashArrowLine(240,78)(264,53){5}
\Text(150,63)[]{$\mr{\cht^{0}_l}$}
\Text(200,54)[]{$\mr{\bar{\nu}_i}$}
\Text(200,20)[]{$\mr{d_k}$}
\Text(180,72)[]{$\mr{\bar{d}_j}$}
\Text(170,40)[]{$\mr{\dnt_{j\al}}$}
\Vertex(240,78){1}
\Vertex(264,53){1}
\ArrowLine(365,78)(420,78)
\ArrowLine(420,78)(465,105)
\ArrowLine(489,26)(444,53)
\ArrowLine(489,80)(444,53)
\DashArrowLine(444,53)(420,78){5}
\Text(277,63)[]{$\mr{\cht^{0}_l}$}
\Text(330,18)[]{$\mr{\bar{\nu}_i}$}
\Text(310,74)[]{$\mr{d_k}$}
\Text(330,58)[]{$\mr{\bar{d}_j}$}
\Text(300,40)[]{$\mr{\dnt_{k\al}}$}
\Vertex(420,78){1}
\Vertex(444,53){1}
\end{picture} 
\begin{picture}(360,80)(0,0)
\SetScale{0.7}
\ArrowLine(5,78)(60,78)
\ArrowLine(105,105)(60,78)
\ArrowLine(84,53)(129,26)
\ArrowLine(129,80)(84,53)
\DashArrowLine(60,78)(84,53){5}
\Text(25,63)[]{$\mr{\cht^{0}_l}$}
\Text(55,72)[]{$\mr{\ell^{+}_i}$}
\Text(75,20)[]{$\mr{d_k}$}
\Text(75,56)[]{$\mr{\bar{u}_j}$}
\Text(45,40)[]{$\mr{\elt_{i\al}}$}
\Vertex(60,78){1}
\Vertex(84,53){1}
\ArrowLine(185,78)(240,78)
\ArrowLine(285,105)(240,78)
\ArrowLine(264,53)(309,26)
\ArrowLine(309,80)(264,53)
\DashArrowLine(240,78)(264,53){5}
\Text(150,63)[]{$\mr{\cht^{0}_l}$}
\Text(200,56)[]{$\mr{\ell^{+}_i}$}
\Text(200,20)[]{$\mr{d_k}$}
\Text(180,72)[]{$\mr{\bar{u}_j}$}
\Text(170,40)[]{$\mr{\upt_{j\al}}$}
\Vertex(240,78){1}
\Vertex(264,53){1}
\ArrowLine(365,78)(420,78)
\ArrowLine(420,78)(465,105)
\ArrowLine(489,26)(444,53)
\ArrowLine(489,80)(444,53)
\DashArrowLine(444,53)(420,78){5}
\Text(277,63)[]{$\mr{\cht^{0}_l}$}
\Text(330,17)[]{$\mr{\ell^{+}_i}$}
\Text(310,75)[]{$\mr{d_k}$}
\Text(330,56)[]{$\mr{\bar{u}_j}$}
\Text(300,40)[]{$\mr{\dnt_{k\al}}$}
\Vertex(420,78){1}
\Vertex(444,53){1}
\end{picture}
\end{center}
\caption{LQD decays of the $\mr{{\tilde\chi}}$.}
\label{fig:LQDneut}
\end{figure}
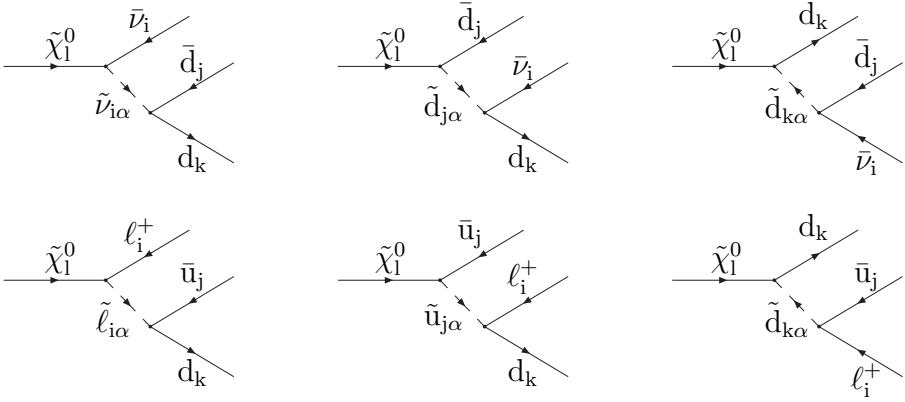

The total three-body decay rate of a photino was first computed in
\cite{Dawson:1985vr} in the limit where the sfermion is much heavier
than the decaying photino and assuming massless final states. In
\cite{Butterworth:1993tc} the general photino matrix element squared
was given, allowing for the computation of final state
distributions. In \cite{Dreiner:1994tj,Baltz:1998gd} this was extended
to the general case of a neutralino. In \cite{Baltz:1998gd} arbitrary
sfermion mixing was included as well.  We have recalculated the rates
with only left/right sfermion mixing (neglecting intergenerational
sfermion mixing).  We 
use a different convention both for the \rpv\  superpotential and the
MSSM Lagrangian, which is more appropriate to the implementation in
HERWIG. The LLE, LQD and UDD decay modes are shown in
Fig.\,\ref{fig:LLEneut}, Fig.\,\ref{fig:LQDneut} and
Fig.\,\ref{fig:UDDneut} respectively.

There are four decay modes
\begin{enumerate}
\item $\mr{ \cht^0 \longrightarrow \bar{\nu}_i \ell^+_j \ell^-_k}$,
\item $\mr{ \cht^0 \longrightarrow \bar{\nu}_i \bar{d}_j d_k}$,
\item $\mr{ \cht^0 \longrightarrow \ell^+_i \bar{u}_j d_k}$,
\item $\mr{ \cht^0 \longrightarrow \bar{u}_i \bar{d}_j \bar{d}_k}$,
\end{enumerate}
as well as their complex conjugates, since the neutralino is a
Majorana fermion.

The spin and colour averaged matrix elements are given below.
\footnote{We have a slight disagreement with \cite{Baltz:1998gd}
concerning the sign of the width of the sfermions. This is numerically
insignificant since when the sfermion is on-shell HERWIG treats this
as a two-body decay. The authors of \cite{Baltz:1998gd} agree with our
signs. We thank Paolo Gondolo for discussion of this point.}
% LLE decay rate
\barr
\lefteqn{|\overline{\me}(\cht^0_l \rightarrow \bar{\nu}_i \ell_j^
+\ell_k^-)|^2 =}  &
   \nonumber \\
%Amp Square pieces
   &&   {\lam}_{ijk}^2 \left[
       \Psi(\nut_i,\ell_j,\ell_k,\nu_i)
     +\alsm|L^{2j-1}_{1\al}|^2\Psi(\elt_{j\al},\nu_i,\ell_k,\ell_j) 
\right. \nonumber \\ 
 &&
+\alsm|L^{2k-1}_{2\al}|^2\Psi(\elt^*_{k\al},\nu_i,\ell_j,\ell_k) 
\nonumber \\
% light\heavy pieces
 &&  +2 L^{2j-1}_{11}L^{2j-1}_{12}\Upsilon(\elt_j,\nu_i,\ell_k,\ell_j) 
   +2 L^{2k-1}_{21}L^{2k-1}_{22}\Upsilon(\elt^*_k,\nu_i,\ell_j,\ell_k)
\nonumber \\
% true interference bits
  && -\alsm2L^{2j-1}_{1\al}\Phi(\elt_{j\al},\nut_i,\nu_i,\ell_k,\ell_j)
   - \alsm 2L^{2k-1}_{2\al}\Phi
(\elt^*_{k\al},\nut_i,\nu_i,\ell_j,\ell_k)  
\nonumber \\
  && \left.
  - \alsm\besm 2L^{2j-1}_{1\al}L^{2k-1}_{2\be}
                        \Phi(\elt^*_{k\be},\elt_{j\al},\ell_j,\nu_i,\ell_k) 
    \right] 
\earr
% First LQD neutralino decay rate
\barr
\lefteqn{|\overline{\me}(\cht^0_l \rightarrow \bar{\nu}_i \bar{d}_j
d_k)|^2 =} & \nonumber \\
%Amp square pieces
 && {\lam'}_{ijk}^2 N_c  \left[ 
   \Psi(\nut_i,d_j,d_k,\nu_i)
  +\alsm|Q^{2j-1}_{1\al}|^2\Psi(\dnt_{j\al},\nu_i,d_k,d_j) \right. \nonumber \\
  && +\alsm|Q^{2k-1}_{2\al}|^2\Psi(\dnt^*_{k\al},\nu_i,d_j,d_k) \nonumber \\
% light/heavy pieces
 && +2Q^{2j-1}_{11}Q^{2j-1}_{12}\Upsilon(\dnt_j,\nu_i,d_k,d_j) 
   +2Q^{2k-1}_{21}Q^{2k-1}_{22}\Upsilon(\dnt^*_k,\nu_i,d_j,d_k)\nonumber \\
% true interference bits
 && - \alsm 2Q^{2j-1}_{1\al}\Phi(\dnt_{j\al},\nut_i,\nu_i,d_k,d_j)
   - \alsm 2Q^{2k-1}_{2\al}\Phi (\dnt^*_{k\al},\nut_i,\nu_i,d_j,d_k)\nonumber \\
 && \left.- \alsm\besm2Q^{2j-1}_{1\al}Q^{2k-1}_{2\be}
                        \Phi(\dnt^*_{k\be},\dnt_{j\al},d_j,\nu_i,d_k) 
    \right] 
\earr
% Second LQD neutralino decay rate
\barr
\lefteqn{|\overline{\me}(\cht^0_l 
\rightarrow \ell^+_i \bar{u}_j d_k)|^2 =} & \nonumber \\
%Amp Square pieces
 && {\lam'}_{ijk}^2 N_c  \left[ 
  \alsm|L^{2i-1}_{1\al}|^2 \Psi(\elt_{i\al},u_j,d_k,\ell_i)
  +\alsm|Q^{2j}_{1\al}|^2 \Psi(\upt_{j\al},\ell_i,d_k,u_j) \right. \nonumber \\
 && +\alsm|Q^{2k-1}_{2\al}|^2 \Psi(\dnt^*_{k\al},\ell_i,u_j,d_k) 
% light/heavy pieces
    +2 L^{2i-1}_{11}L^{2i-1}_{12} \Upsilon(\elt_i,u_j,d_k,\ell_i) \nonumber \\
 &&  +2 Q^{2j}_{11}Q^{2j}_{12} \Upsilon(\upt_j,\ell_i,d_k,u_j) 
    +2 Q^{2k-1}_{21}Q^{2k-1}_{22}\Upsilon(\dnt^*_k,\ell_i,u_j,d_k) \nonumber \\
% true interference bits 
 &&  -\alsm\besm2L^{2i-1}_{1\al}Q^{2j}_{1\be}
                    \Phi(\upt_{j\be},\elt_{i\al},\ell_i,d_k,u_j) \nonumber \\
 && - \alsm\besm2L^{2i-1}_{1\al}Q^{2k-1}_{2\be}
                    \Phi(\dnt^*_{k\be},\elt_{i\al},\ell_i,u_j,d_k)\nonumber \\
 &&  \left.- \alsm\besm 2Q^{2j}_{1\al}Q^{2k-1}_{2\be}
                        \Phi(\dnt^*_{k\be},\upt_{j\al},u_j,\ell_i,d_k) 
    \right]
\earr
% UDD decay rate
\barr
\lefteqn{|\overline{\me}
(\cht^0_l \rightarrow \bar{u}_i \bar{d}_j \bar{d}_k)|^2 =}
 & \nonumber \\
%Amp Square pieces
 && {\lam''}_{ijk}^2 N_c!  \left[  
  \alsm|Q^{2i}_{2\al}|^2\Psi(\upt^*_{i\al},d_j,d_k,u_i)
 +\alsm|Q^{2j-1}_{2\al}|^2\Psi(\dnt^*_{j\al},u_i,d_k,d_j) \right. \nonumber \\
 && +\alsm|Q^{2k-1}_{2\al}|^2\Psi(\dnt^*_{k\al},u_i,d_j,d_k)
% light/heavy pieces
    + 2 Q^{2i}_{21}Q^{2i}_{22}\Upsilon(\upt^*_i,d_j,d_k,u_i) \nonumber \\
  && +2 Q^{2j-1}_{21}Q^{2j-1}_{22}  \Upsilon(\dnt^*_j,u_i,d_k,d_j)
    +2 Q^{2k-1}_{21}Q^{2k-1}_{22} \Upsilon(\dnt^*_k,u_i,d_j,d_k) \nonumber \\
% true interference bits
 && - \alsm\besm 2Q^{2i-1}_{2\al}Q^{2j-1}_{2\be}
                 \Phi(\dnt^*_{j\be},\upt^*_{i\al},u_i,d_k,d_j)\nonumber \\
 &&- \alsm\besm 2Q^{2i-1}_{2\al}Q^{2k-1}_{2\be}
                  \Phi(\dnt^*_{k\be},\upt^*_{i\al},u_i,d_j,d_k)\nonumber \\ 
 &&  \left.- \alsm\besm 2Q^{2j-1}_{2\al}Q^{2k-1}_{2\be}
                        \Phi(\dnt^*_{k\be},\dnt^*_{j\al},d_j,u_i,d_k) 
    \right] 
\earr

% LQD Neutralino Decay feynman diagrams
% End of the figure
\renewcommand{\arraystretch}{2.0}
\begin{table}[htp]
\begin{center}
\begin{tabular}{|l|l|}
\hline
 Coefficient & \\
\hline
 $a( \nut_i)$   & 0   \\
\hline
 $b( \nut_i)$        &  $\frac{g{N'}_{l2}}{2\cw}$ \\
\hline
 $a(\elt_{i\al})$ & $m_{\ell_i}\frac{g N_{l3}}{2M_W \cbe}L^{2i-1}_{1\al}
                  +L^{2i-1}_{2\al}\left( e  {N'}_{l1} 
                  -\frac{g  \ssw {N'}_{l2}}{\cw}\right)$\\
\hline
 $b(\elt_{i\al})$     & $m_{\ell_i}\frac{g N_{l3}}{2M_W \cbe}L^{2i-1}_{2\al}
                        -L^{2i-1}_{1\al}\left( e  {N'}_{l1} 
                       +\frac{ g{N'}_{l2} \left(\frac{1}{2}- \ssw\right) 
                         }{\cw} \right) $  \\
\hline
 $a(\dnt_{i\al})$     & $m_{d_i}\frac{g N_{l3}}{2M_W\cbe}Q^{2i-1}_{1\al}
                        -Q^{2i-1}_{2\al}\left( e e_d {N'}_{l1} 
                        -\frac{g e_d \ssw {N'}_{l2}}{\cw}\right)$ \\
\hline
 $b(\dnt_{i\al})$     & $m_{d_i}\frac{g N_{l3}}{2M_W\cbe}Q^{2i-1}_{2\al}
                         +Q^{2i-1}_{1\al}\left( e e_d {N'}_{l1} 
                      -\frac{g {N'}_{l2}\left(\frac{1}{2}+ e_d \ssw\right) 
                        }{\cw} \right)$\\
\hline
 $a(\upt_{i\al})$ & $m_{u_j}\frac{g N_{l4}}{2M_W \sbe}Q^{2j}_{1\al}
                   -Q^{2j}_{2\al} \left( e e_u {N'}_{l1} 
                    -\frac{g e_u \ssw {N'}_{l2}}{\cw}\right)$ \\
\hline
 $b(\upt_{i\al})$ & $m_{u_i}\frac{g N_{l4}}{2M_W \sbe}Q^{2i}_{2\al}
                    +Q^{2i}_{1\al}\left( e e_u {N'}_{l1} 
                    +\frac{g {N'}_{l2}\left(\frac{1}{2}- e_u \ssw\right) 
                     }{\cw} \right)$ \\
\hline
\end{tabular}
\caption{Couplings for the Neutralino Decays.}
\label{tab:neutcp}
\end{center}
\end{table}

When the neutralino mass matrix is diagonalized it is possible to get
negative eigenvalues in which case the physical field is
$\gamma_5\chi$ rather than $\chi$. This then changes the sign of some
of the coefficients in Table\,\ref{tab:neutcp}: the coefficients
$a(\tilde{c})$ change sign, and hence the coefficients
$b(\tilde{c}^*)$ also change sign.

The partial widths can be obtained from these matrix elements by
integrating over any two of $m^2_{12}$, $m^2_{23}$ and $m^2_{13}$. The
partial width is given by \cite{Caso:1998tx}
\beq
 \Gamma(0 \ra 1,2,3) = \frac1{(2\pi)^3}\frac1{32M^3_0}
             \int^{\left(m^2_{12}\right)_{max}}_{\left(m^2_{12}\right)_{min}}
                            dm^2_{12}
            \int^{\left(m^2_{23}\right)_{max}}_{\left(m^2_{23}\right)_{min}}
                            dm^2_{23}
 				|\overline{\me}|^2,
\eeq
where 
\begin{itemize} 
  \item $\left(m^2_{12}\right)_{max}=(M_0-m_3)^2$, 
  \item $\left(m^2_{12}\right)_{min}=(m_1+m_2)^2$,
  \item $\left(m^2_{23}\right)_{max}=(E^*_2+E^*_3)^2
             -\left(\sqrt{{E^*_2}^2-m^2_2}-\sqrt{{E^*_3}^2-m^2_3}\right)$ 
  \item $\left(m^2_{23}\right)_{min}=(E^*_2+E^*_3)^2
             -\left(\sqrt{{E^*_2}^2-m^2_2}+\sqrt{{E^*_3}^2-m^2_3}\right)$,
  \item $E^*_2=\left(m^2_{12}-m^2_1+m^2_2\right)/2m_{12}$ and
            $E^*_3=\left(M^2_0-m^2_{12}-m^2_3\right)/2m_{12}$
            are the energies of particles 2 and 3 in the $m_{12}$ rest frame.
\end{itemize}

\subsection{Charginos}
  
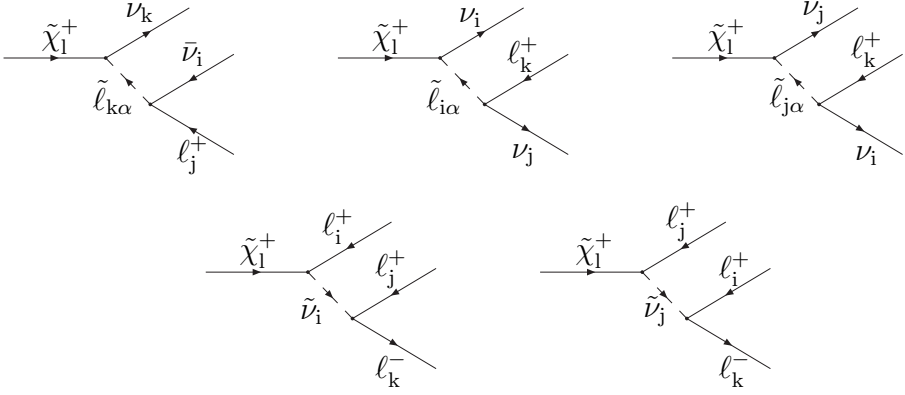
\begin{figure}[htp]
\begin{center} 
\begin{picture}(360,80)(0,0)
\SetScale{0.7}
\ArrowLine(5,78)(60,78)
\ArrowLine(60,78)(105,105)
\ArrowLine(129,26)(84,53)
\ArrowLine(129,80)(84,53)
\DashArrowLine(84,53)(60,78){5}
\Text(25,63)[]{$\mr{\cht^+_l}$}
\Text(55,72)[]{$\mr{\nu_k}$}
\Text(75,18)[]{$\mr{\ell^+_j}$}
\Text(75,56)[]{$\mr{\bar{\nu}_i}$}
\Text(45,40)[]{$\mr{\elt_{k\al}}$}
\Vertex(60,78){1}
\Vertex(84,53){1}
\ArrowLine(185,78)(240,78)
\ArrowLine(240,78)(285,105)
\ArrowLine(264,53)(309,26)
\ArrowLine(309,80)(264,53)
\DashArrowLine(264,53)(240,78){5}
\Text(150,63)[]{$\mr{\cht^+_l}$}
\Text(200,56)[]{$\mr{\ell^+_k}$}
\Text(200,18)[]{$\mr{\nu_j}$}
\Text(180,70)[]{$\mr{\nu_i}$}
\Text(170,40)[]{$\mr{\elt_{i\al}}$}
\Vertex(240,78){1}
\Vertex(264,53){1}
\ArrowLine(365,78)(420,78)
\ArrowLine(420,78)(465,105)
\ArrowLine(444,53)(489,26)
\ArrowLine(489,80)(444,53)
\DashArrowLine(444,53)(420,78){5}
\Text(277,63)[]{$\mr{\cht^+_l}$}
\Text(330,18)[]{$\mr{\nu_i}$}
\Text(310,72)[]{$\mr{\nu_j}$}
\Text(330,58)[]{$\mr{\ell^+_k}$}
\Text(300,40)[]{$\mr{\elt_{j\al}}$}
\Vertex(420,78){1}
\Vertex(444,53){1}
\end{picture} 
\begin{picture}(360,80)(0,0)
\SetScale{0.7}
\SetOffset(-50,0)
\ArrowLine(185,78)(240,78)
\ArrowLine(285,105)(240,78)
\ArrowLine(264,53)(309,26)
\ArrowLine(309,80)(264,53)
\DashArrowLine(240,78)(264,53){5}
\Text(150,63)[]{$\mr{\cht^+_l}$}
\Text(200,56)[]{$\mr{\ell^+_j}$}
\Text(200,18)[]{$\mr{\ell^-_k}$}
\Text(180,72)[]{$\mr{\ell^+_i}$}
\Text(170,40)[]{$\mr{\nut_i}$}
\Vertex(240,78){1}
\Vertex(264,53){1}
\ArrowLine(365,78)(420,78)
\ArrowLine(465,105)(420,78)
\ArrowLine(444,53)(489,26)
\ArrowLine(489,80)(444,53)
\DashArrowLine(420,78)(444,53){5}
\Text(277,63)[]{$\mr{\cht^+_l}$}
\Text(330,18)[]{$\mr{\ell^-_k}$}
\Text(310,74)[]{$\mr{\ell^+_j}$}
\Text(330,56)[]{$\mr{\ell^+_i}$}
\Text(300,40)[]{$\mr{\nut_j}$}
\Vertex(420,78){1}
\Vertex(444,53){1}
\end{picture}
\end{center}
\caption{LLE decays of the $\mr{{\tilde\chi}^+}$.}
\label{fig:LLEchar}
\end{figure}

Most of the chargino \rpv\  decay rates have already been calculated in
\cite{Dreiner:1996dd} in the case of no left/right mixing for the
first two operators in the \rpv\  superpotential. We recalculate these
rates with left/right mixing. First we consider the LLE decays of the
chargino. There are three possible decay modes:
\begin{enumerate}
\item $\mr{ \cht^+ \longrightarrow \bar{\nu}_i \ell^+_j \nu_k}$,
\item $\mr{ \cht^+ \longrightarrow \nu_i \nu_j \ell^+_k}$,
\item $\mr{ \cht^+ \longrightarrow \ell^+_i \ell^+_j \ell^-_k}$.
\end{enumerate}
The Feynman diagrams for these decays are shown in
Fig.\,\ref{fig:LLEchar}.
  The spin averaged matrix elements are given by
\barr
\lefteqn{|\overline{\me}(\cht^+_l \rightarrow \bar{\nu}_i
\ell^+_j\nu_k)|^2 
= } & \nonumber \\
%Amp Square piece
 && \frac{g^2 \lam_{ijk}^2}{2} \left[
  \alsm |L^{2k-1}_{2\al}|^2 \Psi(\elt^*_{k\al},\nu_i,\ell_j,\nu_k)
%Light/Heavy Piece
 +2L^{2k-1}_{21} L^{2k-1}_{22} 
   \Upsilon(\elt^*_k,\nu_i,\ell_j,\nu_k) \right]  
\earr
\barr
\lefteqn{|\overline{\me}(\cht^+_l \rightarrow \nu_i \nu_j \ell^+_k)|^2
   =} 
& \nonumber \\
%Amp Square pieces
 && \frac{g^2 \lam_{ijk}^2}{2} \left[ 
  \alsm |L^{2i-1}_{1\al}|^2 \Psi(\elt_{i\al},\nu_j,\ell_k,\nu_i)
 +\alsm |L^{2j-1}_{1\al}|^2
  \Psi(\elt_{j\al},\nu_i,\ell_k,\nu_j)\right. 
\nonumber \\
%light/heavy pieces
 &&  2L^{2i-1}_{11} L^{2i-1}_{12}\Upsilon(\elt_i,\nu_j,\ell_k,\nu_i)
   +2L^{2j-1}_{11} L^{2j-1}_{12}\Upsilon(\elt_j,\nu_i,\ell_k,\nu_j) 
\nonumber \\
% true interference term
 &&  \left.+\alsm\besm2  L^{2i-1}_{1\al}L^{2j-1}_{1\be} 
       \Phi(\elt_{j\be},\elt_{i\al},\nu_i,\ell_k,\nu_j) \right] 
\earr
\barr
|\overline{\me}(\cht^+_l \rightarrow \ell^+_i \ell^+_j \ell^-_k)|^2 = &
 \frac{g^2 \lam_{ijk}^2}{2} \left[ 
%Amp Square pieces 
 \Psi(\nut_i,\ell_j,\ell_k,\ell_i)
+ \Psi(\nut_j,\ell_i,\ell_k,\ell_j)\right. \nonumber \\
% interference piece
 &\left. +2 \Phi(\nut_j,\nut_i,\ell_i,\ell_k,\ell_j) \right] 
\earr
We go beyond the results of \cite{Dreiner:1996dd} to include the decay
$\mr{ \cht^+ \longrightarrow \bar{\nu}_i \ell^+_j \nu_k}$. 

We now consider the LQD decays of the chargino. There are four
possible decay modes:
\begin{enumerate}
\item $\mr{ \cht^+ \longrightarrow \bar{\nu}_i \bar{d}_j u_k}$,
\item $\mr{ \cht^+ \longrightarrow \ell^+_i \bar{u}_j u_k}$,
\item $\mr{ \cht^+ \longrightarrow \ell^+_i \bar{d}_j d_k}$,
\item $\mr{ \cht^+ \longrightarrow \nu_i u_j \bar{d}_k}$.
\end{enumerate}
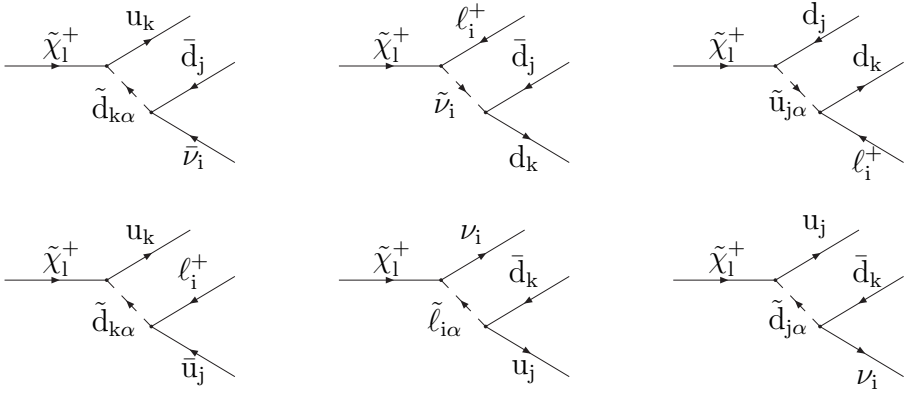
\begin{figure}[htp]
\begin{center} 
\begin{picture}(360,80)(0,0)
\SetScale{0.7}
\ArrowLine(5,78)(60,78)
\ArrowLine(60,78)(105,105)
\ArrowLine(129,26)(84,53)
\ArrowLine(129,80)(84,53)
\DashArrowLine(84,53)(60,78){5}
\Text(25,63)[]{$\mr{\cht^+_l}$}
\Text(55,72)[]{$\mr{u_k}$}
\Text(75,20)[]{$\mr{\bar{\nu}_i}$}
\Text(75,58)[]{$\mr{\bar{d}_j}$}
\Text(45,40)[]{$\mr{\dnt_{k\al}}$}
\Vertex(60,78){1}
\Vertex(84,53){1}
\ArrowLine(185,78)(240,78)
\ArrowLine(285,105)(240,78)
\ArrowLine(264,53)(309,26)
\ArrowLine(309,80)(264,53)
\DashArrowLine(240,78)(264,53){5}
\Text(150,63)[]{$\mr{\cht^+_l}$}
\Text(200,58)[]{$\mr{\bar{d}_j}$}
\Text(200,20)[]{$\mr{d_k}$}
\Text(180,72)[]{$\mr{\ell^+_i}$}
\Text(170,40)[]{$\mr{\nut_i}$}
\Vertex(240,78){1}
\Vertex(264,53){1}
\ArrowLine(365,78)(420,78)
\ArrowLine(465,105)(420,78)
\ArrowLine(489,26)(444,53)
\ArrowLine(444,53)(489,80)
\DashArrowLine(420,78)(444,53){5}
\Text(277,63)[]{$\mr{\cht^+_l}$}
\Text(330,18)[]{$\mr{\ell^+_i}$}
\Text(310,74)[]{$\mr{\bar{d}_j}$}
\Text(330,58)[]{$\mr{d_k}$}
\Text(300,40)[]{$\mr{\upt_{j\al}}$}
\Vertex(420,78){1}
\Vertex(444,53){1}
\end{picture} 
\begin{picture}(360,80)(0,0)
\SetScale{0.7}
\ArrowLine(5,78)(60,78)
\ArrowLine(60,78)(105,105)
\ArrowLine(129,26)(84,53)
\ArrowLine(129,80)(84,53)
\DashArrowLine(84,53)(60,78){5}
\Text(25,63)[]{$\mr{\cht^+_l}$}
\Text(55,72)[]{$\mr{u_k}$}
\Text(75,20)[]{$\mr{\bar{u}_j}$}
\Text(75,58)[]{$\mr{\ell^+_i}$}
\Text(45,40)[]{$\mr{\dnt_{k\al}}$}
\Vertex(60,78){1}
\Vertex(84,53){1}
\ArrowLine(185,78)(240,78)
\ArrowLine(240,78)(285,105)
\ArrowLine(264,53)(309,26)
\ArrowLine(309,80)(264,53)
\DashArrowLine(264,53)(240,78){5}
\Text(150,63)[]{$\mr{\cht^+_l}$}
\Text(200,58)[]{$\mr{\bar{d}_k}$}
\Text(200,20)[]{$\mr{u_j}$}
\Text(180,72)[]{$\mr{\nu_i}$}
\Text(170,40)[]{$\mr{\elt_{i\al}}$}
\Vertex(240,78){1}
\Vertex(264,53){1}
\ArrowLine(365,78)(420,78)
\ArrowLine(420,78)(465,105)
\ArrowLine(444,53)(489,26)
\ArrowLine(489,80)(444,53)
\DashArrowLine(444,53)(420,78){5}
\Text(277,63)[]{$\mr{\cht^+_l}$}
\Text(330,17)[]{$\mr{\nu_i}$}
\Text(310,75)[]{$\mr{u_j}$}
\Text(330,58)[]{$\mr{\bar{d}_k}$}
\Text(300,40)[]{$\mr{\dnt_{j\al}}$}
\Vertex(420,78){1}
\Vertex(444,53){1}
\end{picture}
\end{center}
\caption{LQD decays of the $\mr{{\tilde\chi}^+}$.}
\label{fig:LQDchar}
\end{figure}
  The Feynman diagrams for these decays are shown in Fig.\,\ref{fig:LQDchar}.
  The spin and colour averaged matrix elements are given below

\barr
\lefteqn{|\overline{\me}(\cht^+_l \rightarrow \bar{\nu}_i \bar{d}_j
u_k)|^2 =} & \nonumber \\
%Amp Square piece
 && \frac{g^2 {\lam'}_{ijk}^2 N_c}{2} \left[ 
  \alsm |Q^{2k-1}_{2\al}|^2 \Psi(\dnt^*_{k\al},\nu_i,d_j,u_k)
%Light/Heavy Piece
 +2 Q^{2k-1}_{21}Q^{2k-1}_{22} \Upsilon(\dnt^*_k,\nu_i,d_j,u_k) \right]
\earr
\barr
\lefteqn{|\overline{\me}(\cht^+_l \rightarrow \ell^+_i \bar{u}_j
u_k)|^2 = } & \nonumber \\
%Amp Square piece
 &&\frac{g^2{\lam'}_{ijk}^2N_c }{2}  \left[ 
 \alsm |Q^{2k-1}_{2\al}|^2\Psi(\dnt^*_{k\al},\ell_i,u_j,u_k)
%Light/Heavy Piece
 +2 Q^{2k-1}_{21}Q^{2k-1}_{22} \Upsilon(\dnt^*_k,\ell_i,u_j,u_k) \right]
\earr
\barr
\lefteqn{|\overline{\me}(\cht^+_l \rightarrow \ell^+_i \bar{d}_j
d_k)|^2 =} 
& \nonumber \\
%Amp Square pieces
 && \frac{g^2 {\lam'}_{ijk}^2 N_c}{2}  \left[
  \Psi(\nut_i,d_j,d_k,\ell_i)
 +\alsm|Q^{2j}_{1\al}|^2  \Psi(\upt_{j\al},\ell_i,d_k,d_j)\right. \nonumber \\
%light/heavy piece
 && \left. +2 Q^{2j}_{11} Q^{2j}_{12} \Upsilon(\upt_j,\ell_i,d_k,d_j) 
% true interference term
  +2\alsm  Q^{2j}_{1\al}
       \Phi(\upt_{j\al},\nut_i,\ell_i,d_k,d_j) \right] 
\earr
\barr
\lefteqn{|\overline{\me}(\cht^+_l \rightarrow \nu_i u_j \bar{d}_k)|^2
       =} 
& \nonumber \\
%Amp Square Pieces
 && \frac{g^2 {\lam'}_{ijk}^2 N_c}{2} \left[
  \alsm|L^{2i-1}_{1\al}|^2 \Psi(\elt_{i\al},u_j,d_k,\nu_i)
 +\alsm|Q^{2j-1}_{1\al}|^2 \Psi(\dnt_{j\al},\nu_i,d_k,u_j)\right. \nonumber \\
% light/heavy pieces
 &&  +2 L^{2i-1}_{11}L^{2i-1}_{12}\Upsilon(\elt_i,u_j,d_k,\nu_i) 
   +2 Q^{2j-1}_{11} Q^{2j-1}_{12} \Upsilon(\dnt_j,\nu_i,d_k,u_j) \nonumber \\
% true interference term 
 && \left. +2\alsm\besm L^{2i-1}_{1\al}Q^{2j-1}_{1\be}
       \Phi(\dnt_{j\be},\elt_{i\al},\nu_i,d_k,u_j) \right] 
\earr

We now come to the baryon number violating decays. We do not assume
here that there is only one non-zero \rpv\  coupling. This means that
more than one coupling contributes to these decays. It may seem that
this will only matter in the case where more than one $\lam''$
coupling is taken to be non-zero, however there can be more than one
diagram even with only one coupling non-zero, \eg $\lam''_{112}$ will
give two diagrams for each of the decay modes. In this case one of
these diagrams is obtained from the other simply by crossing the
identical fermions in the final state.

There are two possible decay modes:
\begin{enumerate}
\item $ \cht^+ \longrightarrow u_i u_j d_k $,
\item $ \cht^+ \longrightarrow \bar{d}_i \bar{d}_j \bar{d}_k$.
\end{enumerate}
The Feynman diagrams for these decays are shown in
Fig.\,\ref{fig:UDDchar}. The spin and colour averaged matrix elements
for these processes with left/right sfermion mixing are given below.
\barr
\lefteqn{|\overline{\me}(\cht^+_l \rightarrow u_i u_j d_k)|^2 = } & 
\nonumber \\
%Amp Square Terms
 &&\frac{g^2 N_c!}{2(1+\delta_{ij})} \left[ 
{\lam''}_{jik}^2 \alsm|Q^{2i-1}_{2\al}|^2\Psi(\dnt^*_{i\al},u_j,d_k,u_i)
+{\lam''}_{ijk}^2\alsm|Q^{2j-1}_{2\al}|^2\Psi(\dnt^*_{j\al},u_i,d_k,u_j)
\right.\nonumber \\
%Light/Heavy terms
 &&+2{\lam''}_{jik}^2Q^{2i-1}_{21}Q^{2i-1}_{22}\Upsilon(\dnt^*_i,u_j,d_k,u_i)
  +2{\lam''}_{ijk}^2Q^{2j-1}_{21}Q^{2j-1}_{22}\Upsilon(\dnt^*_j,u_i,d_k,u_j) 
\nonumber \\
% true interference term
 && \left.  + 2{\lam''}_{ijk} {\lam''}_{jik}\alsm\besm Q^{2i-1}_{2\al}
Q^{2j-1}_{2\be} \Phi(\dnt^*_{j\be},\dnt^*_{i\al},u_i,d_k,u_j) 
\right] 
\earr
\barr
\lefteqn{|\overline{\me}(\cht^+_l \rightarrow \bar{d}_i \bar{d}_j 
\bar{d}_k)|^2 =}& \nonumber \\
%Amp Square pieces
 && \frac{g^2 N_c!}{2(1+\delta_{ij}+\delta_{jk}+\delta_{ik})} \left[ 
 {\lam''}_{ijk}^2
 \alsm|Q^{2i}_{2\al}|^2\Psi(\upt^*_{i\al},d_j,d_k,d_i)
\right.\nonumber \\
 &&
 +{\lam''}_{jki}^2\alsm|Q^{2j}_{2\al}|^2\Psi(\upt^*_{j\al},d_i,d_k,d_j)
\nonumber \\
 &&+{\lam''}_{kij}^2\alsm|Q^{2k}_{2\al}|^2\Psi(\upt^*_{k\al},d_i,d_j,d_k)
%Light\Heavy pieces
+2
{\lam''}_{ijk}^2Q^{2i}_{21}Q^{2i}_{22}\Upsilon(\upt^*_i,d_j,d_k,d_i)
\nonumber \\
 &&+2 {\lam''}_{jki}^2Q^{2j}_{21}Q^{2j}_{22}\Upsilon(\upt^*_j,d_i,d_k,d_j)
 +2{\lam''}_{kij}^2Q^{2k}_{21}Q^{2k}_{22}\Upsilon(\upt^*_{k\al},d_i,d_j,d_k)
\nonumber \\
% true interference terms
 && -2{\lam''}_{ijk}{\lam''}_{jki} \alsm\besm Q^{2i}_{2\al}Q^{2j}_{2\be}
                               \Phi(\upt^*_{j\be},\upt^*_{i\al},d_i,d_k,d_j) 
\nonumber \\
 &&  - 2{\lam''}_{ijk}{\lam''}_{kij}\alsm\besm Q^{2i}_{2\al}Q^{2k}_{2\be}
                               \Phi(\upt^*_{k\be},\upt^*_{i\al},d_i,d_j,d_k) 
\nonumber \\
 && \left. - 2{\lam''}_{jki}{\lam''}_{kij}\alsm\besm Q^{2j}_{2\al}Q^{2k}_{2\be}
                               \Phi(\upt^*_{k\be},\upt^*_{j\al},d_j,d_i,d_k)
\right] 
\earr

  The coefficients in the chargino matrix elements are given in 
  Table\,\ref{tab:chargecp} and the partial widths can be obtained by
  integrating the matrix elements in the same way as for the
  neutralino decays.

\renewcommand{\arraystretch}{2.0}
\begin{table}[htp]
\begin{center}
\begin{tabular}{|l|l|l|l|}
\hline
 Coefficient &  & Coefficient  & \\
\hline
 $a(\elt_{i\al})$ & 0  &
 $b(\elt_{i\al})$ & $L^{2i-1}_{1\al} U_{l1}
                    - \frac{U_{l2} L^{2i-1}_{2\al} m_{e_i}}
                            {\sqrt{2}M_W\cos\be}$ \\
\hline
 $a(\nut_i)$ & $-\frac{U_{l2} m_{e_i}}{\sqrt{2}M_W\cos\be}$  &
 $b(\nut_i)$ & $V^*_{l1}$ \\
\hline
 $a(\upt_{i\al})$ & $-\frac{m_{d_i} U_{l2} Q^{2i}_{1\al}}
                      {\sqrt{2}M_W\cos\be}$  &
 $b(\upt_{i\al})$ & $ V^*_{l1}Q^{2i}_{1\al}
                      -\frac{m_{u_i} V^*_{l2} Q^{2i}_{2\al}}
                      {\sqrt{2}M_W\sin\be}$  \\
\hline
 $a(\dnt_{i\al})$ &  $-\frac{m_{u_i} V^*_{l2} Q^{2i-1}_{1\al}}
                      {\sqrt{2}M_W\sin\be}$ &
 $b(\dnt_{i\al})$ & $Q^{2i-1}_{1\al} U_{l1} 
                    -\frac{U_{l2} Q^{2i-1}_{2\al} m_{d_i}}
                            {\sqrt{2}M_W\cos\be}$ \\
\hline
\end{tabular}
\caption{Couplings for the Chargino Decays.}
\label{tab:chargecp}
\end{center}
\end{table}

  Again when the chargino mass matrix is diagonalized negative
  eigenvalues can be obtained and the fields must be rotated. This
  means here that the coefficients  $b(\nut_i)$, $b(\upt_{i\al})$, and
  $b(\dnt_{i\al})$ change sign if the chargino mass is negative.

\subsection{Gluinos}

\begin{figure}[htp]
\begin{center} 
\begin{picture}(360,80)(0,0)
\SetScale{0.7}
\SetOffset(-50,0)
\ArrowLine(185,78)(240,78)
\ArrowLine(285,105)(240,78)
\ArrowLine(264,53)(309,26)
\ArrowLine(309,80)(264,53)
\DashArrowLine(240,78)(264,53){5}
\Text(150,63)[]{$\mr{\tilde{g}}$}
\Text(200,56)[]{$\mr{\bar{\nu}_i}$}
\Text(200,18)[]{$\mr{d_k}$}
\Text(180,74)[]{$\mr{\bar{d}_j}$}
\Text(170,40)[]{$\mr{\dnt_{j\al}}$}
\Vertex(240,78){1}
\Vertex(264,53){1}
\ArrowLine(365,78)(420,78)
\ArrowLine(420,78)(465,105)
\ArrowLine(489,26)(444,53)
\ArrowLine(489,80)(444,53)
\DashArrowLine(444,53)(420,78){5}
\Text(277,63)[]{$\mr{\tilde{g}}$}
\Text(330,18)[]{$\mr{\bar{\nu}_i}$}
\Text(310,74)[]{$\mr{d_k}$}
\Text(330,56)[]{$\mr{\bar{u}_j}$}
\Text(300,40)[]{$\mr{\dnt_{k\al}}$}
\Vertex(420,78){1}
\Vertex(444,53){1}
\end{picture}
\begin{picture}(360,80)(0,0)
\SetScale{0.7}
\SetOffset(-50,0)
\ArrowLine(185,78)(240,78)
\ArrowLine(285,105)(240,78)
\ArrowLine(264,53)(309,26)
\ArrowLine(309,80)(264,53)
\DashArrowLine(240,78)(264,53){5}
\Text(150,63)[]{$\mr{\tilde{g}}$}
\Text(200,56)[]{$\mr{\ell^+_i}$}
\Text(200,18)[]{$\mr{d_k}$}
\Text(180,74)[]{$\mr{\bar{u}_j}$}
\Text(170,40)[]{$\mr{\upt_{j\al}}$}
\Vertex(240,78){1}
\Vertex(264,53){1}
\ArrowLine(365,78)(420,78)
\ArrowLine(420,78)(465,105)
\ArrowLine(489,26)(444,53)
\ArrowLine(489,80)(444,53)
\DashArrowLine(444,53)(420,78){5}
\Text(277,63)[]{$\mr{\tilde{g}}$}
\Text(330,18)[]{$\mr{\ell^+_i}$}
\Text(310,74)[]{$\mr{d_k}$}
\Text(330,56)[]{$\mr{\bar{u}_j}$}
\Text(300,40)[]{$\mr{\dnt_{k\al}}$}
\Vertex(420,78){1}
\Vertex(444,53){1}
\end{picture}
\caption{LQD decays of the $\mr{{\tilde{g}}}$.}
\label{fig:LQDgluino}
\end{center}
\end{figure}
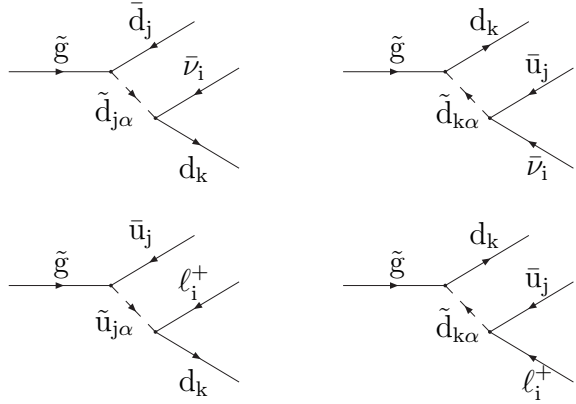

These decay rates are calculated here with left/right mixing. There
are three possible decay modes, two via the LQD operator and one via
the UDD operator:
\begin{enumerate}
\item $\mr{ \tilde{g} \longrightarrow \bar{\nu}_i \bar{d}_j d_k}$,
\item $\mr{ \tilde{g} \longrightarrow \ell^+_i \bar{u}_j d_k }$,
\item $\mr{ \tilde{g} \longrightarrow u_i u_j d_k }$.
\end{enumerate}
Since the gluino is a Majorana fermion the charge conjugate decay
modes are possible as well.  The Feynman diagrams for these processes
are in Fig.\,\ref{fig:LQDgluino} and Fig.\,\ref{fig:UDDgluino},
respectively.  The spin and colour averaged matrix elements with
left/right sfermion mixing are given below.
% First LQD gluino decay rate
\barr
\lefteqn{|\overline{\me}(\tilde{g} \rightarrow \bar{\nu}_i \bar{d}_j
d_k)|^2 =} & \nonumber \\
%Amp Square terms
 && \frac{ {\lam'}_{ijk}^2 g_s^2}{2} \left[ 
 \alsm|Q^{2j-1}_{1\al}|^2\Psi(\dnt_{j\al},\nu_i,d_k,d_j) 
  +\alsm|Q^{2k-1}_{2\al}|^2\Psi(\dnt^*_{k\al},\nu_i,d_j,d_k) \right. 
\nonumber \\
% light/heavy pieces
 && +2Q^{2j-1}_{11}Q^{2j-1}_{12}\Upsilon(\dnt_j,\nu_i,d_k,d_j) 
   +2Q^{2k-1}_{21}Q^{2k-1}_{22}\Upsilon(\dnt^*_k,\nu_i,d_j,d_k)\nonumber \\
% true interference bits
 && \left.- \alsm\besm2Q^{2j-1}_{1\al}Q^{2k-1}_{2\be}
                        \Phi(\dnt^*_{k\be},\dnt_{j\al},d_j,\nu_i,d_k) 
    \right] 
\earr
% Second LQD gluino decay rate
\barr
\lefteqn{|\overline{\me}(\tilde{g} \rightarrow \ell^+_i \bar{u}_j
d_k)|^2 =} & \nonumber \\
 %Amp Square
 &&\frac{{\lam'}_{ijk}^2 g_s^2}{2}  \left[ 
  \alsm|Q^{2j}_{1\al}|^2 \Psi(\upt_{j\al},\ell_i,d_k,u_j) 
  +2 Q^{2j}_{11}Q^{2j}_{12} \Upsilon(\upt_j,\ell_i,d_k,u_j) \right.
\nonumber \\ 
% light/heavy pieces
 &&   +\alsm|Q^{2k-1}_{2\al}|^2 \Psi(\dnt^*_{k\al},\ell_i,u_j,d_k)
    +2 Q^{2k-1}_{21}Q^{2k-1}_{22}\Upsilon(\dnt^*_k,\ell_i,u_j,d_k) \nonumber \\
% true interference bits 
 &&  \left.- \alsm\besm 2Q^{2j}_{1\al}Q^{2k-1}_{2\be}
                        \Phi(\dnt^*_{k\be},\upt_{j\al},u_j,\ell_i,d_k) 
    \right] 
\earr
% UDD decay rate
\barr
\lefteqn{|\overline{\me}(\tilde{g} \rightarrow \bar{u}_i \bar{d}_j 
\bar{d}_k)|^2 = }  & \nonumber \\
%Amp Square Terms
 &&
 \frac{{\lam''}_{ijk}^2 (N_c-1)!}{2}  \left[ \alsm|Q^{2i}_{2\al}|^2\Psi(
\upt^*_{i\al},d_j,d_k,u_i)
 +2 Q^{2i}_{21}Q^{2i}_{22}\Upsilon(\upt^*_i,d_j,d_k,u_i) \right. \nonumber \\
 && +\alsm|Q^{2j-1}_{2\al}|^2\Psi(\dnt^*_{j\al},u_i,d_k,d_j)
% light/heavy pieces
+2 Q^{2j-1}_{21}Q^{2j-1}_{22}  \Upsilon(\dnt^*_j,u_i,d_k,d_j)
     \nonumber \\
  && +\alsm|Q^{2k-1}_{2\al}|^2\Psi(\dnt^*_{k\al},u_i,d_j,d_k) 
    +2 Q^{2k-1}_{21}Q^{2k-1}_{22} \Upsilon(\dnt^*_k,u_i,d_j,d_k) \nonumber \\
% true interference bits
 && + \frac{1}{N_c-1}\alsm\besm Q^{2i}_{2\al}Q^{2j-1}_{2\be}
                        \Phi(\dnt^*_{j\be},\upt^*_{i\al},u_i,d_k,d_j)
\nonumber \\
 && +\frac{1}{N_c-1}\alsm\besm Q^{2i}_{2\al}Q^{2k-1}_{2\be}
                        \Phi(\dnt^*_{k\be},\upt^*_{i\al},u_i,d_j,d_k)
\nonumber \\ 
 &&  +\left. \frac{1}{N_c-1}\alsm\besm Q^{2j-1}_{2\al}Q^{2k-1}_{2\be}
                        \Phi(\dnt^*_{k\be},\dnt^*_{j\al},d_j,u_i,d_k) 
    \right] 
\earr
The coefficients for these matrix elements are given in
Table\,\ref{tab:gluinocp}. As the gluino mass is not obtained by
diagonalising a mass matrix it cannot be negative.

\renewcommand{\arraystretch}{2.0}
\begin{table}[htp]
\begin{center}
\begin{tabular}{|l|l|l|l|}
\hline
 Coefficient &  & Coefficient  & \\
\hline
 $a(\upt_{i\al})$ &  $Q^{2i}_{2\al}$ &
 $b(\upt_{i\al})$ & $-Q^{2i}_{1\al}$ \\
\hline
 $a(\dnt_{i\al})$ & $Q^{2i-1}_{2\al}$ &
 $b(\dnt_{i\al})$ & $-Q^{2i-1}_{1\al}$\\
\hline
\end{tabular}
\caption{Couplings for the Gluino Decays.}
\label{tab:gluinocp}
\end{center}
\end{table}

%
% Hard Processes
%
\section{Hard Processes}

All of the single neutralino, chargino and gluino production cross
sections can be obtained by crossing from the decay matrix elements we
have already presented in Section~\ref{sec-DecayME}. This crossing
will lead to the invariants $m^2_{12}$, $m^2_{23}$, and $m^2_{13}$
being replaced by the usual invariants $s$, $t$, and $u$. There is an
overall sign change due to exchanging fermions between the initial and
final states. It should also be remembered that the decay matrix
elements given have been averaged over the spin and colour of the
initial particle. The cross-sections for the remaining processes are
presented below. In all cases the formulae have been averaged over the
initial spins and colours. The initial state masses have all been set
to zero, except where they appear in a coupling constant. In $t$- and
$u$-channel fermion propagators the fermion masses have been neglected
as well.

%
%LQD processes
%
\subsection{LQD Processes}

%
%  LQD Weak processes
%
\subsubsection{Resonant Slepton followed by Weak Decay}

  There are three processes which can occur via the production of a
resonant slepton followed by a weak decay of this slepton. These are:
\begin{enumerate}
\item $\mr{ d_j \bar{d}_k \longrightarrow \elt^*_i W^- }$,
\item $\mr{ u_j \bar{d}_k \longrightarrow \nut^*_i W^+ }$,
\item $\mr{ u_j \bar{d}_k \longrightarrow \tilde{\tau}_1^* Z^0 }$.
\end{enumerate}
  It should be noted that we have not included processes where the
  resonance is not accessible, \eg
  $\mr{u_j \bar{d}_k \longrightarrow \tilde{\tau}_2^* Z^0}$.
\barr
% sneutrino to charged slepton and W
\lefteqn{|\overline{\me}(d_j \bar{d}_k \ra
\elt^*_{i\al} W^-)|^2 = } & \nonumber\\
 & \frac{g^2{\lam'}_{ijk}^2 |L^{2i-1}_{1\al}|^2}{2 M^2_W N_c}\left[
	\sh^2 p^2_{cm} R(\nut_i,\sh) \right.
	+\frac{1}{4\uh^2} \left(2M^2_W(\uh\tha-M^2_{\elt_{i\al}} M^2_W)
						+\uh^2\sh\right) \nonumber \\ 
 & \left.+\frac{\sh\left(\sh-M^2_{\nut_i}\right)R(\nut_i,\sh)}{2\uh}
	\left(M^2_W(2M^2_{\elt_{i\al}}-\uh)+\uh(\sh-M^2_{\elt_{i\al}})
\right)\right]\\
% charged slepton to sneutrino and W
\lefteqn{|\overline{\me}(u_j \bar{d}_k   \rightarrow \nut^*_i W^+)|^2
 =} & 
 \nonumber\\
& \frac{g^2{\lam'}_{ijk}^2}{2 M^2_W N_c}\left[
      \alsm  |L^{2i-1}_{1\al}|^4 \sh^2 p^2_{cm} R(\elt_{i\al},\sh) 
     +2|L^{2i-1}_{11}|^2 |L^{2i-1}_{12}|^2 \sh^2 p^2_{cm} S(\elt_{i1},
\elt_{i2},\sh,\sh)
     \right. \nonumber \\
&+\frac{1}{4\uh^2} \left(2M^2_W(\uh\tha-M^2_{\nut_i}M^2_W) +\uh^2\sh\right) 
  \nonumber \\ 
& \left.  +\frac{|L^{2i-1}_{1\al}|^2
	\sh\left(\sh-M^2_{\elt_{i\al}}\right) R(\elt_{i\al},\sh)}{2\uh} 
  \left(M^2_W(2M^2_{\nut_i}-\uh)+\uh(\sh-M^2_{\nut_i})\right)   \right] \\
% stau_2 to stau_1 and Z 
\lefteqn{|\overline{\me}(u_j \bar{d}_k \rightarrow  \elt^*_{i1}
  Z^0)|^2 =} &
  \nonumber\\
  &		\frac{g^2{\lam'}_{ijk}^2}{N_c M_Z^2\cos^2 \theta_w} \left[
    \alsm |L^{2i-1}_{1\al}|^2 |Z^{\al1}_{\ell_i}|^2 \sh^2 p^2_{cm} 
R(\elt_{i\al},\sh^2)
  +2L^{2i-1}_{11}L^{2i-1}_{12}Z^{11}_{\ell_i}Z^{21}_{\ell_i} \sh^2 p^2_{cm}
    S(\elt_{i1},\elt_{i2},\sh,\sh) \right. \nonumber \\
 & +\frac{|L^{2i-1}_{11}|^2 Z_{u_L}^2}{\uh^2}
	\left(2M_Z^2(\uh\tha-M^2_{\elt_{i1}}M_Z^2) +\uh^2\sh\right)
   +\frac{|L^{2i-1}_{11}|^2 Z_{d_R}^2}{\tha^2}
	\left(2M_Z^2(\uh\tha-M^2_{\elt_{i1}}M_Z^2) +\tha^2\sh\right) 
\nonumber \\
  & +\alsm \frac{L^{2i-1}_{1\al} L^{2i-1}_{11}Z^{\al1}_{\ell_i} Z_{u_L}
        \sh \left(\sh-M^2_{\elt_{i\al}}\right)R(\elt_{i\al},\sh^2) }{\uh}
	\left( M_Z^2(2M^2_{\elt_{i1}}-\uh) +\uh
  (s-M^2_{\elt_{i1}})\right) \nonumber \\
  & +\alsm \frac{L^{2i-1}_{1\al} L^{2i-1}_{11}Z^{\al1}_{\ell_i} Z_{d_R}
        \sh  \left(\sh-M^2_{\elt_{i\al}}\right)R(\elt_{i\al},\sh^2)}{\tha}
	\left( M_Z^2(2M^2_{\elt_{i1}}-\tha) +\tha(s-M^2_{\elt_{i1}})
\right)\nonumber\\
  &\left. -\frac{2 |L^{2i-1}_{11}|^2  Z_{u_L} Z_{d_R}}{\uh\tha}
	\left( 2M_Z^2(M^2_{\elt_{i1}}-\tha)(M^2_{\elt_{i1}}-\uh) 
-\sh\tha\uh \right)
	\right]
\earr
where in all the above equations
\beq
   p_{cm}^2 = \frac{1}{4\sh^2}
           \left[\sh-(m_1+m_2)^2\right]
           \left[\sh-(m_1-m_2)^2\right] \nonumber,
\eeq
and $m_1$, $m_2$ are the masses of the final state particles.

\begin{table}[htp]
\begin{center}
\begin{tabular}{|c|c|c|c|}
\hline
\multicolumn{4}{|c|}{Slepton Couplings} \\
\hline
 $ Z^{\al\be}_{\nu_i}$ & $-\frac{1}{2}\delta_{\al=1,\be=1}$ &
 $ Z^{\al\be}_{\ell_i}$ & $\frac{1}{2}\left(L^{2i-1}_{1\al}L^{2i-1}_{1\be}
			-2\ssw\delta_{\al\be}\right)$ \\
\hline
\multicolumn{4}{|c|}{Squark Couplings} \\
\hline
 $ Z^{\al\be}_{u_i}$ & $\frac{1}{2}\left(-Q^{2i}_{1\al}Q^{2i}_{1\be}
			+2e_u\ssw\delta_{\al\be}\right)$ &
 $ Z^{\al\be}_{d_i}$ & $\frac{1}{2}\left(Q^{2i-1}_{1\al}Q^{2i-1}_{1\be}
			+2e_d\ssw\delta_{\al\be}\right)$ \\
\hline
\multicolumn{4}{|c|}{Quark Couplings} \\
\hline
 $Z_{u_L}$ & $-\frac{1}{4}\left(1-2e_u\ssw\right)$ & 
 $Z_{d_L}$ & $ \frac{1}{4}\left(1+2e_d\ssw\right)$ \\ 
 $Z_{u_R}$ & $ \frac{1}{2}e_u\ssw$ & 
 $Z_{d_R}$ & $\frac{1}{2}e_d\ssw$\\ 
\hline
\end{tabular}
\end{center}
\caption{Couplings of Sleptons, Squarks and Quarks to the $Z^0$.}
\end{table}
%
% LQD Rslash processes
%
\subsubsection{Resonant Slepton followed by \boldmath{\rpv}\  Decay}
  There are four process it which we can produce a resonant slepton
via \rpv\   which then decays back to Standard Model particles via a \rpv\   
decay. These are:
\begin{enumerate}
\item $\mr{ d_j \bar{d}_k \longrightarrow d_l \bar{d}_m} $,
\item $\mr{ u_j \bar{d}_k \longrightarrow u_l \bar{d}_m} $,
\item $\mr{ d_j \bar{d}_k \longrightarrow \ell^-_l \ell^+_m}$,
\item $\mr{ u_j \bar{d}_k \longrightarrow \nu_l \ell^+_m} $.
\end{enumerate}
  The first two of these process only require non-zero LQD couplings
whereas the second two require both non-zero LQD and LLE couplings.
 The matrix elements are presented below for an arbitrary number of
non-zero \rpv\  couplings
% Resonant sneutrino to quarks
\barr
|\overline{\me}(d_j \bar{d}_k \rightarrow d_l \bar{d}_m)|^2 &= &
 \frac{1}{4}\displaystyle\sum_{i,n=1,3}
 {\lam'}_{ijk}{\lam'}_{ilm}{\lam'}_{njk}{\lam'}_{nlm} 
S(\nut_i,\nut_n,\sh,\sh)
   \sh\left(\sh-m^2_{d_l}-m^2_{d_m} \right) \nonumber \\
&& +\frac{1}{4}\displaystyle\sum_{i,n=1,3} 
  {\lam'}_{ijl}{\lam'}_{ikm}{\lam'}_{njl}{\lam'}_{nkm}
  \frac{(m^2_{d_l}-\hat{t})(m^2_{d_m}-\hat{t})}
{(\hat{t}-M^2_{\nut_i})(\hat{t}-M^2_{\nut_n})} \\
&& \nonumber \\
% Resonant slepton to quarks
|\overline{\me}(u_j \bar{d}_k \rightarrow u_l \bar{d}_m)|^2& = &
 \frac{1}{4}\displaystyle \sum_{\al,\be=1,2}\displaystyle\sum_{i,n=1,3}
 {\lam'}_{ijk}{\lam'}_{ilm}{\lam'}_{njk}{\lam'}_{nlm}
 |L^{2i-1}_{1\al}|^2|L^{2n-1}_{1\be}|^2 \nonumber \\
&&  
  S(\elt_{i\al},\elt_{n\be},\sh,\sh)
   \sh\left(\sh-m^2_{u_l}-m^2_{d_m} \right) \\
&& \nonumber \\
% Resonant sneutrino to leptons
|\overline{\me}(d_j \bar{d}_k \rightarrow \ell^-_l \ell^+_m)|^2 &= &
 \frac{1}{4N_c}\displaystyle\sum_{i,n=1,3}
 {\lam'}_{ijk}{\lam'}_{njk}{\lam}_{ilm}{\lam}_{nlm} 
S(\nut_i,\nut_n,\sh,\sh)
   \sh\left(\sh-m^2_{\ell_m}-m^2_{\ell_l} \right) \nonumber\\&&\\
&& \nonumber \\
% Resonant slepton to leptons
|\overline{\me}(u_j \bar{d}_k \rightarrow \nu_l \ell^+_m)|^2 &= &
 \frac{1}{4N_c}\displaystyle \sum_{\al,\be=1,2}\displaystyle\sum_{i,n=1,3}
 {\lam'}_{ijk}{\lam'}_{njk}{\lam}_{ilm}{\lam}_{nlm}
 |L^{2i-1}_{1\al}|^2|L^{2n-1}_{1\be}|^2  \nonumber \\
& & 
  S(\elt_{i\al},\elt_{n\be},\sh,\sh)
   \sh\left(\sh-m^2_{\ell_m} \right)
\earr

%
% LQD Higgs Processes
% 
\subsubsection{Resonant Slepton Production followed by Higgs Decay}

There are a number of processes which can occur via the production of a
resonant slepton which can then decay to either a neutral or charged
Higgs:
\begin{enumerate}
\item $\mr{ u_j \bar{d}_k \longrightarrow \elt^*_{i\beta} h_0/H_0/A_0}$,
\item $\mr{ d_j \bar{d}_k \longrightarrow \elt^*_{i\alpha} H^-}$,
\item $\mr{ u_j \bar{d}_k \longrightarrow \nut^*_i H^+}$.
\end{enumerate}
As we only include processes where there is a possibility of a
resonant production mechanism, the process $\mr{ d_j \bar{d}_k
\longrightarrow \nut^*_i h_0/H_0/A_0}$ is not included. For the same
reason we also have not included the processes $\mr{ u_j \bar{d}_k
\longrightarrow \elt^*_{iL} h_0/H_0/A_0}$ for the first two slepton
generations. This is because HERWIG does not include mixing for the first two generation
 sleptons and
the initial state only couples to the left-handed slepton. The process $\mr{ u_j
\bar{d}_k \longrightarrow \elt^*_{i2} h_0/H_0/A_0}$ is also not
included for the third generation ($i=3$) as there is no accessible
resonance.

\begin{table}
\begin{center}
\begin{tabular}{|c|c|c|c|}
\hline
 $U^1_i$ & $ \frac{m_{u_i}\ca}{2M_W \sbe}$ &
 $D^1_i$ & $-\frac{m_{d_i}\sa}{2M_W\cbe}$ \\
 $U^2_i$ & $ \frac{m_{u_i}\sa}{2M_W \sbe}$ & 
 $D^2_i$ & $ \frac{m_{d_i}\ca}{2M_W\cbe}$ \\
 $U^3_i$ & $ \frac{m_{u_i}\cot\be}{2M_W}$ &
 $D^3_i$ & $ \frac{m_{d_i}\tan\be}{2M_W}$ \\
 $U^c_i$ & $ \frac{m_{u_i}\cot\be}{2\sqrt{2}M_W}$ &
 $D^c_i$ & $ \frac{m_{d_i}\tan\be}{2\sqrt{2}M_W}$ \\
\hline
\end{tabular}
\end{center}
\caption{Higgs Couplings to quarks.}
\label{tab:higgsqk}
\end{table}

The matrix elements for these processes are given below. Since the
matrix elements have the same form for all the neutral Higgs processes
we use the notation $\mr{H^l_0}$ where $l$=1,2,3 is $\mr{ h_0}$,
$\mr{H_0}$ and $\mr{A_0}$.
\barr
% sneutrino to charged slepton and H
\lefteqn{|\overline{\me}(d_j \bar{d}_k \rightarrow  \elt^*_{i\al} 
H^-)|^2 =}\nonumber\\
 & \frac{g^2 {\lam'}_{ijk}^2}{4N_c}\left[
	 |H^c_{\nut\elt_{i\al}}|^2 \sh  R(\nut_i,\sh)
	+\frac{4|L^{2i-1}_{1\al}|^2|D^c_j|^2}{\uh^2}
	 \left(\uh\tha-M^2_{ \elt_{i\be}}M^2_{H^-}\right) \right] \\
 & \nonumber \\
% charged slepton to sneutrino and H
\lefteqn{|\overline{\me}(u_j \bar{d}_k   \rightarrow \nut^*_i H^+)|^2
=} \nonumber\\
\nopagebreak[4]
 & \frac{g^2{\lam'}^2_{ijk}\sh}{4 N_c}\left[
	\alsm |L^{2i-1}_{i\al}|^2 |H^c_{\nut\elt_{i\al}}|^2
		\sh  R(\elt_{i\al}) 
	+ 2  L^{2i-1}_{i1}L^{2i-1}_{i2} H^c_{\nu\elt_{i1}}H^c_{\nu\elt_{i2}}
		\sh  S(\elt_{i1},\elt_{i2},\sh,\sh) 
\right. \nonumber \\
 & \left. +\frac{4|U^c_j|^2}{\uh^2}\left(\uh\tha-M^2_{\nut_i}M^2_{H^+}
 \right)\right] \\
 & \nonumber \\
% stau_2 to stau_1 and H 
\lefteqn{|\overline{\me}(u_j \bar{d}_k \rightarrow \elt^*_{i\be}
H^l_0)|^2 =}
\nonumber\\
 & \frac{g^2 {\lam'}_{ijk}^2}{4N_c}\left[
	\alsm  |L^{2i-1}_{i\al}|^2  |H^l_{\elt_{i\al}\elt_{i\be}}|^2
		\sh  R(\elt_{i\al})
	+2  L^{2i-1}_{i1}L^{2i-1}_{i2}
		 H^l_{\elt_{i1}\elt_{i\be}}H^l_{\elt_{i2}\elt_{i\be}}
		\sh S(\elt_{i1},\elt_{i2},\sh,\sh) 
 \right .\nonumber \\
 & \left. +\frac{|L^{2i-1}_{1\be}|^2|D^l_j|^2}{\uh^2}
		\left(\uh\tha-M^2_{ \elt_{i\be}}M^2_{H^l_0}\right)
 	  +\frac{|L^{2i-1}_{1\be}|^2|D^l_k|^2}{\tha^2}
		 \left(\uh\tha-M^2_{ \elt_{i\be}}M^2_{H^l_0}\right) \right]
\earr
The couplings involved in the various processes can be found in
Tables\,\ref{tab:higgsqk} and \ref{tab:LQDhiggs}.

\begin{table}[htp]
\begin{center}
\begin{tabular}{|l|l|}
\hline
 Coefficient &   \\
\hline
  $H^1_{\elt_{i\al}\elt_{i\be}}$ &
  $ -\frac{M_Z\sin(\al+\be)}{\cw} \left[ 
    L^{2i-1}_{1\al}L^{2i-1}_{1\be}( \frac{1}{2}-\ssw )
    +\ssw L^{2i-1}_{2\al}L^{2i-1}_{2\be} \right]$ \\
 & $ +\frac{m_{e_i}^2\sa}{M_W \cbe}\left[
     L^{2i-1}_{1\al}L^{2i-1}_{1\be}+ L^{2i-1}_{2\al}L^{2i-1}_{2\be}
 \right] $ \\
 & $ -\frac{m_{e_i}}{2M_W \cbe}\left( \mu\ca+A_{e_i}\sa\right)
    \left[ L^{2i-1}_{2\al}L^{2i-1}_{1\be}+ L^{2i-1}_{1\al}L^{2i-1}_{2\be}
 \right] $  \\
\hline
  $H^2_{\elt_{i\al}\elt_{i\be}}$ &
  $ \frac{M_Z\cos(\al+\be)}{\cw} \left[ 
    L^{2i-1}_{1\al}L^{2i-1}_{1\be}( \frac{1}{2}-\ssw )
    +\ssw L^{2i-1}_{2\al}L^{2i-1}_{2\be} \right]$ \\
 & $ -\frac{m_{e_i}^2\ca}{M_W \cbe}\left[
     L^{2i-1}_{1\al}L^{2i-1}_{1\be}+ L^{2i-1}_{2\al}L^{2i-1}_{2\be}
 \right] $ \\
 & $ -\frac{m_{e_i}}{2M_W \cbe}\left( \mu\sa-A_{e_i}\ca\right)
    \left[ L^{2i-1}_{2\al}L^{2i-1}_{1\be}+ L^{2i-1}_{1\al}L^{2i-1}_{2\be}
 \right] $  \\
\hline
  $H^3_{\elt_{i\al}\elt_{i\be}}$ &
  $ \de_{\al\neq\be}\frac{m_{e_i}}{2M_W}\left(\mu+A_{e_i}\tan\beta\right) $  \\
\hline
  $H^c_{\nut\elt_{i\al}}$ &
   $\frac{1}{\sqrt{2}M_W}\left[
	L^{2i-1}_{1\al}\left(m_{e_i}^2\tan\beta-M_W^2\sin2\beta\right)
   	-L^{2i-1}_{2\al}m_{e_i}\left(\mu+\tan\beta A_{e_i}\right)\right]$\\
\hline
\end{tabular}
\caption{Higgs couplings to Sleptons.} 
\label{tab:LQDhiggs}
\end{center}
\end{table}

\subsection{UDD Processes}
\subsubsection{Resonant Squark followed by Weak Decay}

   There are four process which can occur via the production of a
resonant squark followed by a weak decay of this squark:
\begin{enumerate}
\item $\mr{ d_j d_k \longrightarrow \dnt^*_i W^- }$,
\item $\mr{ d_j d_k \longrightarrow \tilde{t}^*_1 Z^0 }$,
\item $\mr{ u_i d_j \longrightarrow \upt^*_k W^+ }$,
\item $\mr{ u_i d_j \longrightarrow \tilde{b}^*_1 Z^0 }$.
\end{enumerate}
  Again we do not include processes where the resonance is not
  accessible, \ie  $\mr{ d_j d_k \longrightarrow \tilde{t}^*_2 Z^0 }$ and
  $\mr{ u_i d_j \longrightarrow \tilde{b}^*_2 Z^0}$.
  So the matrix elements for these processes are given by
\barr
% sup to sdown and W
\lefteqn{|\overline{\me}(d_j d_k \rightarrow \dnt^*_{i\be} W^-)|^2 
=} & \nonumber\\
 & \frac{g^2 {\lam''}_{ijk}^2 (N_c-1)!|Q^{2i-1}_{1\be}|^2\sh^2 p^2_{cm}}
   {2 N_c M^2_W}\left[
    \alsm |Q^{2i}_{2\al}|^2 |Q^{2i}_{1\al}|^2
  R(\upt_{i\al},\sh) \right. \nonumber \\
 & \left.+2 Q^{2i}_{21}Q^{2i}_{22}Q^{2i}_{11}Q^{2i}_{12}
  S(\upt_{i1},\upt_{i2},\sh,\sh) \right]\\
 & \nonumber \\
% sdown to sup and W
\lefteqn{|\overline{\me}(u_i d_j \rightarrow \upt^*_{k\be} W^+)|^2 =}
& 
\nonumber \\
 & \frac{g^2 {\lam''}_{ijk}^2 (N_c-1)!\sh^2 p^2_{cm}|Q^{2i}_{1\be}|^2}
  {2 N_c M^2_W} \left[
 |Q^{2i-1}_{2\al}|^2 |Q^{2i-1}_{1\al}|^2 R(\dnt_{k\al},\sh) \right.
\nonumber \\
 &  \left.+2Q^{2i-1}_{21}Q^{2i-1}_{22}Q^{2i-1}_{11}Q^{2i-1}_{12}
 S (\dnt_{k1},\dnt_{k2},\sh,\sh)  \right] \\
 & \nonumber \\
% stop_2 to stop_1 and Z 
\lefteqn{|\overline{\me}(d_j d_k \rightarrow \upt^*_{i1} Z^0)|^2 =}& 
\nonumber \\
 & \frac{g^2 {\lam''}_{ijk}^2 (N_c-1)!}{N_c M_Z^2\cos^2\theta_w}\left[
	\alsm |Q^{2i}_{2\al}|^2|Z^{\al1}_{u_i}|^2 \sh^2 p^2_{cm} 
R(\upt_{i\al},\sh^2)
 	+2 Q^{2i}_{21}Q^{2i}_{22}Z^{11}_{u_i}Z^{21}_{u_i}\sh^2 p^2_{cm}
	 S(\upt_{i1},\upt_{i2},\sh,\sh) \right.  \nonumber \\
 & +\frac{|Q^{2i}_{21}|^2Z_{d_R}^2}{\uh^2}
	\left(2M_Z^2(\uh\tha-M^2_{\upt_{i1}}M_Z^2) +\uh^2\sh\right)
       +\frac{|Q^{2i}_{21}|^2Z^2_{d_R}}{\tha^2}
	\left(2M_Z^2(\uh\tha-M^2_{\upt_{i1}}M_Z^2) +\tha^2\sh\right) 
\nonumber \\
 & +\alsm\frac{Q^{2i}_{2\al}Q^{2i}_{21}Z^{\al1}_{u_i}Z_{d_R}}{\uh}
	 \sh(\sh-M^2_{\upt_{i\al}})R(\upt_{i\al},\sh^2)
	 \left(M_Z^2(2M^2_{\upt_{i1}}-\uh) +\uh(\sh-M^2_{\upt_{i1}})\right)
 \nonumber \\
 &  +\alsm\frac{Q^{2i}_{2\al}Q^{2i}_{21}Z^{\al1}_{u_i}Z_{d_R}}{\tha}
	 \sh(\sh-M^2_{\upt_{i\al}})R(\upt_{i\al},\sh^2)
	 \left(M_Z^2(2M^2_{\upt_{i1}}-\tha) +\tha(\sh-M^2_{\upt_{i1}})\right)
 \nonumber\\
 & \left.-\frac{2|Q^{2i}_{21}|^2}{\uh\tha}
  	 \left(2M_Z^2(M^2_{\upt_{i1}}-\uh)(M^2_{\upt_{i1}}-\tha)-
\sh\tha\uh\right)
 \right] \\
% bottom_2 to bottom_1 and Z 
\lefteqn{|\overline{\me}(u_i d_k \rightarrow \dnt^*_{j1} Z^0)|^2 =} & 
\nonumber \\
 & \frac{g^2 {\lam''}_{ijk}^2 (N_c-1)!}{N_c M_Z^2 \cos^2\theta_w}\left[
 	\alsm|Q^{2j-1}_{2\al}|^2|Z^{\al1}_{d_j}|^2  \sh^2 p^2_{cm} 
R(\dnt_{j\al},\sh^2) 
        +2Q^{2j-1}_{21}Q^{2j-1}_{22}Z^{11}_{d_j}Z^{21}_{d_j}\sh^2 p^2_{cm}
	 S(\dnt_{j1},\dnt_{j2},\sh^2,\sh^2)
\right.\nonumber \\
 & +\frac{|Q^{2i-1}_{21}|^2Z_{u_R}^2}{\uh^2}
	\left(2M_Z^2(\uh\tha-M^2_{\dnt_{j1}}M_Z^2)+\uh^2\sh\right)
+\frac{|Q^{2i-1}_{21}|^2Z_{d_R}^2}{\tha^2}
	\left(2M_Z^2(\uh\tha-M^2_{\dnt_{j1}}M_Z^2)+\tha^2\sh\right)
\nonumber \\
 & +\alsm\frac{Q^{2i-1}_{2\al}Q^{2i-1}_{21}Z^{\al1}_{d_j}Z_{u_R}}{\uh}
	\sh(\sh-M^2_{\dnt_{j\al}})R(\dnt_{j\al},\sh^2)
	\left(M_Z^2(2M^2_{\dnt_{j1}}-\uh) +\uh(\sh-M^2_{\dnt_{j1}})\right)
\nonumber \\
 & +\alsm\frac{Q^{2i-1}_{2\al}Q^{2i-1}_{21}Z^{\al1}_{d_j}Z_{d_R}}{\tha}
	\sh(\sh-M^2_{\dnt_{j\al}})R(\dnt_{j\al},\sh^2)
	\left(M_Z^2(2M^2_{\dnt_{j1}}-\tha) +\tha(\sh-M^2_{\dnt_{j1}})\right)
\nonumber \\
 & \left.-\frac{2|Q^{2i-1}_{21}|^2Z_{u_R}Z_{d_R}}{\uh\tha}
	\left(2M_Z^2(M^2_{\dnt_{j1}}-\uh)(M^2_{\dnt_{j1}}-\tha)-
\sh\tha\uh\right)\right]
\earr
   where $\sh$ and $p_{cm}$ are as before.

\subsubsection{Resonant Squark followed by \boldmath{\rpv}\  Decay}

  There are two processes in which a resonant squark is produced via
the \bv\  term in the superpotential and these squarks then decay to
Standard Model particles:
\begin{enumerate}
\item $\mr{ d_j d_k \longrightarrow  d_l d_m   }$,
\item $\mr{ u_i d_j \longrightarrow  u_l d_m  }$.
\end{enumerate}
  The matrix elements are given by
\barr
% sup to quarks
|\overline{\me}(d_j d_k \rightarrow d_l d_m)|^2 = 
 & \frac{(N_c-1)!^2}{4N_c}\displaystyle
 \sum_{\al,\be=1,2}\displaystyle
\sum_{i,n=1,3}
  {\lam''}_{ijk}{\lam''}_{ilm}{\lam''}_{njk}{\lam''}_{nlm}
 |Q^{2i}_{2\al}|^2|Q^{2n}_{2\be}|^2 \\
&  S(\upt_{i\al},\upt_{n\be},\sh,\sh)
   \sh\left(\sh-m^2_{d_l}-m^2_{d_m} \right)\nonumber  \\
 & \nonumber \\
% sdown to quarks 
|\overline{\me}(u_i d_j \rightarrow u_l d_m)|^2 = 
 & \frac{(N_c-1)^2}{4N_c}\displaystyle \sum_{\al,\be=1,2}\displaystyle
\sum_{i,n=1,3}
  {\lam''}_{ijk}{\lam''}_{lmk}{\lam''}_{ijn}{\lam''}_{lmn}
 |Q^{2k-1}_{2\al}|^2|Q^{2n-1}_{2\be}|^2
 \\
& S(\dnt_{i\al},\dnt_{n\be},\sh,\sh)
   \sh\left(\sh-m^2_{u_l}-m^2_{d_m} \right)   \nonumber
\earr
\subsubsection{Resonant Squark Production followed by Higgs Decay}
  
There are a number of processes which occur via the production of a
resonant squark which subsequently decays to either a neutral or
charged Higgs. Again we only consider those processes for which a
resonance is possible, \ie we neglect the processes 
 $\mr{ d_j d_k \longrightarrow \tilde{u}^*_{iR} h_0/H_0/A_0 }$ and
 $\mr{ u_i d_j \longrightarrow \tilde{d}^*_{iR}  h_0/H_0/A_0 }$
  for the first two generations and the processes 
 $\mr{ d_j d_k \longrightarrow \tilde{t}^*_{i2} h_0/H_0/A_0 }$ and
 $\mr{ u_i d_j \longrightarrow \tilde{b}^*_{i2}  h_0/H_0/A_0 }$
for the third generation, where we consider left/right
mixing as these process cannot occur via a resonant diagram:
\begin{enumerate}
\item $\mr{ d_j d_k \longrightarrow \dnt^*_i H^- }$,
\item $\mr{ d_j d_k \longrightarrow \tilde{u}^*_{i1} h_0/H_0/A_0 }$,
\item $\mr{ u_i d_j \longrightarrow \upt^*_k H^+ }$,
\item $\mr{ u_i d_j \longrightarrow \tilde{d}^*_{i1}  h_0/H_0/A_0 }$.
\end{enumerate}
The matrix elements for these processes are given below. Due to our
notation for the squark mixing matrices in the case of no left/right
mixing the right squark is denoted as the second mass eigenstate.
\begin{table}[htp]
\begin{center}
\begin{tabular}{|l|l|} \hline
 Coefficient &   \\
\hline
  $H^1_{\dnt_{i\al}\dnt_{i\be}}$ &
  $ -\frac{M_Z\sin(\al+\be)}{\cw} \left[ 
    Q^{2i-1}_{1\al}Q^{2i-1}_{1\be}( \frac{1}{2}+e_d\ssw )
    -e_d \ssw Q^{2i-1}_{2\al}Q^{2i-1}_{2\be} \right]$ \\
 & $ +\frac{m_{d_i}^2\sa}{M_W \cbe}\left[
     Q^{2i-1}_{1\al}Q^{2i-1}_{1\be}+ Q^{2i-1}_{2\al}Q^{2i-1}_{2\be} 
 \right] $ \\
 & $ -\frac{m_{d_i}}{2M_W \cbe}\left( \mu\ca+A_{d_i}\sa\right)
    \left[ Q^{2i-1}_{2\al}Q^{2i-1}_{1\be}+
Q^{2i-1}_{1\al}Q^{2i-1}_{2\be}
\right] $  \\
\hline
  $H^1_{\upt_{i\al}\upt_{i\be}}$ &
  $ \frac{M_Z\sin(\al+\be)}{\cw} \left[ 
    Q^{2i}_{1\al}Q^{2i}_{1\be}( \frac{1}{2}-e_u\ssw )
    +e_u \ssw Q^{2i}_{2\al}Q^{2i}_{2\be} \right]$ \\
 & $ -\frac{m_{u_i}^2\ca}{M_W \sbe}\left[
     Q^{2i}_{1\al}Q^{2i}_{1\be}+ Q^{2i}_{2\al}Q^{2i}_{2\be}  \right] $ \\
 & $ +\frac{m_{u_i}}{2M_W \sbe}\left( \mu\sa+A_{u_i}\ca\right)
    \left[ Q^{2i}_{2\al}Q^{2i}_{1\be}+ Q^{2i}_{1\al}Q^{2i}_{2\be}\right] $  \\
\hline
  $H^2_{\dnt_{i\al}\dnt_{i\be}}$ &
  $ \frac{M_Z\cos(\al+\be)}{\cw} \left[ 
    Q^{2i-1}_{1\al}Q^{2i-1}_{1\be}( \frac{1}{2}+e_d\ssw )
    -e_d\ssw Q^{2i-1}_{2\al}Q^{2i-1}_{2\be} \right]$ \\
 & $ -\frac{m_{d_i}^2\ca}{M_W \cbe}\left[
     Q^{2i-1}_{1\al}Q^{2i-1}_{1\be}+ Q^{2i-1}_{2\al}Q^{2i-1}_
{2\be} \right] $ \\
 & $ -\frac{m_{d_i}}{2M_W \cbe}\left( \mu\sa-A_{d_i}\ca\right)
    \left[ Q^{2i-1}_{2\al}Q^{2i-1}_{1\be}+
Q^{2i-1}_{1\al}Q^{2i-1}_{2\be} \right] $  \\
\hline
  $H^2_{\upt_{i\al}\upt_{i\be}}$ &
  $ -\frac{M_Z\cos(\al+\be)}{\cw} \left[ 
    Q^{2i}_{1\al}Q^{2i}_{1\be}( \frac{1}{2}-e_u\ssw )
    +e_u \ssw Q^{2i}_{2\al}Q^{2i}_{2\be} \right]$ \\
 & $ -\frac{m_{u_i}^2\sa}{M_W \sbe}\left[
     Q^{2i}_{1\al}Q^{2i}_{1\be}+ Q^{2i}_{2\al}Q^{2i}_{2\be}  \right] $ \\
 & $ -\frac{m_{u_i}}{2M_W \sbe}\left( \mu\sa-A_{u_i}\ca\right)
    \left[ Q^{2i}_{2\al}Q^{2i}_{1\be}+Q^{2i}_{1\al}Q^{2i}_{2\be}\right] $  \\
\hline
  $H^3_{\dnt_{i\al}\dnt_{i\be}}$ &
  $ \de_{\al\neq\be}\frac{m_{d_i}}{2M_W}\left(\mu+A_{d_i}\tan\beta\right) $  \\
\hline
  $H^3_{\upt_{i\al}\upt_{i\be}}$ &
  $ \de_{\al\neq\be}\frac{m_{u_i}}{2M_W}\left(\mu+A_{u_i}\cot\beta\right) $  \\
\hline
  $H^c_{\upt_{i\al}\dnt_{i\be}}$ &
   $\frac{1}{\sqrt{2}M_W}\left[Q^{2i}_{1\al}Q^{2i-1}_{1\be} \left(
      m_{d_i}^2\tan\beta +m_{u_i}^2\cot\beta-M_W^2\sin2\beta\right)\right.$ \\
 & $+Q^{2i}_{2\al}Q^{2i-1}_{2\be}m_{u_i}m_{d_i}\left(\cot
\beta+\tan\beta\right)$ \\
 & $ \left.-Q^{2i}_{1\al}Q^{2i-1}_{2\be}m_{d_i} \left(\mu+ A_{d_i}\tan
\beta \right)
     -Q^{2i}_{2\al}Q^{2i-1}_{1\be}m_{u_i} \left(\mu+ A_{u_i}\cot\beta 
\right) \right]$\\
\hline
\end{tabular}
\caption{Higgs couplings to Squarks.}
\label{tab:UDDhiggs}
\end{center}
\end{table}
\barr
% sup to sdown and H^-
\lefteqn{|\overline{\me}(d_j d_k \rightarrow \dnt^*_{i\be} H^-)|^2 
= } & \nonumber \\
 & \frac{g^2 (N_c-1)!}{4N_c} \left[ 
	\alsm  {\lam''}_{ijk}^2|Q^{2i}_{2\al}|^2|H^c_{\upt_{i\al}\dnt_
{i\be}}|^2 
		\sh R(\upt_{i\al},\sh) 
	+ 2{\lam''}_{ijk}^2 Q^{2i}_{21}Q^{2i}_{22}
		H^c_{\upt_{i1}\dnt_{i\be}}H^c_{\upt_{i2}\dnt_{i\be}}
 	       	\sh S(\upt_{i1},\upt_{i2},\sh,\sh)
\right. \nonumber \\
 & \left.+\frac{4 {\lam''}_{jik}^2|U^c_j|^2|Q^{2i-1}_{2\be}|^2}{\uh^2}
		\left(\uh\tha-M^2_{\dnt_{i\be}}M^2_{H^-}\right)
  +\frac{4 {\lam''}_{kij}^2|U^c_k|^2|Q^{2i-1}_{2\be}|^2}{\tha^2}
		\left(\uh\tha-M^2_{\dnt_{i\be}}M^2_{H^-}\right)
 \right]\\
 & \nonumber \\
% sdown to sup and H^+
\lefteqn{|\overline{\me}(u_i d_j \rightarrow \upt^*_{k\be} H^+)|^2 = }
& \nonumber \\
 & \frac{g^2(N_c-1)!}{4N_c} \left[ 
	\alsm {\lam''}_{ijk}^2|Q^{2k-1}_{2\al}|^2H^c_{\upt_{k\be}\dnt_
{k\al}}|^2
		\sh  R(\dnt_{k\al},\sh)
	+2{\lam''}_{ijk}^2 Q^{2k-1}_{21} Q^{2k-1}_{22}
          	H^c_{\upt_{k\be}\dnt_{k1}}H^c_{\upt_{k\be}\dnt_{k2}}
            	\sh S(\dnt_{k1},\dnt_{k2},\sh,\sh)
\right. \nonumber \\
 & \left.+\frac{4 {\lam''}_{kij}^2|D^c_i|^2|Q^{2k-1}_{2\be}|^2}{\uh^2}
		\left(\uh\tha-M^2_{\dnt_{k\al}}M^2_{ H^+}\right)
\right] \\
 & \nonumber \\
% stop_2 to stop_1 and h_0 
\lefteqn{|\overline{\me}(d_j d_k \rightarrow \upt^*_{i1} H^l_0)|^2 = }
& \nonumber \\
 & \frac{g^2{\lam''}_{ijk}^2(N_c-1)!}{4N_c}\left[
	\alsm|Q^{2i}_{2\al}|^2|H^l_{\upt_{i\al}\upt_{i1}}|^2 \sh 
R(\upt_{i\al},\sh)
	+2Q^{2i}_{21}Q^{2i}_{22}H^l_{\upt_{i1}\upt_{i1}}H^l_
{\upt_{i2}\upt_{i1}}
 	 \sh S(\upt_{i1},\upt_{i2},\sh,\sh) \right.
\nonumber \\
 & \left. +\frac{|Q^{2i}_{21}|^2|D^l_j|^2}{\tha^2}
		\left(\uh\tha-M^2_{\upt_{i1}}M_{ H^l_0}^2\right)
	  +\frac{|Q^{2i}_{21}|^2|D^l_k|^2}{\uh^2}
		\left(\uh\tha-M^2_{\upt_{i1}}M_{ H^l_0}^2\right)\right] \\
 & \nonumber \\
% bottom_2 to bottom_1 and h_0 
\lefteqn{|\overline{\me}(u_i d_k \rightarrow \dnt^*_{j1} H^l_0)|^2 = 
} & \nonumber \\
 & \frac{g^2{\lam''}_{ijk}^2(N_c-1)!}{4 N_c}\left[
	\alsm |Q^{2j-1}_{2\al}|^2|H^l_{\dnt_{j\al}\dnt_{j1}}|^2 
\sh R(\dnt_{j\al},\sh)
+2Q^{2j-1}_{21}Q^{2j-1}_{22}H^l_{\dnt_{j1}\dnt_{j1}}H^l_{\dnt_{j2}\dnt_{j1}}
	 \sh S(\dnt_{j1},\dnt_{j2},\sh,\sh) \right.
\nonumber \\
 & \left. +\frac{|Q^{2j-1}_{21}|^2|U^l_i|^2}{\tha^2}
		\left(\uh\tha-M^2_{\dnt_{j1}}M_{ H^l_0}^2\right)
	  +\frac{|Q^{2j-1}_{21}|^2|D^l_k|^2}{\uh^2}
		\left(\uh\tha-M^2_{\dnt_{j1}}M_{ H^l_0}^2\right)\right]  
\earr

  The coefficients for the various processes can be found in 
  Tables\, \ref{tab:higgsqk} and \ref{tab:UDDhiggs}.

%\bibliography{Rpaper}

\end{document}